\documentclass[mnsc,nonblindrev]{informs3_SSRN}

\OneAndAHalfSpacedXI

\usepackage{amsmath,amssymb,amsfonts}

\usepackage{natbib}
\bibpunct[, ]{(}{)}{,}{a}{}{,}%
\def\bibfont{\small}%

\usepackage{rotating}
\usepackage{fancyvrb}

\TheoremsNumberedThrough     
\ECRepeatTheorems
\JOURNAL{Operations Research}

\usepackage{multimedia}
\usepackage{multicol}
\usepackage{bbm}
\usepackage{graphics}
\usepackage{graphicx}
\usepackage{booktabs}
\usepackage{romannum}
\usepackage{grffile}
\usepackage{float}
\usepackage{url}
\usepackage{mathabx}
\usepackage{supertabular}
\usepackage{rotating}
\usepackage{pdflscape}
\usepackage{verbatim}
\usepackage{listings}
\usepackage{xcolor}
\usepackage{multicol}
\usepackage[shortlabels]{enumitem} 
\usepackage{caption,subcaption}
\usepackage{epstopdf}
\usepackage{algorithm}
\usepackage{algorithmic}
\usepackage{enumitem}

\usepackage{xr}
\makeatletter
\let\XRBibcite\bibcite
\let\bibcite\@gobbletwo
\let\bibcite\XRBibcite
\makeatother

\newcommand{\dd}{\mathrm{d}}
\newcommand{\one}{\mathbbm{1}}
\newcommand{\E}{\mathbb{E}}
\newcommand{\p}{\mathbb{P}}
\newcommand{\q}{\mathbb{Q}}

\newcommand{\f}{\mathcal{F}}

\EquationsNumberedThrough    

\MANUSCRIPTNO{OPRE-2024-05-1034.R3}

\begin{document}
	\pagenumbering{arabic}
	
	\RUNAUTHOR{Fuh, Jia, and Kou}
	
	\RUNTITLE{Importance Sampling with Markov Random Walks}
	
	\TITLE{A General Framework for Importance Sampling with Markov Random Walks}
	
	\ARTICLEAUTHORS{%
		\AUTHOR{Cheng-Der Fuh}
		
		\AFF{Graduate Institute of Statistics, National Central University, Taoyuan City 320317, Taiwan  \EMAIL{ cdffuh@gmail.com}}
		
		\AUTHOR{Yanwei Jia}
		\AFF{Department of Systems Engineering and Engineering Management, The Chinese University of Hong Kong, Shatin, N.T., Hong Kong \EMAIL{yanweijia@cuhk.edu.hk}} 
	
	\AUTHOR{Steven Kou}
	\AFF{Department of Finance, Questrom School of Business, Boston University, Boston, MA 02215, USA \EMAIL{ kou@bu.edu}}
}

\ABSTRACT{%
	Although stochastic models driven by latent Markov processes are widely used, the classical importance sampling methods based on the exponential tilting for these models suffer from difficulties in computing associated eigenvalues and  eigenfunctions and the plausibility of the indirect asymptotic large deviation regime for the variance of the estimator. We propose an importance sampling framework that twists the observable and latent processes separately using a link function that directly minimizes the estimator's variance. An optimal link function is chosen within the locally asymptotically normal family. We show the logarithmic efficiency of the estimator. As applications, we estimate an overflow probability under a pandemic model and the CoVaR, a measurement of the co-dependent financial systemic risk; both are beyond the scope of traditional importance sampling methods due to nonlinear features.
}


\SUBJECTCLASS{Simulation: systems dynamics, applications; Probability: Markov processes
}

\AREAOFREVIEW{Simulation.}
\maketitle

\section{Introduction}
\label{sec:1}

Stochastic models with latent processes
are ubiquitous.\footnote{From a simulation perspective, nothing is really ``latent'' since we need to simulate the whole system. We use this terminology to highlight that a ``latent process'' is not our target.} 
For example, when estimating the risk measure of a portfolio, the portfolio value, as the process of interest, may be affected by the volatilities of each asset, which are latent. Other examples with latent variables include production modes in supply chains and inventory management, and unobserved sub-populations during a pandemic; see, e.g., \cite{tang2008markov}. Calculating small probabilities
of rare events, e.g., default and overflow events, is indispensable to risk management and public policy issues. 
Due to the complexity of such stochastic models, simulation methods are useful.\footnote{Simulation-based methods are widely used in derivative pricing \citep{glasserman1999asymptotically,haugh2004pricing}, portfolio evaluation \citep{glasserman2002portfolio,haugh2006evaluating}. For a review of applications in financial engineering, see \cite{haugh2003duality,glasserman2004monte}. Applications in other fields include inventory planning and routing \citep{caceres2012combining}, and epidemiology \citep{hamra2013markov}.}
Since plain Monte Carlo methods are inefficient in computing small probabilities because the estimator's variance gradually dominates its mean, importance sampling was proposed by \citet{siegmund1976importance} as a variance reduction technique in rare-event simulation for the case of independently and identically distributed (i.i.d.) random variables. The basic idea is to sample under an alternative distribution by changing the probability measure; see, e.g., \cite{asmussen2007stochastic} for a comprehensive introduction. 

The research question of this paper is: Can we design a general importance sampling algorithm to do rare-event simulation for stochastic models with latent processes? For tractability, we shall restrict to Markovian systems and attempt to provide a general importance sampling framework that is easy to implement for models known as \textit{Markov random walks}. Moreover, we are especially interested in events of a moderately small probability 
($10^{-2}$ to $10^{-4}$) that are common in finance, economics, hypothesis testing, and epidemic applications. The basic idea is to enlarge the search space of the sampling distribution by tilting the latent and observed processes separately and using a suitably chosen link function.

The simulation time for estimating a moderately small probability is crucial for many applications \citep{glasserman2002portfolio, juneja2006rare, BassJuneja2008, fuh2011efficient}. For example, a financial firm needs to evaluate the risk profile associated with thousands of portfolio positions on a daily basis to meet the stress-testing requirement of Basel regulations, while the probabilities of interest are often moderately small. Hence, even a moderate improvement in either variance reduction or computational time in each simulation can have a huge impact because the number of simulations needed can be massive.

{\bf Motivation and Our Contribution.}
For general Markov random walks, 
\cite{collamore2002importance} proposes a method to approximate the optimal change of measure by using an exponential tilting method based on the large deviation (LD) 
principle.
See \cite{ney1987markov1,ney1987markov2} for the LD principle.
However, two challenges remain. (1) Since the classical LD-based tilting only aims at maximizing the exponential decay rate of the variance,
not the original variance or asymptotic variance, the tilting measure given by the LD may not be close to the optimal one for the moderately small ($10^{-2}$ to $10^{-4}$) probabilities. (2) It is generally difficult to calculate the eigenvalues and eigenfunctions required by the exponential tilting method. 
If the latent Markov process has a finite state space, one needs to find the eigenvalue of a (non-negative) matrix. 
However, it becomes a functional eigenvalue problem in an infinite-dimensional space when the underlying latent Markov process has a continuous general state space.
Only affine processes are known to admit log-affine eigenfunctions analytically (cf. \citealt{GlassermanKim2010, zhang2015affine}).

The contributions of the paper are twofold.
First, theoretically, we propose a duo-exponential tilting framework to minimize the estimator’s variance directly (instead of only the exponential decay rate of the variance).  The idea is to change the distribution of the underlying and observed processes {\it separately} via a link function, instead of changing the two distributions {\it jointly} as in the LD-based tilting methods. The proposed tilting family is large enough to include the classical exponential tilting (e.g., \citealt{collamore2002importance}) as a special case,
implying that, in theory, the variance reduction of our method should be as good as existing methods. 
We provide an asymptotically optimal link function under the local asymptotic normal (LAN) family.
Since  the asymptotically optimal link function is related to the derivative of the eigenfunction in the classical exponential tilting algorithms at only one point (zero),  there is no need to find the whole eigenfunction. 
We show the logarithmic efficiency of the proposed estimator under the LAN family, and discuss how to choose the link function outside the LAN family. \footnote{This paper is also related to other strands of literature. First, 
	the study of moderate deviation rare event simulations also aims to minimize the estimator's variance directly instead of only the exponential decay rate of the variance.
	See \cite{do1991importance,fuh2004efficient, fuh2011efficient} for i.i.d. random variables, and \cite{andradottir1995choice, andradottir1996potentially, fuh2007estimation} for finite-state Markov chains. They are special cases of our general framework because we consider Markov random walks. Second, when searching for the desirable tilting parameters, our algorithm minimizes the estimator's variance using a pre-sampling stochastic approximation algorithm. The stochastic approximation has been used in importance sampling for some special Markov processes; see, e.g., 
	\citet{egloff2005optimal, ahamed2006adaptive, kawai2009optimal}. 
	We complement this strand of literature by studying general state space Markov processes and treating the search step as a separate stage (like \citealt{rubinstein2004cross}) rather than an adaptive procedure. 
}
The LAN-based criterion and the role of the Poisson equation are related to the moderate-deviation importance sampling literature, especially \citet{fuh2007estimation}. Our contribution is to construct a link function that induces a larger duo-exponential tilting family and then to optimize within that family, rather than to identify a single sampling distribution.

Second, in terms of application, the generality of our family enables us to study complicated nonlinear problems to which the current exponential tilting method cannot be applied. 
(1) We apply our method to a compartmental pandemic model (SIRD model, short for susceptible-infectious-recovered-deceased, cf., \citealt{allen2017primer}) to estimate the overflow probability of the hospitalized population, where the evolution of the (observed) quantity of interest
(infectious) also depends on other latent processes (susceptible) in a nonlinear way. Due to the nonlinear structure, the usual eigenfunction-based importance sampling techniques are not applicable. (2) We apply our method to compute the conditional value-at-risk (CoVaR) in \cite{adrian2016covar} under the vector autoregression model with generalized autoregressive conditional heteroskedasticity (VAR-GARCH) to study systemic co-dependence motivated by the 2008 financial crisis. Due to the conditioning and multivariate features, the joint event may not be in the standard form of large deviation theory. However, since the model is within the LAN family, we can provide the asymptotically optimal link function for importance sampling. To the best of our knowledge, this is the only paper that addresses importance sampling techniques for these two applications with complicated models.

The managerial insights are twofold. First, our pandemic model simulation suggests that the infected population's overflow probability is sensitive to the transmission rate. 
Even a one-basis-point change can dramatically affect the probability. Thus, any marginal effort to limit transmission may have a huge impact, and this calls for more accurate estimation of such parameters to better evaluate the risk associated with a pandemic. Second, our CoVaR simulation confirms the empirical finding in \cite{adrian2016covar} that VaR and CoVaR have no significant cross-sectional relationship at a fixed quantile level. Hence, institutions' exposure to systemic risk may not align with individual risk measures. However, 
our study points to a new observation regarding tail dependence: An institution's VaR and CoVaR may follow a similar trend with respect to the level of risk, but such a trend varies across different institutions, suggesting that an individual-specific factor could approximate the cross-sectional dependence well.

\textbf{Comparison to the State-Dependent Importance Sampling Methods.}
There is a large body of literature that studies the state-dependent importance sampling methods; see, e.g., \citet{dupuis2004importance,dupuis2005dynamic,dupuis2007subsolutions,dupuis2007dynamic,blanchet2006state,blanchet2008state,blanchet2012lyapunov}. For a general review, see \citet{blanchet2012state}. 
However, there are still difficulties. (1) Finding a subsolution to an Isaacs equation can be challenging; often, the subsolution is non-smooth (a piecewise affine function, as suggested in \citealt{dupuis2007subsolutions}) and needs to be modified. (2) The exponential tilting family is still the building block of the state-dependent importance sampling method. Thus, when applied to Markov random walk models, such as the ones considered in this paper, it still requires the eigenvalues and eigenfunctions to be analytically known or easily computable. Meanwhile, there is a middle ground between fully explicit subsolutions and complete intractability: Approximate subsolutions and related numerical approximations can also be used to define effective state-dependent changes of measure; see, e.g., \citet{djehiche2022importance,hult2024deep,nusken2021solving}. A comparison with such approaches is given in EC.7.
(3) It requires the underlying Markov chain to be uniformly recurrent (cf. \citealt{dupuis2005dynamic}), which is stronger than all assumptions we made in this paper. 
(4) Computationally, it is usually more time-consuming to calculate the state-dependent tilting parameters and the subsolutions. 
(5) When the event of interest is simple (e.g., being in a convex set), the state-dependent importance sampling method degenerates to the conventional LD-based exponential tilting with a state-independent tilting parameter. 
Thus, for the Heston model in Section \ref{subsec:sv example}, it coincides with \citet{collamore2002importance}. When the models are complicated, e.g.,  the applications in Section \ref{sec:5},  
the state-dependent importance sampling methods are not applicable because the eigenvalues and eigenfunctions are not available. Our proposed method outperforms a state-dependent IS method for a well-known tandem queue model \citep{glasserman1995analysis,dupuis2005dynamic,dupuis2007dynamic,
	dupuis2007subsolutions,blanchet2012lyapunov} as shown in Section \ref{subsec:tandem queue}. 

\textbf{Comparison to the Importance Sampling for Diffusion Processes.} There are studies on the importance sampling methods for diffusion processes, such as \citet{newton1994variance,fournie1997small,dupuis2012importance}, and those specifically for finance applications include \citet{guasoni2008optimal,fouque2002variance}. 
(1) These methods assume the diffusion is in a specific ``small noise" regime so that its limit approaches some deterministic system. 
The resulting Hamilton-Jacobi-type of first-order partial differential equation (PDE) is derived by approximating an elliptic PDE associated with the original diffusion process (cf. \citealt{fournie1997small,dupuis2012importance}). 
Such an approximation may not be valid beyond the above-mentioned limiting regime. 
For example, the SIRD diffusion model considered in our paper does not belong to this class of diffusion processes with the required scaling property. 
(2) The subsolution approach (e.g., \citealt{dupuis2004importance,dupuis2012importance}) is still difficult to apply for complex models because even finding a subsolution to a nonlinear first-order PDE can be computationally challenging. 
Instead, our approach uses a parametric family to directly approximate the solution to the original, unscaled, second-order PDE. 
For the particular SIRD model, the linkage with the subsolution approaches is given in E-Companion \ref{appendix_control}.

\section{Basic Setting and Review of the Classical Exponential Tilting }
\label{sec:2}

\subsection{The framework of Markov random walks}\label{sec2.1}

Consider a general framework 
of a discrete-time latent Markov process, called the Markov random walk, cf. 
\cite{ney1987markov1}, and \cite{meyn2012markov}. The observable process is a random walk: $S_n=\sum_{i=1}^n Y_i \in {\mathbb R}^d$, where the distribution of increment is governed by a latent Markov process 
$\{X_n,n\geq 0\}$ on a general state space $\mathcal{X}$, with transition probability kernel $p(x,\dd x')$. For $A\subset \mathcal{X}, B\subset \mathbb{R}^d$,
\[ \p(X_{n+1}\in A | X_n=x) = \int_A p(x,\dd x'),\ \p(Y_{n+1}\in B|X_n=x,X_{n+1}=x') = \int_B \rho(y|x,x')\dd y , \]
where $\rho(y|x,x')$ is the conditional probability density of $Y_{n+1}$ given $(X_n, X_{n+1})$ with respect to 
the Lebesgue measure on $\mathbb{R}^d$.
Then $\{(X_n,S_n), n \geq 0\}$ is a Markov process with transition probability kernel
\begin{equation}
	\label{eq:mcrw model}
	(X_{n+1},S_{n+1})\in (\dd x',\dd s') |X_n=x,S_n=s \sim p(x,\dd x')\rho(s'-s|x,x')\dd s' .
\end{equation}

The problem of interest is to estimate the expectation of a functional with an initial distribution $Law(X_0,S_0)=\nu \otimes \delta_0 $ (or $\delta_x \otimes \delta_0$, where $\otimes$ stands for the product measure), i.e.,
\begin{eqnarray}\label{exp} 
	\E_{\nu}[F(S_{\tau})]\ (\text{ or }\E_{x}[F(S_{\tau})] ), 
\end{eqnarray}
where $\nu$ is an initial distribution of the Markov chain $\{X_n, n \geq 0\}$, $F$ is a measurable function on $\mathbb{R}^d$, and $\tau$ is a stopping time with respect to the natural filtration of $\{(X_n,S_n), n \geq 0\}$. In this paper, $\tau$ is assumed to be a first passage time or a fixed constant, and $F(S_{\tau})$ is called ``the event of interest''.

In most applications, the problem of interest depends only on the observed process (random walk) $S_n$, e.g., the events $S_n>B$ (a \textit{fixed-time event}) or 
$\tau_b < T$ (a  \textit{first-passage-time event}), where $\tau_b = \inf\{n\geq 0:S_n>b\}$. These two quantities are related to overflow probability or bankruptcy. We use these two events to 
illustrate our theory, though our framework applies to more general events.
Although $X_n$ may not explicitly relate to the objective, it still governs the evolution of the quantities of interest.
Thus, we call $X_n$ a ``latent process'' and $S_n$ the ``observed process'' because it is directly relevant to the quantity of interest. 
For instance, $X_n$ can be a hidden regime that affects the quantities of interest or an adjoint process like the volatility process in computing VaR.

\subsection{Review of the classical exponential tilting family}
\label{sec2.2}

Let $X_n$ be the latent process (underlying Markov chain), and $Y_n$ be the observed process (additive part). 
Consider the eigenvalues--eigenfunctions framework in \cite{ney1987markov1}. 
More precisely,  we define an operator as 
\begin{equation}
	\label{eq:operator}
	(\mathcal{P}_\alpha f)(x) = \E_x[e^{\alpha^\top Y_1}f(X_1)] = \int p(x,\dd x')f(x')\int e^{\alpha^\top y}\rho(y|x,x')\dd y = \int f(x')e^{\psi(x,x',\alpha)}p(x,\dd x'),
\end{equation}
where $f(x)$ is a function to which the operator applies; $\top$ denotes transpose, and $\psi(x,x',\alpha) = \log \E[e^{\alpha^\top Y_1}|X_0 = x, X_1 = x'] =\log\int e^{\alpha^\top y}\rho(y|x,x')\dd y$ is the conditional cumulant generating function of the additive part, given the state $(x,x')$. Here $\alpha\in \mathbb{R}^d$ has the same dimension as the observed variable $Y_1$ and 
$\alpha^\top Y_1$ is the inner product.
Under some regularity conditions to be specified in Section \ref{ec:technical conditions}, let $e^{\Lambda(\alpha)}$ be the largest eigenvalue of $\mathcal{P}_\alpha$, and $r(\cdot,\alpha)$ be the associated positive right eigenfunction
such that
\begin{equation}
	\label{eq:eigenvalue}
	E_x[e^{\alpha^\top Y_1}r(X_1,\alpha)] = e^{\Lambda(\alpha)}r(x,\alpha).
\end{equation}

Following \cite{collamore2002importance}, 
we can define the induced tilting probability $\p_{\alpha}$ as
\begin{eqnarray}\label{tilt}
	\frac{\dd \p_{\alpha}}{\dd \p} = \frac{r(X_n,\alpha)}{r(X_0,\alpha)} e^{\alpha^\top S_n - n\Lambda(\alpha)}.
\end{eqnarray}
Under the new probability measure $\p_{\alpha}$ in (\ref{tilt}), 
the transition probability is 
\begin{equation}
	\label{eq:ld prob}
	\p_{\alpha}(\dd x',\dd s'|x,s) = \frac{r(x',\alpha)}{r(x,\alpha)}e^{-\Lambda(\alpha)+\alpha^{\top} (s'-s)}p(x,\dd x')\rho(s' - s|x,x')\dd s',
\end{equation} 
and an unbiased estimator of (\ref{exp}) is
\begin{eqnarray*}
	\label{eq:ld estimator}
	\E_{\nu}^{\p_{\alpha}}\bigg[F(S_{\tau}) \prod _{i=1 }^{\tau}\frac{r(X_{i-1},\alpha)}{r(X_i,\alpha)}e^{-\alpha^\top Y_i + \Lambda(\alpha)}\bigg] = \E_{\nu}^{\p_{\alpha}}\bigg[F(S_{\tau}) \frac{r(X_{0},\alpha)}{r(X_{\tau},\alpha)}e^{-\alpha^\top S_{\tau} + \tau \Lambda(\alpha)}\bigg].
\end{eqnarray*}

Under the tilting probability 
$\p_{\alpha}$, 
\begin{equation*}
	\label{eq:ld prob separate}
	\begin{aligned}
		& X_{n+1}\in \dd x'|X_n = x \sim \frac{r(x',\alpha)}{r(x,\alpha)}e^{-\Lambda(\alpha)+\psi(x,x',\alpha)}p(x,\dd x'), \\
		& Y_{n+1}\in \dd y|X_n=x,X_{n+1}=x' \sim e^{\alpha^\top y - \psi(x,x',\alpha)} \rho(y|x,x')\dd y.
	\end{aligned}
\end{equation*}
Note that the exponential tilting family is characterized by one parameter: The change of transition probability of the underlying Markov process and the conditional probability is only indexed by $\alpha$.

A widely used way to select the tilting parameter is based on the large deviation (LD) principle \citep{collamore2002importance}, denoted by $\alpha_{LD}$. More precisely, the probability of interest is embedded in a regime where $F(S_{\tau})$ converges to a constant when a particular parameter tends to infinity (e.g., $\frac{S_n}{n}$ converges due to the law of large numbers as $n\to \infty$) and the rate of convergence is described by a rate function. Usually, the LD-based methods could achieve asymptotic optimality in terms of either logarithmic efficiency or bounded relative error \citep{asmussen2007stochastic}. 
The LD theory and the rate function are profoundly related to the eigenvalue problem \eqref{eq:eigenvalue}.

There are two main drawbacks of applying the LD-based exponential tilting. First, the LD-based exponential tilting only aims to maximize the variance's exponential decay rate, not the original variance or asymptotic variance. When the probability of the event of interest is only moderately small ($10^{-2}$ to $10^{-4}$), the tilting probability measure given by the LD may not be close to the optimal one.
Second, in general, it is difficult to calculate the eigenvalues and eigenfunctions \eqref{eq:eigenvalue} required by the classical exponential tilting method, except when the 
latent Markov process has finite state space or is affine (cf. \citealt{GlassermanKim2010, zhang2015affine}). A functional eigenvalue problem is typically very challenging to solve. To avoid these drawbacks, we propose a broader tilting family that is more flexible and easier to implement so that importance sampling can be performed for more complex applications.

\section{A General Duo-Exponential Tilting Family}
\label{sec:3}

In our proposed algorithm, {\it it is not necessary to compute eigenvalues and eigenfunctions}; hence the computational cost can be reduced. 
It also performs importance sampling for more general stochastic models, to which the eigenvalues are unavailable. 
Furthermore, the algorithm tries to minimize the importance sampling estimator's variance directly rather than the asymptotic exponential rate function. Such optimization will incur more computational costs but will also provide a more refined approximation to the optimal variance-minimizing tilting distributions. The proposed algorithm tilts the transition probability and the distribution of innovations separately. 

For given $\theta \in \Theta \subseteq \mathbb{R}^d $, define the following exponential embedding for the random walk part. 
\begin{equation}
	\label{eq:innovation}
	\rho_{\theta}(y|x,x')\dd y = e^{\theta^\top y - \psi(x,x',\theta)}\rho(y|x,x')\dd y,
\end{equation}
where $\psi(x,x',\theta)$ is the same function defined after \eqref{eq:operator} by replacing $\alpha$ by $\theta$, and $\Theta = \{ \theta\in \mathbb{R}^d: \psi(x,x',\theta) < \infty, \text{ for all }x,x'\in \mathcal{X} \}$.
Next, we apply exponential tilting for the transition probability kernel of the latent process $X_n$. For given $\eta \in H \subseteq \mathbb{R}^m$, a given parameter space, define
\begin{equation}
	\label{eq:transition}
	p_{\eta}(x,\dd x') = e^{ k(x,x',\eta)-\phi(x,\eta)}p(x,\dd x'),
\end{equation} 
where $\phi(x,\eta) = \log \E_x[e^{ k(x,X_1,\eta)}]$,
and $H =\{ \eta\in \mathbb{R}^m: \phi(x,\eta) < \infty,\text{ for all }x\in \mathcal{X} \}$.
We call $k(x,x',\eta)$ a \textit{link function} with parameter $\eta$ that maps $\mathcal{X} \times \mathcal{X}$ to $\mathbb{R}$ and will be specified later.\footnote{When the link function is independent of $x'$, we call it a trivial link function since it does not tilt the transition probability of the underlying Markov process. In special cases, a trivial link function is already a good choice and can recover some existing method. For an example, see E-Companion \ref{appendix:example of markov intensity jump}.} 
Generally, the choice of $k(x,x',\eta)$ should be tailored to specific models and objectives. It determines which exponential family to be embedded for the transition probability. We will discuss some prototypical choices of $k(x,x',\eta)$ and its linkage to some well-known exponential families in Section \ref{sec:3 lan}.
Combining \eqref{eq:innovation} and \eqref{eq:transition} enables us to define the new transition probability $\p_{\theta,\eta}$ for Markov process $\{(X_n,S_n),n\geq 0\}$ as
\begin{equation}
	\label{eq:new prob}
	\p_{\theta,\eta}(\dd x',\dd s'|x,s) = e^{\theta^\top (s'-s) + k(x,x',\eta)   - \psi(x,x',\theta)- \phi(x,\eta)}p(x,\dd x')\rho(s'-s|x,x')\dd s'.
\end{equation}
Thus, an unbiased estimator of $\E_{\nu}[F(S_{\tau})]$ is given by
\[ \E_{\nu}^{\p_{\theta,\eta}}[F( S_{\tau}) e^{-\theta^\top S_{\tau} + \sum_{i=1}^{\tau} - k(X_{i-1},X_i,\eta) + \psi(X_{i-1},X_{i},\theta)+\phi(X_{i-1},\eta)}]. \]
Since the estimator is unbiased, it suffices to minimize the second moment of 
the  estimator, i.e.,
\begin{equation}
	\label{eq:2moment}
	\begin{aligned}
		G(\theta,\eta)=&\E_{\nu}^{\p_{\theta,\eta}}[F^2(S_{\tau}) e^{-2\theta^\top S_{\tau} + \sum_{i=1}^{\tau} - 2k(X_{i-1},X_i,\eta) + 2\psi(X_{i-1},X_{i},\theta)+2\phi(X_{i-1},\eta)}] \\
		=& \E_{\nu}[F^2(S_{\tau}) e^{-\theta^\top S_{\tau} + \sum_{i=1}^{\tau} - k(X_{i-1},X_i,\eta) + \psi(X_{i-1},X_{i},\theta)+\phi(X_{i-1},\eta)}].
	\end{aligned}
\end{equation}
In (\ref{eq:2moment}) $\E_{\nu}$ is the expectation related to the original
probability $\p_\nu$. The optimal parameters $(\theta,\eta)$ can be found by solving 
$
\label{eq:minimize variance}
\min_{\theta,\eta} G(\theta,\eta) .
$
We summarize all necessary technical definitions and assumptions common for Markov random walks in Section \ref{ec:technical conditions}. 
At a high level, the assumptions used in this paper play different roles. The exponential-moment conditions guarantee that the tilted transition probabilities are well defined. The recurrence and related stability conditions provide the probabilistic structure needed to apply Markov random walk arguments. The Poisson-equation and eigenfunction assumptions are mainly used to establish the theoretical guarantees and to connect our framework to the existing LAN- and LD-based importance sampling literature. In particular, some of these assumptions are analytical devices for the theory rather than prerequisites for implementing the algorithm in a specific application.

\begin{theorem}
	\label{thm:link}
	Under Assumptions \ref{ass:convex} to \ref{ass:eigenvalue exist}, the family of probability distributions induced by \eqref{eq:ld prob} belongs to the family of probability distributions defined in \eqref{eq:new prob}, with $k(x,x',\eta) = \psi(x,x',\eta) + \log r(x',\eta)$. Thus, our duo-exponential tilting reduces to the special case of the classical exponential family when $\theta = \eta$.
\end{theorem}

The proof is given in \ref{ec:proof link}. 
Theorem \ref{thm:link} shows that we can embed the classical exponential tilting family in \eqref{eq:ld prob} into our general duo-exponential tilting family through a particular link function.\footnote{ 
	Note that \eqref{eq:ld prob} is based on the exponential tilting developed in \cite{ney1987markov1}, under the Harris recurrent
	condition in Assumption \ref{ass:recurrent}, which also guarantees the finiteness of the growth condition and the tail probability of the quantities of interest. These properties are useful for defining the exponential embedding in Theorem \ref{thm:link}. Further details can be found in E-Companion \ref{appendix:regularity conditions}.}

\subsection{Basic properties of link functions}\label{sec:3.1}

To minimize the second moment in \eqref{eq:2moment}, we calculate the first-order derivatives in the next Lemma, which rigorously justifies interchanging the order between differentiation and expectation.
\begin{lemma}
	\label{lemma:interchange order}
	Under Assumptions \ref{ass:convex} to \ref{ass:rigorous eigenvalues}, when $\theta \in \Theta$ is sufficiently small such that $-\theta \in \Theta$ and both $\theta$ and $-\theta$ lie in the interior region of $\Theta$, the derivatives with respect to $(\theta,\eta)$ of \eqref{eq:2moment} can be represented via:
	\begin{equation}
		\label{eq:foc 1}
		\begin{aligned}
			\nabla_{\theta} G(\theta,\eta) = \E_{\nu}[ & F^2(S_{\tau}) e^{-\theta^\top S_{\tau} + \sum_{i=1}^{\tau} - k(X_{i-1},X_i,\eta) + \psi(X_{i-1},X_{i},\theta)+\phi(X_{i-1},\eta)} \\
			& \times \{ -S_{\tau} + \sum_{i=1}^{\tau}\frac{\partial \psi}{\partial \theta}(X_{i-1}, X_i,\theta) \}],
		\end{aligned}
	\end{equation}
	\begin{equation}
		\label{eq:foc 2}
		\begin{aligned}
			\nabla_{\eta} G(\theta,\eta) = \E_{\nu}[ & F^2(S_{\tau}) e^{-\theta^\top S_{\tau} + \sum_{i=1}^{\tau} - k(X_{i-1},X_i,\eta) + \psi(X_{i-1},X_{i},\theta)+\phi(X_{i-1},\eta)} \\
			& \times \sum_{i=1}^{\tau}  \{-\frac{\partial k}{\partial \eta}(X_{i-1},X_i,\eta) + \frac{\partial \phi}{\partial \eta}(X_{i-1},\eta)\} ].
		\end{aligned}
	\end{equation}
\end{lemma}

\begin{lemma}
	\label{thm:optimal foc}
	When the link function is linear in terms of the tilting parameter for the latent Markov process $\eta$, i.e., $k(x,x',\eta) = \eta^\top \tilde{k}(x,x')$, under Assumption \ref{ass:convex}, consider a function of $(\theta,\eta)\in (\Theta,H)$ with the following form: 
	\[ G(\theta,\eta) = \E_{\nu}\bigg[F^2(S_{\tau}) \exp\bigg\{ -\theta^\top S_{\tau} + \sum_{i=1}^{\tau} -\eta^\top \tilde{k}(X_{i-1},X_i) + \psi(X_{i-1},X_i,\theta) +\phi(X_{i-1},\eta) \bigg\}\bigg], \]
	where $\tau$ is a stopping time with $\E_{\nu}[\tau]<\infty$, and $F$ is a function with $\E_{\nu}[F^2(S_{\tau})]<\infty$.
	Then $G(\theta,\eta)$ is a convex function with a unique global minimum. In particular, the optimal tilting parameter can be found via solving first-order conditions \eqref{eq:foc 1} and \eqref{eq:foc 2}. 
\end{lemma}

The proof of Lemma \ref{thm:optimal foc} will be given in \ref{ec:proof foc}. Lemma \ref{thm:optimal foc} specifies the form of the link function to be linear in the tilting parameter $\eta$ and indicates that under Assumption \ref{ass:convex}, $G(\theta,\eta)$ is a convex function so that the first order condition yields the global minimum. 

However, the choice of link function in Theorem \ref{thm:link}, which reduces our duo-exponential family to the classical exponential family, is generally not a linear function in $\eta$. Hence, searching for the tilting parameter that minimizes the variance is also more difficult, even if the eigenvalue-eigenfunction can be computed. Next, we provide a corollary (as an example) where the eigenvalues-eigenfunctions are solvable under affine models and give a linear link function satisfying conditions in Lemma \ref{thm:optimal foc}. 

\begin{corollary}
	\label{coro:affine}
	Under Assumption \ref{ass:convex}, suppose the model is affine in the sense that the transition probability is affine in $x$, i.e., $\log \E_x[e^{\eta^\top X_1}] = C_1(\eta)^\top x+C_2(\eta)$, and the additive part is also affine, i.e., $\psi(x,x',\theta) = D_0(\theta)^\top x'+D_1(\theta)^\top x+D_2(\theta)$. Then the eigenvalues-eigenfunctions problem in \eqref{eq:eigenvalue} has a semi-explicit solution, in the sense that the eigenfunction is $r(x,\theta) = e^{A(\theta)^\top x}$, where $A(\theta)$ solves $C_1\big(A(\theta) + D_0(\theta)\big) + D_1(\theta) = A(\theta)$, and the associated eigenvalue is $e^{\Lambda(\theta)}$ with $\Lambda(\theta) = C_2\big(A(\theta) + D_0(\theta)\big) + D_2(\theta)$.
	Moreover, the classical one-parameter exponential tilting family \eqref{eq:ld prob} belongs to the probability family defined in \eqref{eq:new prob} with $k(x,x',\eta) = \eta^\top x'$ and $\eta = A(\theta) + D_0(\theta)$.
\end{corollary}

For affine models, by simply choosing a linear link function, the tilting family \eqref{eq:new prob} contains the classical exponential tilting family \eqref{eq:ld prob} as a subset since the log-eigenfunction is a linear function. Thus, the proposed algorithm can be at least as good as the traditional importance sampling estimator for affine models in terms of variance reduction. 
Many popular models are affine, e.g., Heston's stochastic volatility (SV) model. 
In Section \ref{sec:illustration}, we will use this SV model to show that enlarging the tilting family is non-trivial and can contribute to further variance reduction. 
As we shall see in Section \ref{sec:3 lan}, choosing a linear link function is not only suitable for affine models but also 
asymptotically optimal within a LAN family.

\subsection{A two-stage algorithm for duo-exponential tilting importance sampling}
\label{sec:3.2}

Once the link function is determined, we describe our importance sampling algorithm through a two-stage process. The first stage involves searching for the optimal tilting parameters, and the second stage entails sampling under the alternative probability. In the first stage, we  directly minimize \eqref{eq:2moment} using the stochastic gradient descent algorithm as in \cite{kawai2009optimal} to. Alternatively, other stochastic optimization algorithms can also be applied here, e.g., sample average approximation \citep{kim2015guide}.\footnote{Optimizing tilting parameters for rare-event problems typically require adaptive algorithms or warm-start, to gradually tilt the distribution in the optimization stage to enhance the search efficiency, as in the cross-entropy method \citep{rubinstein2004cross}, and more in the adaptive importance sampling by \citet{bardou2009computing,lemaire2010unconstrained,he2024adaptive,aolaritei2025stochastic}.
	Which optimization algorithm is best in the first stage is worth studying in the future research. In our implementation of Duo-CE, we indeed used a recursive optimization procedure in \citet{de2005tutorial}.} The second stage follows the standard importance sampling procedure and is straightforward once the tilting parameters are fixed. In summary, we present the proposed Algorithm \ref{algo:1} as follows and refer to this method as ``\textit{Duo-V}".

\begin{algorithm}[!htbp]
	\label{algorithm1}
	\caption{Importance sampling to estimate $\E_{\nu}[F(S_{\tau})]$ }
	\textbf{Stage 0} Preparation 
	
	Specify the parametric link function $k(x,x',\eta)$ (for example, $k(x,x',\eta) = \eta^\top x'$), and compute the conditional cumulant generating function $\phi(x,\eta)$ and $\psi(x,x',\theta)$.)
	
	\textbf{Stage 1} Searching the optimal tilting parameters 
	
	Find $(\theta^*,\eta^*) = \arg\min_{\theta,\eta}G(\theta,\eta)$ by stochastic gradient descent to minimize $G(\theta,\eta)$.
	
	\textbf{Stage 2} Importance sampling
	\begin{algorithmic}
		\STATE By using the optimal $(\theta^*,\eta^*)$ from stage 1, compute the conditional distribution of the additive increment $\rho_{\theta^*}(y|x,x')$ and the alternative transition probability $x'|x\sim p_{\eta^*}(x,\dd x')$ based on \eqref{eq:innovation} and \eqref{eq:transition}.
		\REPEAT \STATE{Generate initial state $x_0\sim \nu$ and underlying Markov process $x_{k+1}|x_k \sim p_{\eta^*}(x_k,\cdot)$, $y_{k+1}|x_k,x_{k+1}\sim \rho_{\theta^*}(y|x_k, x_{k+1})$, for $k=0,\cdots,\tau-1$, and compute $S_{\tau} = \sum_{i=1}^{\tau}y_i$, to get one estimator 
			$F(S_{\tau}) e^{-{\theta^*}^\top S_{\tau} + \sum_{i=1}^{\tau} - k(x_{i-1},x_i,\eta^*) + \psi(x_{i-1},x_{i},\theta^*)+\phi(x_{i-1},\eta^*)}.$}
		\UNTIL{we have sufficient samples.}
	\end{algorithmic}
	\label{algo:1}
\end{algorithm}

Because stage 1 is a stochastic program, the solution accuracy also depends on the sample size we draw, which is proportional to the computational cost. The accuracy of the numerical approximation of the tilting parameters may affect the variance reduction that can be achieved in the second stage. Thus, two cases may arise. (1) Optimizing over $\theta,\eta$ can significantly improve the variance reduction, hence the extra computation cost is worth it. (2) The variance reduction is insensitive to $\theta,\eta$; hence, it is not necessary to spend too much time to calculate the optimal parameters. Such a trade-off is further demonstrated in Section \ref{sec:illustration}, measured by the efficiency ratio.

Besides minimizing the variance, an alternative optimization objective is to optimize the tilting parameters via the cross-entropy (CE) method by \citet{rubinstein2004cross}. 
We incorporate this optimization objective into our duo-exponential tilting framework in E-Companion \ref{appendix:cross entropy}. 
However, the cross-entropy method is not a standalone simulation method; rather, it proposes to minimize a cross-entropy objective function to approximate the minimum-variance change-of-measure within a parametric tilting family. 
Unlike the LD- and MD-based exponential tilting, the cross-entropy method itself does not specify which tilting family to choose. 
Thus, CE can be used as an alternative to our first stage in Algorithm \ref{algo:1}, and the second stage remains the same. 
Therefore, the cross-entropy method searches for tilting parameters within the same family as our proposed method, differing only in the objective function to be minimized.  
In all numerical examples, we also implement this approach to demonstrate the superiority of the proposed duo-exponential tilting family, 
following the implementation in \citet{de2005tutorial}, referred to as ``\textit{Duo-CE}".

\section{Choice of Link Functions}
\label{sec:link}

This section studies how to construct link functions generally and in what sense they are (or approximate) the optimal ones. 
In particular, a class of link functions is asymptotically optimal in the sense of Definition \ref{def:asymptotic optimal lan}; see Theorem \ref{thm:lan optimal linear} for details.
Here is a brief road map:
(i) We construct a link function in Theorem \ref{thm:lan optimal linear} that achieves the optimal asymptotic variance reduction under the asymptotic normal regime. This link function is based on the solution of the Poisson equation \eqref{eq:poisson 0}, 
which is much easier to solve than the eigenvalues-eigenfunctions problem \eqref{eq:eigenvalue}.
Moreover, it gives the link function that satisfies the condition in Lemma \ref{thm:optimal foc} and hence the resulting objective is guaranteed to be convex and simplify the minimization procedure. 
When the model is affine (like the Heston model in Section \ref{sec:illustration}), the solution to \eqref{eq:poisson 0} is simply a linear function.
The Poisson equation \eqref{eq:poisson 0} may be solvable explicitly in non-affine cases (e.g., the VAR-GARCH model in our second main application in Section \ref{subsec:covar}). 
(ii) We discuss how to choose link functions by approximating the optimal exponential martingale that attempts to approximate the zero-variance tilting 
when the stochastic model is given by diffusion processes in Section \ref{sec:diffusion}. The condition in Lemma \ref{thm:optimal foc} is also satisfied. We illustrate this choice by our first main application related to the SIRD model in Section \ref{subsec:sir}.

\subsection{Optimal link function within local asymptotic normal family}
\label{sec:3 lan}

We shall select the optimal link function asymptotically by using 
the concept of the LAN family \citep{le2012asymptotics}\footnote{The definition of the  LAN family can be 
	extended significantly. For example, a family of probability measures $\{\p_{\theta,n}, \theta\in \Theta\}$ can be 
	said in LAN, if for every $\theta\in \Theta$, $\log\frac{\dd \p_{\theta + \delta_n h, n}}{\dd \p_{\theta, n}} \to \mathcal{N}(-\frac{1}{2}|h|^2,|h|^2)$ in distribution; see \citet{le2012asymptotics}.}. 
Using the LAN family to justify the asymptotic optimality of importance sampling was first suggested in \cite{fuh2004efficient} for i.i.d. random variables.
Later \cite{fuh2007estimation} study the asymptotic optimality for finite-state Markov
random walks. This subsection extends their results to general state Markov random walks in terms of a link function.

\begin{definition}[LAN Family]
	\label{def:lan}
	We say the twisted probability $\tilde{\p}$ with transition probability $\tilde{p}(x,\dd x')\tilde{\rho}(y|x,x')$ belongs to the LAN family of $\p$, if
	\begin{enumerate}[(i)]
		\item for a fixed-time event when $\tau = n$ is deterministic, under the original probability $\p$, $\sum_{i=1}^n \log\frac{\tilde{p}(X_{i-1},X_i)\tilde{\rho}(Y_i|X_{i-1},X_i)}{p(X_{i-1},X_i)\rho(Y_i|X_{i-1},X_i)} \to N_{\tilde{\p}}$ in distribution, as $n\to \infty$, or
		\item for a first-passage-time event when $\tau=\tau_b=\inf\{n\geq 0:S_n>b\}$, under the original probability $\p$, $\sum_{i=1}^{\tau_b} \log\frac{\tilde{p}(X_{i-1},X_i)\tilde{\rho}(Y_i|X_{i-1},X_i)}{p(X_{i-1},X_i)\rho(Y_i|X_{i-1},X_i)} \to N_{\tilde{\p}} $ in distribution, as $b\to \infty$,
	\end{enumerate}
	where $N_{\tilde{\p}} \sim \mathcal{N}(-\frac{1}{2}\sigma_{\tilde{\p}}^2, \sigma_{\tilde{\p}}^2)$ for some $\sigma_{\tilde{\p}}^2 > 0$. Furthermore, two twisted distributions $\tilde{\p}_1$ and $\tilde{\p}_2$ in the LAN family are asymptotic equivalent if $\sigma^2_{\tilde{\p}_1} = \sigma^2_{\tilde{\p}_2}$.
\end{definition}

\begin{definition}[Asymptotic Normal Regime]
	\label{def:normal regime}
	The event of interest $F(S_{\tau})$ is under the asymptotic normal regime if it converges to $\bar{F}(N)$ in distribution for some function $\bar{F}$ and a normally distributed random variable $N\sim \mathcal{N}(\mu_N,\Sigma_N)$ under the original probability $\p$, as
	(i)  $n\to \infty$, for a fixed-time event $\tau = n$; or
	(ii) $b\to \infty$, for a first-passage-time event $\tau = \tau_b = \inf\{n\geq 0:S_n>b\}$.

\end{definition}

For example, consider the event of interest $\{S_n > B\}$ for a fixed $n$. A typical event under the asymptotic normal regime arises when $B = n\E_{\pi}[Y_1] + \sqrt{n} c$ for a given $c$, where $\E_{\pi}[Y_1]$ is the expectation of the increment under the stationary distribution $\pi$. In contrast, a typical event under the LD regime arises when $B = n\E_{\pi}[Y_1] + n c$. 
That is why the probability converges to a positive number under the asymptotic normal regime, while it converges to 0 under the LD regime.

The asymptotic normal regime
event $\{S_n > B\}$ with $B = n\E_{\pi}[Y_1] + \sqrt{n} c$ belongs to the domain of moderate deviation.
Such events (with probability around $10^{-2} \sim 10^{-4}$) arise in risk management, construction of confidence regions, etc. In contrast, the LD rare events often occur in the study of queuing networks, reliability systems, or statistical physics, where the probability can be even as small as $10^{-7} \sim 10^{-50}$, see \citet{juneja2006rare} for more introductions.
As the probabilities get smaller, although the variance reduction of our proposed method relative to the standard Monte Carlo simulation is also bigger, 
it is better to use LD-based simulation methods, if available, as the focus of our method is on the rare events belong to the domain of moderate (not large) deviation. However, LD-based simulation methods
may not be available due to the difficulties in computing the eigenfunctions and eigenvalues.

Given the above asymptotic behavior, the importance sampling estimator can be chosen as $\E^{\tilde{\p}}[\bar{F}(N)e^{-N_{\tilde{\p}}}]$, whose variance is
$\E^{\tilde{\p}}[\bar{F}^2(N)e^{-2N_{\tilde{\p}}}] - \big( \E^{\tilde{\p}}[\bar{F}(N)e^{-N_{\tilde{\p}}}] \big)^2 = \E[\bar{F}^2(N)e^{-N_{\tilde{\p}}}] - \big( \E[\bar{F}(N)] \big)^2. $
Different choices of twisted distributions $\tilde{\p}$ yield different asymptotic joint distributions between $N_{\tilde{\p}}$ and $N$, while the marginal distribution of $N$ is independent of the twisted distributions and solely be determined by the event of interest. 
\begin{definition}[Asymptotic Optimality within the LAN Family]
	\label{def:asymptotic optimal lan}
	The asymptotic optimal tilting distribution within the LAN family of $\p$ for an event under the asymptotic normal regime is defined to be a twisted distribution $\tilde{\p}$ that belongs to the LAN family such that $\E[\bar{F}^2(N)e^{-N_{\tilde{\p}}}]$ is minimized.
\end{definition}

The following theorem provides one optimal link function that achieves the minimum asymptotic variance under the asymptotic normal regime. In Theorem \ref{thm:lan optimal linear}, we consider either fixed sample size $n$ or a first passage time $\tau_b$ crossing a boundary $b$. The asymptotic 
is defined as $n \to \infty$ in the first case, while  $b \to \infty$ for the second case of the first passage time $\tau_b$. 

\begin{theorem}
	\label{thm:lan optimal linear}
	Under Assumptions \ref{ass:convex}, \ref{ass:recurrent}, and \ref{ass:rigorous}, suppose the event of interest is under the asymptotic normal regime. When\footnote{
		In Theorem \ref{thm:lan optimal linear}, we consider the case $\theta = O(\frac{1}{\sqrt{n}})$ and 
		$\eta = O(\frac{1}{\sqrt{n}})$.  A more general 
		setting based on moderate deviation principle, which considers $\frac{1}{b_n} \sum_{j=1}^n (X_j - E X_j)$,
		where $\frac{b_n}{n} \downarrow 0$ but $\frac{b_n}{\sqrt{n}} \uparrow \infty$, 
		instead of using the central limit theorem,
		was proposed by \cite{deAcosta1992} for the sum of independent random vectors. 
		The extension to Markov random walks simulation is an interesting open problem.} 
	$\theta = O(\frac{1}{\sqrt{n}})$ and $\eta = O(\frac{1}{\sqrt{n}})$, the proposed tilting family with the link function $k(x,x',\eta) = \eta\top \tilde{k}(x,x')$, belongs to LAN, where 
	\begin{equation}
		\label{eq:md choose k}
		\tilde{k}(x,x') = \frac{\partial \psi}{\partial \theta}(x,x',0) +g(x') ,
	\end{equation}
	and $g(x)$ is the solution to the Poisson equation 
	\begin{equation}
		\label{eq:poisson 0}
		\big( (I-\mathcal{P}_0)g\big)(x) = \E_x[Y_1] - \E_{\pi}[Y_1],
	\end{equation} 
	where $\pi$ is the stationary distribution of the underlying Markov chain $\{X_n,n \geq 0\}$, $I$ is the identity operator and $\mathcal{P}_0$ is the operator defined in \eqref{eq:operator} with $\theta=0$.
	The asymptotic optimal tilting distribution\footnote{
		Different twisted probabilities may be asymptotic equivalent in the LAN family, 
		which all achieve the same, smallest asymptotic variance in the sense of Definition \ref{def:asymptotic optimal lan}. Hence the optimal link function may not be unique. In Theorem \ref{thm:lan optimal linear}, a specific link function is given that induces the asymptotic optimal tilting distribution within the LAN family, which is tractable and whose finite sample variance is easy to minimize. } 
	within the LAN family for an event under the asymptotic normal regime is achieved when $\theta-\eta = o(\frac{1}{\sqrt{n}})$. 
\end{theorem}

Here $\frac{\partial \psi}{\partial \theta}$ 
stands for the partial derivative with respect to the third argument, recalling their definition in \eqref{eq:innovation}. 
The first term  in \eqref{eq:md choose k} is the conditional mean of the increment: $\frac{\partial \psi}{\partial \theta}(x,x',0) = \E[Y_1|X_0 = x, X_1 = x']$. The second term is characterized by the solution to a Poisson equation rather than an eigenvalues-eigenfunctions problem. Assumption \ref{ass:rigorous} may not be needed if we can get the explicit solutions directly. The properties of LAN family can be proved only under Assumptions \ref{ass:convex} and \ref{ass:recurrent} in the existing literature (see \citealt{alsmeyer1999sMRW} for details).
As an immediate corollary, for affine models in Corollary \ref{coro:affine}, the log-eigenfunction $\log r(x,\theta)$ is linear in $x$, hence the solution to \eqref{eq:poisson 0} is also a linear function in $x$. Therefore, Theorem \ref{thm:lan optimal linear} designates setting the function $k(x,x',\eta)$ as a linear function in $x'$ for all affine models.

After choosing this link function, we know there exist specific values of $\theta,\eta$ such that the asymptotic variance of the importance sampling estimator can be minimized. By optimizing over the parameters $\theta$ and $\eta$ for a finite-time event by stochastic approximation, we can achieve better variance reduction than the asymptotically optimal one because our two-stage method (Algorithm \ref{algo:1}) aims to minimize the exact variance directly, within the family of asymptotically optimal link functions, with Lemma \ref{thm:optimal foc} ensuring convexity of the optimization objective. We require $\theta = O(\frac{1}{\sqrt{n}})$, $\eta = O(\frac{1}{\sqrt{n}})$ and $\theta-\eta = o(\frac{1}{\sqrt{n}})$ in Theorem \ref{thm:lan optimal linear} to achieve the minimum asymptotic variance, while this condition may be relaxed when optimizing finite-time variance.\footnote{
	A way to enforce these conditions is by letting $\theta = \frac{\bar{\theta}}{\sqrt{n}}$ and $\eta = \frac{\bar{\theta}}{\sqrt{n}} + \frac{\bar{\eta}}{n}$ and then optimizing over $\bar{\theta},\bar{\eta}$.}

The asymptotic optimal link function \eqref{eq:md choose k} can only be obtained analytically if the Poisson equation \eqref{eq:poisson 0} can be solved explicitly\footnote{Note that the solution to the Poisson equation \eqref{eq:poisson 0} is a function of $x$, which may lie in general (continuous) state space. If efficient numerical methods exist to approximate its solution as a function of $x$, then we may also adopt such numerical solutions. However, this equation is usually a (possibly high dimensional) integral equation, which is difficult to solve numerically.}. 
We can solve \eqref{eq:poisson 0} explicitly in some cases, such as the Heston model and the VAR-GARCH model in Sections \ref{sec:illustration} and \ref{sec:5}, respectively.

In our examples, the link function $k(x,x',\eta)$ is a function of the previous state $x$, the current state $x'$, and a tilting parameter $\eta$. 
For many models in this paper, the link function is either $k(x,x',\eta) = \eta^\top x'$ (in the Heston model, VAR-GARCH model, affine models, and tandem queue model, based on the solution to the Poisson equation), or $k(x,x',\eta) = \eta^\top B(x) x'$ (in the SIRD model, based on the structure of the model), which are all linear functions in $x'$. This is more likely a coincidence, as the underlying Markov process $X$ also has a ``random-walk-like" structure in all these examples. 
Hence, the apparent success of linear link functions in our numerical examples should be interpreted as an empirical observation rather than a general theorem. Identifying concrete classes of models in which nonlinear link functions are essential remains an open problem.
For a general Markov process, e.g., the Markov switching model illustrated in Example \ref{eg:markov swtiching} in \ref{appendix:two additional examples}, the link function is given by the solution to a (finite dimensional) linear equation, and, thus, is not necessarily linear in $x'$. 
It is also interesting to note that the constructed subsolutions in \citet{dupuis2007subsolutions} are often linear in the previous state $x$. However, the subsolutions to the SIRD model presented in E-Companion \ref{appendix_control} are not linear in $x$;  
the corresponding link function $k(x,x',\eta) = \eta^\top B(x) x'$ is linear in $x'$ but is not linear in $x$.

\begin{lemma}
	\label{lemma:poisson}
	Under Assumptions \ref{ass:convex}, \ref{ass:recurrent}, and \ref{ass:eigenvalue exist}, $\Delta(x,\theta) := \frac{\partial r}{\partial \theta}(x,\theta)$ solves the following equation:
	\begin{equation}
		\label{eq:poisson general}
		\E^{\p_{\theta}}_x[\Delta(X_1,\theta) - \Delta(x,\theta) + Y_1] = \Lambda'(\theta).
	\end{equation}
	In particular, Assumption \ref{ass:rigorous} holds with the solution to \eqref{eq:poisson 0} given by
	\begin{equation}
		\label{eq:poisson general1}
		\Delta(x,0) = \frac{\partial r}{\partial \theta}(x,0)
	\end{equation} 
	
\end{lemma}

The proof of Lemma \ref{lemma:poisson} will be given in \ref{ec:proof poisson}. The second term in \eqref{eq:md choose k} reflects the derivatives of eigenfunction and hence can be regarded as a first-order or linear approximation of the unknown eigenfunction. 
Moreover, by \eqref{eq:poisson general1}, solving the Poisson equation \eqref{eq:poisson 0} means simply finding the derivative of the eigenfunction $r(x,\theta)$ at one point $\theta = 0$, which is much simpler than finding the entire eigenfunction $r(x,\theta)$, thus significantly reducing computational burden in the searching of the optimal tilting probability. We give Example EC.2 with the analytical solution of the Poisson equation available, while it is difficult to find the eigenfunction to be used for the LD-based method.

The link function \eqref{eq:md choose k} in Theorem \ref{thm:lan optimal linear} is a first-order approximation to the one that recovers the classical exponential tilting in Theorem \ref{thm:link}. More precisely, if we apply Taylor's expansion, with respect to $\eta$, to the general link function in Theorem \ref{thm:link},  then we obtain
\begin{equation}
	\psi(x,x',\eta) + \log r(x',\eta) \approx \eta^\top [ \frac{\partial \psi}{\partial \theta}(x,x',0) + \frac{\partial r}{\partial \theta}(x',0)] = \eta\top \tilde{k}(x,x'),
\end{equation}
when $\eta$ is close to 0, and the eigenfunction $r(x,0)$ can be normalized to be $1$ and $\psi(x,x',0) = 0$. Thus, the LAN family in Theorem \ref{thm:lan optimal linear} is also useful in terms of approximating the enlarged tilting family which contains the conventional LD-based tilting distribution. Even if we know function $\psi$, we can still choose to approximate it by its first-order derivative so that the constructed link function is linear in $\eta$. Based on Lemma \ref{thm:optimal foc}, the global optimal point can be obtained by solving the first-order condition. Moreover, using the first-order derivative to approximate this term still preserves the asymptotic property. Combining \eqref{eq:md choose k} and \eqref{eq:poisson 0}, we can indeed find a function $k(x,x',\eta)$, which can be relatively easily calculated and approximates the classical exponential family without solving the eigenfunction problem in \eqref{eq:eigenvalue}.

Before discussing the differences, we emphasize that the LAN-based choice of link function in Theorem \ref{thm:lan optimal linear} is built on the same moderate-deviation/Poisson-equation insight as \citet{fuh2007estimation}. The difference is not in replacing that insight, but in using it to construct a link function and to optimize within a larger duo-exponential tilting family.
First, to the best of our knowledge, the general duo-exponential tilting family induced through a link function is new to the literature, and it includes the classical LD-based exponential tilting family when the link function is chosen in a particular way and $\theta=\eta$, as shown in Theorem \ref{thm:link}. Hence, we enlarge the search space to a larger distribution family.
By contrast, \cite{fuh2007estimation} still aims to approximate the classical LD-based exponential tilting family without the enlargement. Second, Theorem \ref{thm:lan optimal linear} suggests a particular choice of the link function relying on the notions of the LAN family under the asymptotic normal regime. In terms of optimizing over the choice of link function, the criterion used in Theorem \ref{thm:lan optimal linear} is the same as \cite{fuh2007estimation}, and they both recover the role of the Poisson equation. However, the difference is that \cite{fuh2007estimation} focuses on finding one sampling distribution, whereas we find a function that further induces a duo-exponential tilting family. In other words, \cite{fuh2007estimation} starts sampling upon solving the Poisson equation, but our proposed method still needs to search for the optimal tilting parameters before the sampling stage, as described in Section \ref{sec:3.2}. In addition, our framework of LAN and the asymptotic normal regime extends the fixed-time event in \cite{fuh2007estimation} to include the first-passage-time event. Third, we also develop a theoretical concept of the logarithmic efficiency for rare-event simulation under the asymptotic normal regime and verify the desired efficiency in Section \ref{subsec:efficiency log main text}, which is missing in \cite{fuh2007estimation}. Fourth, our link function is more flexible, as shown in Section \ref{sec:diffusion}.

\subsection{Approximating the optimal exponential martingale for diffusion processes}
\label{sec:diffusion}

For complex stochastic models, if the Poisson equation \eqref{eq:poisson 0} in Theorem \ref{thm:lan optimal linear} is difficult to solve numerically, 
how do we choose link functions? We answer this question for diffusion processes in this section. In short, we can construct a link function as an approximation to a Poisson equation, by using a linear partial differential equation (PDE) associated with the generator of the stochastic system.

Consider the system $ \dd X_t = b(X_t)\dd t + \sigma(X_t)\dd W_t,\ X_0 = x_0$. The problem of interest is to estimate $\E_{x_0}[F(X_{\tau})]$ for a stopping time $\tau$, where $X_t\in \mathbb{R}^d$, $W_t\in \mathbb{R}^m$, $b:\mathbb{R}^d \to \mathbb{R}^d$, and $\sigma:\mathbb{R}^d \to \mathbb{R}^{d\times m}$. To simulate the continuous-time diffusion processes, we employ the Euler-Maruyama scheme so that the underlying Markov process has a conditional normal transition probability:
$
X_{n+1}|X_n \sim \mathcal{N}\left( X_n + b(X_n)\Delta t,\sigma\sigma^\top(X_n)\Delta t \right) .
$
Hence, these processes still belong to our general framework of Markov random walks \eqref{eq:mcrw model} with a degenerate additive part $S_n = X_n$.

For this class of stochastic models, it is challenging to do LAN analysis because its stationary distribution is usually not easy to obtain. Moreover, neither the eigenvalue--eigenfunction problems nor the Poisson equation can be solved explicitly. 
However, we can provide a useful exponential tilting formula by approximating a zero variance importance sampling: $\frac{\dd \q^*}{\dd \p}|_{\f_T} = L_T$, where $L_t$ satisfies
$\dd L_t =  L_t {U^*_t}^\top\dd W_t,\  L_0 = 1$, and $U^*(t,X_t) = \sigma^\top(X_t)g^*(t,X_t)$ for a function $g^*:\mathbb{R}_+\times \mathbb{R}^d \to \mathbb{R}^d$ that is related to the solution to a linear PDE (for details, see E-Companion \ref{appendix_control}). 
Thus, to sample $X_{\tau}$ under the zero-variance tilting measure $\q^*$, after the discretization, is to sample 
$
\label{eq:sde q measure}
X_{n+1}|X_n \sim \mathcal{N}\left( X_n + b(X_n)\Delta t + \sigma\sigma^\top(X_{n}) g^*(t_n,X_n) \Delta t,\sigma\sigma^\top(X_n)\Delta t \right).
$

Although it may be impractical to solve the PDE, we can still approximate its solution within a linear space spanned by some basis functions, 
so that $g^*(t,x) \approx B(x)\eta$, where $B:\mathbb{R}^d \to \mathbb{R}^{d\times L}$ and $\eta\in \mathbb{R}^L$. 
The basis function is typically taken to be linear or quadratic, or based on the original functional form of $b,\sigma$. 
Hence, to approximate the zero-variance tilting measure $\q^*$, we sample under
$
\label{eq:sde q approximate}
X_{n+1}|X_n \sim  \mathcal{N}\left(X_n + b(X_n)\Delta t + \sigma\sigma^\top(X_n)B(X_n)\eta\Delta t,\sigma\sigma^\top(X_n)\Delta t \right).
$
Although this is a heuristic approximation, it may still achieve significant variance reduction; see the SIRD model in Section \ref{subsec:sir}, 
where we also explain how to choose these basis functions for this particular case. 
Its success depends on how well the chosen basis family captures the main structure of the zero-variance tilting measure. At present, we do not have a general theorem translating approximation error in the PDE/Poisson step into logarithmic or near-logarithmic efficiency of the resulting importance sampling estimator. Therefore, this diffusion construction should be interpreted as a principled but approximation-dependent way to choose tractable link functions beyond the small-noise setting.
The approximation is equivalent to defining the tilting probability measure using the link function $ k(x,x',\eta) = \eta^\top B^\top(x) x'$. 
More precisely, the change of measure \eqref{eq:transition} becomes
$ p_{\eta}(x,\dd x') = \exp\{\eta^\top B^\top(x) x' - \phi(x,\eta)\} p(x,\dd x'), $
where $\phi(x,\eta) = \eta^\top B^\top(x)\left( x + b\left(x \right)\Delta t \right) + \frac{1}{2}\Delta t\eta^\top B^\top(x)\sigma\sigma^\top(x) B(x) \eta  $.

\section{Efficiency Criteria and an Illustrative Example}
\label{sec:illustration}

\subsection{Theoretical efficiency}
\label{subsec:efficiency log main text}

In the conventional rare event simulation, we consider a sequence of events indexed by the rarity, and the probabilities of such events tend to zero under a particular limiting regime of the index. Some theoretical efficiency criteria, 
such as bounded relative error or logarithmic efficiency, can be used; see \citet[Chapter 6]{asmussen2007stochastic} for more details.
To discuss the theoretical efficiency criteria in the context of the rare events and the asymptotic normal regime, we need to re-index the event of interest and give the rational of defining {\it logarithmic efficiency.}, by letting both $n\to \infty$ (or $b \to \infty$) and $\ell\to \infty$.
Note that as \citet[Chapter VI, page 160]{asmussen2007stochastic} put it, a limiting rarity regime is just a `'mathematical formalism'' because
rarity could occur in more than one way in practice.

\begin{definition}[Rarity in Asymptotic Normal Regime]
	\label{def:rare normal}
	Under the asymptotic normal regime as in Definition \ref{def:normal regime}, suppose the event of interest can be written as (i)
	$\p(S_n > B) = \p\left( \Sigma_N^{-1/2}\frac{S_n - \mu_N}{\sqrt{n}} > \ell \right)$, for a fixed-time event; or 
	(ii) $\p(\tau_b < T) = \p\left( \Sigma_N^{-1/2}\frac{\tau_b - \mu_N}{\sqrt{b}} < \ell \right)$ for a first-passage-time event with $\tau_b = \inf\{n\geq 0: S_n > b\}$.
	We say the sequence of fixed-time events indexed by $(n,\ell)$ are LAN-rare if $n\to \infty$, $\ell\to \infty$, and $\ell/\sqrt{n} \to 0$. We say the sequence of first-passage-time events indexed by $(b,\ell)$ are LAN-rare if $b\to \infty$, $\ell\to -\infty$, and $\ell/\sqrt{b} \to 0$.
\end{definition} 

The limiting regime defined in Definition \ref{def:rare normal} is reasonable for the asymptotic normal regime that we considered throughout this paper. To be more specific, it includes the following special regime: we first take limit $n \to \infty$ ($b \to \infty$) to ensure the event of interest is in the asymptotic normal regime and then consider
the probabilities of the limiting normally distributed variables go to zero, indexed by $\ell$, in the usual sense. In other words, the events are rare not because they deviate from their asymptotic normality but because the tail probability of an asymptotic normal distribution is small. Other limiting regimes, such as $\lim_{n \to \infty} \lim_{\ell \to \infty}$, are beyond the scope of the theoretical consideration of the LAN family.
In fact, a typical LD regime corresponds to the regime $\ell \to \infty, n\to \infty, \ell/\sqrt{n} \to c > 0$ or $\ell \to -\infty, b \to \infty, \ell/\sqrt{b} \to -c < 0$. 
For more comparison with the LD regime, see E-Companion \ref{appendix:log efficiency}.
The rarity in Definition \ref{def:rare normal} motivates us to define the logarithmic efficiency under the asymptotic normal regime as follows.

\begin{definition}[Logarithmic Efficiency in Asymptotic Normal Regime]
	\label{def:log efficiency normal main text}
	
	(i). For a fixed-time event indexed by $\ell$ ($\ell \to \infty$)  under the asymptotic normal regime indexed by $n$ ($n \to  \infty$), suppose the probability of the event, denoted as
	$z(n,\ell)$, satisfying $\lim_{n \to \infty} z(n,\ell) = \Phi(-\ell)$, where $\Phi(\cdot)$ is the cumulative distribution function of a standard normal distribution. Then an unbiased estimator $\hat{Z}(n,\ell)$ is said to have logarithmic efficiency in the asymptotic normal regime, if,
	for all $\varepsilon > 0$,
	\begin{equation}
		\label{eq:log efficiency md fixed time main}
		\limsup_{\ell \to \infty, n\to \infty, \ell/\sqrt{n} \to 0} \frac{\operatorname{Var}\left( \hat Z(n,\ell) \right)}{\left( \Phi(-\ell)\right)^{2-\varepsilon} }= 0.
	\end{equation}

	(ii)  For a first-passage-time event $\ell$ ($\ell \to - \infty$) under the asymptotic normal regime indexed by $b$ ($b \to  \infty$), suppose the probability of the event is represented by $z(b,\ell)$, with $\lim_{b \to \infty} z(b,\ell) = \Phi(\ell)$. Then an unbiased estimator $\hat{Z}(b,\ell)$ is said to have logarithmic efficiency in the asymptotic normal regime if,
	for all $\varepsilon > 0$,
	\begin{equation}
		\label{eq:log efficiency md stopping time main}
		\limsup_{\ell \to -\infty, b\to \infty, \ell/\sqrt{b} \to 0}	 \frac{ \operatorname{Var}\left( \hat Z(b,\ell) \right)}{\left( \Phi(\ell)\right)^{2-\varepsilon} }= 0.
	\end{equation}
\end{definition}

This definition is adapted from the conventional definition of logarithmic efficiency in the rare event simulation to fit the asymptotic normal regime within
the LAN family. Hence, in the denominator, the probability of the event is replaced by the corresponding probability in a normal distribution, and in the numerator, we examine the variance of the estimator. In contrast to the conventional definition of logarithmic efficiency in the LD literature, the difference in Definition \ref{def:log efficiency normal main text} is that we replace 
\begin{equation}
	\label{eq:log efficiency ld main text}
	\frac{\operatorname{Var}\left( \hat Z(n,\ell) \right)}{\left( z(n,\ell)\right)^{2-\varepsilon} },\text{ for a fixed-time event; or }  \frac{\operatorname{Var}\left( \hat Z(n,\ell) \right)}{\left( z(b,\ell)\right)^{2-\varepsilon} } \text{ for a first-passage-time event,}
\end{equation}
by \eqref{eq:log efficiency md fixed time main} and \eqref{eq:log efficiency md stopping time main}.
In other words, the denominator of the probability of the event of interest is replaced by a normal tail probability. 
The restriction essentially requires 
that the rarity goes to zero at the same rate as that of the tail probability of a normal distribution. 
Just as the classical exponential tilting can achieve logarithmic efficiency (which usually requires stronger conditions), we can show that the optimal tilting within the LAN family is also logarithmic efficient in the sense of Definition \ref{def:log efficiency normal main text}. 

\begin{theorem}
	\label{thm:log efficiency}
	Under Assumptions \ref{ass:convex}, \ref{ass:recurrent}, \ref{ass:rigorous eigenvalues}, \ref{ass:eigenvalue exist} and \ref{ass:eigenvalue bounded}, 
	for a fixed-time event or a first-passage-time event under the asymptotic normal regime, the importance sampling estimator proposed in Theorem \ref{thm:lan optimal linear} has logarithmic efficiency in the sense of \eqref{eq:log efficiency md fixed time main} and \eqref{eq:log efficiency md stopping time main}.
\end{theorem}

\subsection{Numerical efficiency}

There is a trade-off between accuracy (as measured by the standard deviation) and time consumption
(as measured by the CPU time),
when evaluating the numerical performance of different simulation methods,
Following  \citet{glynn1989importance}, we shall use the following ratio to strike a balance between the two criteria.\footnote{
	Since other criteria (e.g.,  difficulty in implementation, robustness to model parameters, adaptiveness to model specifications) 
	are difficult to measure, we do not compare algorithms based on them.}
The efficiency ratio of an algorithm relative to the plain Monte Carlo  on the same sample size is 
\[ \text{Efficiency ratio} := \frac{\text{standard deviation reduction ratio}}{\sqrt{\text{time consumption ratio}}} = \frac{\text{standard deviation}_{MC}/\text{standard deviation}_{a}}{\sqrt{\text{time consumption}_{a}/\text{time consumption}_{MC}}},\]
as increasing the sample size by $N$ times reduces the standard deviation by $\sqrt{N}$ times.

For all the numerical experiments, we focus on moderately small probability events ($10^{-2}\sim 10^{-5}$) and 
pick a sufficiently large $N$ such that the plain Monte Carlo method can obtain an estimate with about 1\% standard deviation. 
In the numerical results, the standard deviation of our method is typically 0.1\% of the estimated probability, which means in practice, we do not need that many samples to achieve the same accuracy as the plain Monte Carlo method. 
Thus, we mainly focus on comparing the efficiency ratio. For our proposed method, the number of samples that are used for the first-stage optimization is taken as $N/100$ for the examples in Sections \ref{subsec:sv example} and \ref{subsec:sir}, and $N/10$ for the example in Section \ref{subsec:covar}. For general problems, based on our experience in these experiments, the number of samples required for the optimization stage is much smaller than the required number of samples for the plain Monte Carlo method.
This also explains why the simulation budgets may appear conservative relative to the rarity of the events: part of the numerical effort is used to stabilize the first-stage search so that the second-stage comparison is meaningful across methods. In practice, if $G(\theta,\eta)$ is relatively flat near its minimizer, a coarse search, a warm start, or an adaptive refinement may already capture most of the benefit at a lower cost. Overall, relative to the whole computational time used by our algorithm, stage 1 typically takes around 1\% to 20\% of the computational cost in our implementation, while this cost is controlled by the number of samples used in stage-1 optimization versus stage-2 sampling. How to achieve the best balance between these two stages remains an open question worth studying in future work. 
\subsection{Illustrative example (I): Heston model}
\label{subsec:sv example}
To give a numerical example, consider the Heston stochastic volatility (SV) model: 
$$
\dd S_t = \mu S_t\dd t + \sqrt{X_t}S_t\dd W^S_t,  ~~~
\dd X_t = \kappa (\alpha - X_t)\dd t+\sigma\sqrt{X_t}\dd W^{X}_t,
$$
where $\dd W^S_t \dd W^{X}_t=\rho \dd t$, $2\kappa\alpha \geq \sigma^2$.
We consider discretization via the Euler-Maruyama scheme
and regard the system as a Markov random walk 
by taking $(X_{t_k})$ as the underlying Markov chain:
\[ X_{t_{k+1}}|X_{t_k} \sim \frac{\sigma^2(1 - e^{-\kappa \Delta t})}{4\kappa} \chi^2_{\frac{4\kappa \alpha}{\sigma^2}}\left( \frac{4\kappa e^{-\kappa \Delta t}}{\sigma^2(1 - e^{-\kappa \Delta t})} X_{t_k}\right) ,  \]
and the additive part
\begin{equation*}
	Y_{t_{k+1}} = [\mu-\frac{1}{2}X_{t_k}-\frac{\rho}{\sigma}\kappa(\alpha-X_{t_k})]\Delta t + \frac{\rho}{\sigma}(X_{t_{k+1}}-X_{t_k})+\sqrt{(1-\rho^2)X_{t_k} \Delta t}~\epsilon_{k+1} ,
\end{equation*} 
where $\epsilon_{k}\sim N(0,1)$ are all i.i.d. random variables, and $\chi^2_d(\lambda)$ stands for a non-central
chi-square distribution with degree of freedom $d$ and non-centrality $\lambda$. $Y_{t_{k+1}} = \log S_{t_{k+1}}-\log S_{t_k}$ represents the log-return, and $t_k = k\Delta t$ is the time-discretization. 
This is a special case of a discrete-time affine model in \cite{le2010discrete}. 

We test the numerical performance of the proposed algorithm in the SV model for computing
the tail probability in the form of $\p(S_T > c | S_0 = 1, X_0 = \alpha )$. The E-Companion \ref{appendix:heston} and \ref{appendix:verify assumptions} give the resulting tilting probability and verify the regularity conditions. All
our theorems are applicable to this example.

To use Theorem \ref{thm:lan optimal linear}, we need to compute the solution to the Poisson equation for this affine model. By Corollary \ref{coro:affine} and Lemma \ref{lemma:poisson}, we know the eigenfunction for affine models is a log-linear function, and the derivative of the eigenfunction is the solution to the Poisson equation \eqref{eq:poisson 0}. Hence, the solution to \eqref{eq:poisson 0} is a linear function. Thus, we choose the link function as $k(x,x',\eta) = \eta x'$. See Example \ref{eg:ec affine} in the E-Companion for the explicit calculation of the Poisson equation for a wide class of affine models. 
Theorem \ref{thm:lan optimal linear} can be applied to justify that our choice of link function is asymptotically optimal within the LAN family.
Theorem \ref{thm:log efficiency}
can be applied to justify that the importance sampling estimator proposed in Theorem \ref{thm:lan optimal linear} has logarithmic efficiency in the sense of \eqref{eq:log efficiency md fixed time main}.

\begin{table}[!htbp]
	\centering
	\caption{\textbf{Numerical results of different methods}. Estimation of  $\mathbb{P}_{\alpha}(S_T > b)$, where $(S_t,X_t)$ follows Heston's stochastic volatility model. $T = \frac{1}{12}$. Discretization is taken as Euler-Maruyama scheme with $\Delta t=\frac{1}{120}$. The sample size is $10^6,10^7,10^8,10^8$ respectively.  $\mu=0.02$, $\kappa=3$, $\alpha=0.015$, $\sigma=0.25$ and $\rho = 0.05$.}
	
	{\small
		
		\begin{tabular}{llcccc}
			\hline
			& $b/S_0$ & 1.08  & 1.12  & 1.15 & 1.18 \\
			\hline
			Plain MC & mean & 1.79E-02 & 1.47E-03 & 1.83E-04 & 2.11E-05 \\
			& standard deviation   & 1.33E-04 & 1.21E-05 & 1.35E-06 & 4.60E-07 \\
			
			& elapsed time & 0.50  & 0.92  & 7.82  & 7.81 \\
			
			\cite{collamore2002importance}   
			& sd reduction ratio & 1.71  & 2.23  & 2.76  & 3.48 \\

			& time consumption ratio & 1.18  & 1.26  & 1.03  & 1.01 \\

			& {\bf efficiency ratio} & 1.58  & 1.99  & 2.72  & 3.46 \\

			\cite{fuh2007estimation}  
			& sd reduction ratio & 1.72  & 2.27  & 2.80  & 3.44 \\
			
			& time consumption ratio & 1.03  & 1.61  & 1.33  & 1.08 \\
			
			&{\bf  efficiency ratio} & 1.69  & 1.79  & 2.43  & 3.31 \\

			\cite{fouque2002variance}  
			& sd reduction ratio & 4.78  & 8.46  & 12.71 & 18.78 \\

			& time consumption ratio & 9.67  & 36.12 & 38.49 & 39.45 \\
			
			&{\bf  efficiency ratio} & 1.54  & 1.41  & 2.05  & 2.99 \\
			
			Duo-CE & sd reduction ratio & 4.21  & 10.96 & 24.70 & 60.11 \\

			& time consumption ratio & 1.18  & 2.27  & 2.91  & 3.28 \\
			
			&{\bf  efficiency ratio} & 3.88  & 7.27  & 14.48 & 33.20 \\
			
			& stage-1 time proportion & 0.17 & 0.34  & 0.46  & 0.41 \\
			Duo-V 
			& sd reduction ratio & 4.33  & 11.33 & 24.72 & 60.13 \\

			& time consumption ratio & 1.32  & 1.87  & 1.68  & 1.63 \\
			
			& {\bf efficiency ratio} & 3.76  & 8.29  & 19.05 & 47.15 \\
			& stage-1 time proportion & 0.49 & 0.22  &  0.06 & 0.06 \\
			\hline
		\end{tabular}%
		
	}

	\label{tab:sv toy}%
\end{table}%

For comparison, we include three other methods:
(i) the LD-based exponential tilting method in \cite{collamore2002importance}; (ii)
the moderate deviation (MD) based exponential tilting method in \cite{fuh2007estimation}; (iii) the method in \cite{fouque2002variance} derived from the diffusion models that is specialized for SV models (note that this method belongs to the class of state-dependent IS methods). Our proposed duo-exponential tilting is implemented by minimizing variance (Duo-V) and minimizing the cross entropy (Duo-CE). 

In Table \ref{tab:sv toy}, we use the plain Monte Carlo method as a benchmark to compare the four methods.
The importance sampling method in \cite{fouque2002variance} 
appears to have the largest variance reduction when the probability of the event is not extreme. However, it always takes considerable computational time because the tilting probability there has to be recomputed via an expansion at each time step. In contrast, in our method, the tilting parameters are first computed via an optimization procedure and then fixed in the sampling stage; our method has lower computational costs. 
The proposed algorithm seems to outperform the others in terms of the efficiency ratio, with the probability being around $10^{-2}\sim 10^{-5}$. 

Our algorithm outperforms the LD- and MD-based exponential tilting methods in \citet{collamore2002importance, fuh2007estimation} because the family of sampling distributions embedded in their one-parameter exponential tilting (induced by the eigenvalues and eigenfunctions) is not large enough. 
More precisely, the duo-exponential family in our algorithm includes theirs as special cases and strictly enlarges the search space of alternative distributions. Thus, it is not surprising that our algorithm performs better. Minimizing the cross-entropy (Duo-CE) leads to a similar variance reduction to our method.
The variance reduction of the cross-entropy is typically slightly worse than Duo-V since it does not directly minimize the variance of the estimator. In addition, we notice that the cross-entropy method usually takes longer time to find the optimal tilting parameters.

\subsection{Illustrative example (II): Tandem queue}
\label{subsec:tandem queue}

Consider a tandem queue model with $d$ stations, with service rates $\mu_i$ and arrival rate $\mu_0>0$. We are interested in the probability that the total number of jobs reaches a threshold $K$ before the queue is empty, a well-known example analyzed by \citet{glasserman1995analysis} and many others. When $\mu_0<\min_{1\leq i \leq d}\mu_i$, this probability is small if $K$ is large. Without loss of generality, assume $\sum_{i=0}^d \mu_i=1$. The problem can be embedded into a Markov chain $\{X_n,n\geq 0\}$, where $X_n = (X_n^1,\cdots,X_n^d)^\top\in \mathbb R^d$ is the number of jobs at each station, and whose transition probability is described as follows: Let $A\in \mathbb{R}^{d \times (d+1)}$ be
\[A =
\begin{pmatrix}
	1 & -1  & \cdots & 0 & 0\\
	0 & 1 & \cdots & 0 & 0\\
	\vdots &  \vdots & \ddots & \vdots & \vdots\\
	0 &  0 & \cdots & 1 & -1
\end{pmatrix}  = : (\alpha_0,\cdots,\alpha_d),
\]
where $\alpha_i$'s are column vectors in $A$ and represent all possible directions; given $X_{n}=x$, $\p(X_{n+1}= x + \alpha_i|X_{n}=x)=\mu_i$ if $x$ is an interior point; and $\p(X_{n+1}= x + \alpha_i|X_{n}=x)=\frac{\mu_i}{\sum_{j:} \mu_j}$, if $x$ is on the boundary, where $\sum_{j:}$ means the summation over index $j$ such that  $x+\alpha_j$ is attainable from $x$. We are interested in the probability $\p_{e_1}(T_K<T_0)$, where $T_k = \inf\{n>0: \sum_{i=1}^d X_n^{i} = k\}$, $e_1=(1,0,\cdots,0)^\top \in \mathbb{R}^d$.

This queuing example is a well-known case showing the 
superiority of the state-dependent importance sample methods for a Markov model (e.g., \citealt{dupuis2007subsolutions,dupuis2007dynamic,blanchet2012lyapunov}). We compare our method with
them under two different configurations, 
one ``easy" setting and one ``difficult" setting as identified in \citet{glasserman1995analysis}. In particular, we implement the method in \citet{blanchet2012lyapunov} who give the explicit expressions of the tilting distribution.

\paragraph{\textbf{Our tilting family}} $X_n$ is the underlying Markov chain and $S_n = X_n$, meaning that the additive part degenerates. We can verify that the Poisson equation \eqref{eq:poisson 0} still has a solution in the form of $g(x) = - x$ since $(I - \mathcal{P}_0)g(x) = -x + \E_x\left[ X_1 \right]$, and $\E_x\left[ X_1 - X_0 \right] - \E_{\pi}\left[X_1 - X_0 \right] = \E_x\left[ X_1\right] - x$. Thus, Theorem \ref{thm:lan optimal linear} suggests a linear link function $k(x,x',\eta) = \eta^\top x'$, and the tilting distribution is 
$\frac{\dd \p_{\eta}}{\dd \p} = \exp\{  \eta^\top X_{n+1} - \phi(X_n,\eta) \}, $
\[\phi(x,\eta) = \log \E_x\left[ e^{\eta^\top X_1} \right] = \left\{  
\begin{aligned}
	& \eta^\top x + \log \sum_{i=0}^{d} \mu_i e^{\eta^\top \alpha_i} , \text{ if $x$ is an interior point;} \\
	& \eta^\top x + \log \sum_{i=0}^{d} \frac{\mu_i}{\sum_{j:}\mu_j} e^{\eta^\top \alpha_i} , \text{ if $x$ is on the boundary.}
\end{aligned}
\right.   \]
Under the tilted distribution $\p_{\eta}$, given $X_{n}=x$, $\p_{\eta}(X_{n+1}= x + \alpha_i |X_{n}=x )=\frac{\mu_i e^{\eta^\top \alpha_i}}{\sum_{j=0}^d \mu_j e^{\eta^\top \alpha_j} }$ if $x$ is an interior point; $\p_{\eta}(X_{n+1}= x + \alpha_i | X_{n}=x)=\frac{\mu_i e^{\eta^\top \alpha_i}}{\sum_{j:}  \mu_j e^{\eta^\top \alpha_j} }$ if $x$ is on the boundary and $x+\alpha_i$ is attainable from $x$.

Although our aim is not to deduce any 
form of state-dependent change-of-measure and our tilting parameter $\eta$ remains a constant,  the induced tilting family is indeed state-dependent via a link function. Depending on the current position of the state $x$ (boundaries or interior), the expression of $\phi$ and the resulting transition probability $\p_{\eta}$ are different. Finally, we optimize over $\eta\in \mathbb R^{d}$ via the two-stage procedure as in Section \ref{sec:3.2}.

\paragraph{\textbf{Importance sampling in \citet{glasserman1995analysis}}} \citet{glasserman1995analysis} suggest simulating from the system with $(\mu_{0}',\mu_1',\cdots,\mu_{d-1}',\mu_d') = (\mu_{d},\mu_1,\cdots,\mu_{d-1},\mu_0)$ and then adjusting by the likelihood ratio. That is, under the tilted distribution $\tilde \p$, given $X_{n}=x$, $\tilde \p(X_{n+1}= x + \alpha_i |X_{n}=x )=\frac{\mu_i' }{\sum_{j=0}^d \mu_j' }$ if $x$ is an interior point; and $\tilde \p(X_{n+1}= x + \alpha_i | X_{n}=x)=\frac{\mu_i'}{\sum_{j:}  \mu_j' }$ if $x$ is on the boundary and $x+\alpha_i$ is attainable from $x$.

\paragraph{\textbf{Importance sampling in \citet{blanchet2012lyapunov}}} The change of measure in \citet[Section 6.1]{blanchet2012lyapunov} is constructed by $\bar{\p}(X_{n+1}\in x+ \alpha_i |X_n = x) = \sum_{j=0}^{d}w^{(j)}(x)\rho_{\theta_j}( \alpha_i|x)$, where $\theta_j\in \mathbb R^d$ and each entry is given by $\theta_j^i = \gamma \one_{\{1\leq i \leq d-j\}}$, for $1\leq i\leq d,0\leq j\leq d$, $\gamma = -\log\frac{\mu_0}{\min_{1\leq i\leq d}\mu_{i}}$. For a fixed $\theta$, $\rho_{\theta}(\alpha_i| x)=\frac{\mu_i\exp\{\theta^\top \alpha_i - \psi(x, \theta)\}}{\sum_{i:}  \mu_i}$, where $\psi(x,\theta) =\log \E_x[e^{\theta^\top (X_1-x)}] = \log\sum_{i:} \frac{\mu_i e^{\theta^\top \alpha_i}}{\sum_{i:} \mu_i} = \phi(x,\theta) - \theta^\top x$. The weight $w^{(j)}(x) = \frac{\bar{w}^{(j)}(x)}{ \sum_{k=1}^d \bar{w}^{(k)}(x) } $, where $\bar{w}^{(j)}(x) = \exp\{ \theta_j^\top x/\Delta -\gamma/\Delta -j \log\Delta \}$. Here $\Delta$ is a scaling parameter and is taken as $\Delta = 1$ in our implementation.

For the numerical illustration, we consider $d=2$ under two different configurations. The first configuration is $(\mu_0,\mu_1,\mu_2) = (0.1, 0.6, 0.3)$, which is identified as the ``easy" case by \citet{glasserman1995analysis}. The second configuration is $(\mu_0,\mu_1,\mu_2) = (0.2, 0.4, 0.4)$, which is identified as the ``difficult" case by \citet{glasserman1995analysis}. The numerical results are presented in Table \ref{tab:tandem queue}. 

Panel (a) stands for the ``easy" case, in which the method in \citet{glasserman1995analysis} already achieves significant variance reduction. Compared to \citet{glasserman1995analysis}, all other methods can achieve smaller variance, but they differ in the time consumption because of computational complexity. In particular, \citet{blanchet2012lyapunov} is the most computationally intensive. Our proposed Duo-V has the largest efficiency ratio and has significantly less computation time compared to \citet{blanchet2012lyapunov}. Duo-CE has the second-best performance and is close to Duo-V (with slightly higher computational time). 
Panel (b) is for the ``difficult" case, in which the method suggested in  \citet{glasserman1995analysis} only has limited variance reduction. \citet{blanchet2012lyapunov} achieves much smaller variance, while the proposed tilting family (both Duo-V and Duo-CE) yields even higher standard deviation reduction. The computational cost of \citet{blanchet2012lyapunov} remains the highest. Our method has the highest variance reduction and efficiency.

\begin{table}[htbp]
	\centering
	\caption{\textbf{Numerical results of different methods}. Estimation of $\p_{e_1}(T_K<T_0)$, where $T_k = \inf\{n>0: X_n^{1} + X_n^2 = k\}$, $X_t$ is a two-station tandem queue, and $e_1=(1,0)^\top$. The sample size for each column is $10^6,10^7,10^8,10^8$ respectively. }
	
	{\small
		
		\begin{subtable}[t]{\textwidth}
			\centering
			\caption{The ``easy" case with model configuration $\mu_0 = 0.1$, $\mu_1 = 0.6$, and $\mu_2 = 0.3$.}
			\begin{tabular}{llcccc}
				\toprule
				& $K$     & 5     & 7     & 9     & 11 \\
				\midrule
				Plain MC & mean  & 1.59E-02 & 1.80E-03 & 2.01E-04 & 2.29E-05 \\
				& standard deviation    & 1.25E-04 & 1.34E-05 & 1.42E-06 & 4.78E-07 \\
				& elapsed time & 4.15  & 34.51 & 531.76 & 285.28 \\
				\citet{glasserman1995analysis}	& sd reduction ratio & 11.30 & 30.62 & 87.46 & 259.48 \\
				& time consumption ratio & 2.11  & 3.50  & 2.82  & 6.71 \\
				& \textbf{efficiency ratio} & 7.78  & 16.36 & 52.12 & 100.17 \\
				\citet{blanchet2012lyapunov}	& sd reduction ratio & 9.65  & 28.96 & 87.05 & 264.17 \\
				& time consumption ratio & 5.92  & 9.39  & 7.98  & 17.46 \\
				& \textbf{efficiency ratio} & 3.97  & 9.45  & 30.82 & 63.22 \\
				Duo-CE	& sd reduction ratio & 12.10 & 30.87 & 87.50 & 259.27 \\
				& time consumption ratio & 2.50  & 3.48  & 3.02  & 6.72 \\
				& \textbf{efficiency ratio} & 7.66  & 16.54 & 50.31 & 99.98 \\
				& stage-1 time proportion & 0.12 & 0.01  & 0.01  & 0.01 \\
				Duo-V	& sd reduction ratio & 12.82 & 34.78 & 98.44 & 287.96 \\
				& time consumption ratio & 2.51  & 3.51  & 2.85  & 6.68 \\
				& \textbf{efficiency ratio} & 8.09  & 18.56 & 58.33 & 111.43 \\
				& stage-1 time proportion & 0.19 & 0.03  & 0.01  & 0.01 \\
				\bottomrule
			\end{tabular}%
		\end{subtable}
		
		\vspace*{2mm}
		
		\begin{subtable}[t]{\textwidth}
			\centering
			\caption{The ``difficult" case with model configuration $\mu_0 = 0.2$, $\mu_1 = 0.4$, and $\mu_2 = 0.4$.}
			\begin{tabular}{llcccc}
				\toprule
				& $K$     & 7     & 11    & 15    & 19 \\
				\midrule
				Plain MC & mean  & 4.70E-02 & 4.86E-03 & 4.26E-04 & 3.36E-05 \\
				& standard deviation    & 2.12E-04 & 2.20E-05 & 2.06E-06 & 5.80E-07 \\
				& elapsed time & 7.05  & 78.15 & 758.43 & 702.56 \\
				\citet{glasserman1995analysis} & sd reduction ratio & 1.55  & 2.35  & 1.42  & 5.12 \\
				& time consumption ratio & 1.75  & 3.06  & 4.38  & 6.14 \\
				& \textbf{efficiency ratio} & 1.17  & 1.34  & 0.68  & 2.07 \\
				\citet{blanchet2012lyapunov} & sd reduction ratio & 3.92  & 7.41  & 17.26 & 33.32 \\
				& time consumption ratio & 4.85  & 7.48  & 10.86 & 15.68 \\
				& \textbf{efficiency ratio} & 1.78  & 2.71  & 5.24  & 8.41 \\
				Duo-CE & sd reduction ratio & 4.18  & 10.00 & 24.10 & 59.38 \\
				& time consumption ratio & 2.21  & 3.57  & 5.54  & 8.63 \\
				& \textbf{efficiency ratio} & 2.81  & 5.29  & 10.23 & 20.21 \\
				& stage-1 time proportion & 0.21 &  0.02 & 0.01  & 0.01 \\
				Duo-V & sd reduction ratio & 4.43  & 10.12 & 24.13 & 59.98 \\
				& time consumption ratio & 2.01  & 3.47  & 5.55  & 7.80 \\
				& \textbf{efficiency ratio} & 3.12  & 5.43  & 10.25 & 21.47 \\
				& stage-1 time proportion & 0.23 & 0.22  & 0.06  & 0.05 \\
				\bottomrule
			\end{tabular}%
		\end{subtable}
		
	}

	\label{tab:tandem queue}%
\end{table}%

\section{Two Applications}
\label{sec:5}

We consider two applications, modeling pandemics in Section \ref{subsec:sir} and measuring financial systemic risk in Section \ref{subsec:covar},
which are both challenging for the classical importance sampling methods for two reasons.
First, the system dynamics are not affine. Thus, the eigenfunctions are not easy to compute and usual regularity conditions may not be satisfied. Second, the quantity of interest may not be standard, and applying the large deviation principle is difficult. 
For example, to study systemic financial risk, one is interested in the risk of one institute conditioned on the other institute's defaults, 
which involves estimating the probability of a joint event of one first passage time and the terminal value of another correlated stochastic process.

\subsection{An overflow probability in a pandemic model}
\label{subsec:sir}

We consider a stochastic variant of the celebrated SIRD model in epidemics as the first application.
The total population is classified into four types: susceptible ($S_t$), infected ($I_t$), recovered ($R_t$), and deceased ($D_t$), 
\begin{equation}
	\label{eq:stochastic sir}
	\left\{
	\begin{aligned}
		\dd S_t & = -\frac{\alpha S_t I_t}{N_t}\dd t - \sqrt{\frac{\alpha S_t I_t}{N_t}} \dd W^1_t, \\
		\dd I_t & = (\frac{\alpha S_t I_t}{N_t} - (\beta + \gamma) I_t)\dd t + \sqrt{\frac{\alpha S_t I_t}{N_t}} \dd W^1_t - \sqrt{\beta I_t}\dd W^2_t - \sqrt{\gamma I_t}\dd W^3_t,\\
		\dd R_t & = \beta I_t\dd t + \sqrt{\beta I_t}\dd W^2_t,\\
		\dd D_t & = \gamma I_t \dd t + \sqrt{\gamma I_t}\dd W^3_t ,\\
		\dd N_t & = - \dd D_t.
	\end{aligned}\right.
\end{equation}
Here $\alpha$ stands for the transmission rate,  $\beta$  the recovery rate, $\gamma$ the death rate, and $N_t$ the number of the total population,  $ W^1_t, W_t^2, W^3_t$ are three independent Brownian motions. 
In the standard deterministic SIRD model, Brownian motions $ W^1_t, W_t^2, W^3_t$ disappear.
The deterministic counterpart without the deceased group is first introduced in \cite{kermack1927contribution} and has been extended to the stochastic case. For a review on stochastic epidemic models and the volatility structure in \eqref{eq:stochastic sir}, see \cite{allen2017primer}.

The SIRD model and its variants have been widely adopted in pandemic studies to predict the number of infected individuals; see, for example, \citet{caccavo2020chinese,athayde2022forecasting,sebbagh2022ekf}. We consider a particular quantity of this type: the time at which the number of existing cases exceeds the maximum capacity of a given medical system. Mathematically, it is the first passage time $\tau_{c} = \inf\{t \geq 0: I_t \geq c\}$ for a given $c > 0$. In particular, we want to estimate the overflow probability $\p(\tau_c < T)$ for a fixed number $T<\infty$ given initial condition $(N_0-I_0,I_0,0,0)$, where $I_0$ is the number of initial cases and $N_0$ is the total population. Note that $I = 0$ is an absorbing state.
The model parameters are taken from those in \citet[Table 1, Scenario II]{anastassopoulou2020data} with $(\alpha,\beta,\gamma) =  (0.319,0.1, 0.00147)$ and the reproduction number 
$R_0 = \frac{\alpha}{\beta + \gamma}=3.1438.$ 

Since $N_t = S_t + I_t + R_t$ and $\dd D_t = - \dd N_t$, the above system can be represented by a three-dimensional vector $X_t = (S_t,I_t,R_t)^\top\in \mathbb{R}^3$, such that 
$\dd X_t = b(X_t)\dd t + \sigma(X_t)\dd W_t$, where
\[b(X_t) = \begin{pmatrix}
	-\frac{\alpha S_t I_t}{N_t} \\
	\frac{\alpha S_t I_t}{N_t} - \beta I_t - \gamma I_t\\
	\beta I_t
\end{pmatrix},\ 
\sigma\sigma^\top(X_t) = \begin{pmatrix}
	\frac{\alpha S_t I_t}{N_t} & -\frac{\alpha S_t I_t}{N_t} & 0\\
	-\frac{\alpha S_t I_t}{N_t} & \frac{\alpha S_t I_t}{N_t} + \beta I_t + \gamma I_t & -\beta I_t \\
	0 & -\beta I_t & \beta I_t
\end{pmatrix} . \]

Due to  the nonlinear structure in the drift term, it is infeasible to compute eigenfunctions and eigenvalues or to solve the Poisson equation. 
However, our method only requires a link function $k(x,x',\eta) = \eta^\top B^\top(x) x'$, as a linear approximation to the optimal exponential martingale based on the discussion in Section \ref{sec:diffusion}.\footnote{Lemma \ref{thm:optimal foc} shows that the second moment is a convex function in $\eta$. It suffices to solve the first-order condition. Other algorithms, like sample average approximation, may also be applicable.}  
Under the tilted probability, the drift term becomes $\tilde b(X_t) = b(X_t) + \sigma\sigma^\top(X_t)B(X_t)\eta$.

Next, we explain the heuristics on how to choose the basis function based on the structure of this process. 
We aim to find the link function such that sampling is done under a similar stochastic system with a different parameter configuration. In particular, we use the following drift term with distinct parameters $(\alpha_{+},\alpha_-,\beta_{+},\beta_-,\gamma_-)$:
$ \tilde{b}(X_t) = 
\left( - \frac{\alpha_- S_t I_t}{N_t}, 
\frac{\alpha_+ S_t I_t}{N_t} - \beta_- I_t - \gamma_- I_t , \beta_+ I_t \right)^\top$. Let $\eta = (\alpha_- - \alpha, \alpha_+-\alpha,\beta_--\beta,\gamma_--\gamma,\beta_+-\beta)^\top\in \mathbb{R}^5$.
This yields $\tilde{b}(X_t) - b(X_t) = \sigma\sigma^\top(X_t)B(X_t) \eta$, where
\[B(X_t) = 
\begin{pmatrix}
	\frac{\alpha S_t I_t}{N_t} & -\frac{\alpha S_t I_t}{N_t} & 0\\
	-\frac{\alpha S_t I_t}{N_t} & \frac{\alpha S_t I_t}{N_t} + \beta I_t + \gamma I_t & -\beta I_t \\
	0 & -\beta I_t & \beta I_t
\end{pmatrix}^{-1}
\begin{pmatrix}
	-\frac{S_t I_t}{N_t} & 0 & 0 & 0 & 0\\
	0 & \frac{S_t I_t}{N_t} & -I_t & -I_t & 0\\
	0 & 0 & 0 & 0 & I_t
\end{pmatrix}.  \]

The details to derive the tilting probability and the likelihood ratio are included in the E-Companion \ref{appendix:sird}. 
The variance reduction is achieved by a numerical minimization procedure within the chosen exponential family. However, there is no guarantee in the asymptotic sense based on Definitions \ref{def:asymptotic optimal lan} or \ref{def:log efficiency normal main text} due to the lack of regularity conditions in the original SIRD model. To the best of our knowledge, there is no discussion of how classical importance sampling 
can be applied to the SIRD model due to the same technical difficulties. Our framework provides 
at least one feasible solution to it within our framework of importance sampling methodologies.

Simulation results are presented in Table \ref{tab:sird}, with discretization size $\Delta t= 1$.
In our SIRD model, the number of infected individuals is not equivalent to the number of hospitalized or quarantined individuals. Simulating the hospitalized population requires more refined compartmental epidemiology models, for example, to account for different levels of severity, which is not considered here. Simulating the number of infected individuals with the SIRD model may provide guidance to policymakers in understanding the potential consequences of the epidemic, such as how to set up agile hospital capacity to admit patients.
Although the computational time of our methods is almost double, partly due to the searching procedure relying on additional matrix operations, the proposed method has a much better efficiency ratio than the plain Monte Carlo method due to significant variance reduction.

\begin{table}[!htbp]
	\centering
	\caption{Estimation results of the plain MC estimator and our IS estimator. In the simulation, we fix $N_0=5\times 10^6$ (a city with a population of 5 million) and $I_0 = 100$ (initial infected cases). The parameters in the SIRD model are taken as $(\alpha,\beta,\gamma) = (0.319,0.1, 0.00147)$. We estimate the probability $\p(\tau_c < T)$ with $T = 100$. The sample size is taken as $N=10^6,10^7,10^8,10^8$ for each instance.}
	
	{\small
		
		\begin{tabular}{llcccc}
			\hline
			& $c/N_0$     & 0.3325 & 0.3329 & 0.3331 & 0.3333 \\
			\hline
			Plain MC & Mean  & 3.56E-02 & 1.99E-03 & 3.11E-04 & 3.69E-05 \\
			& standard deviation   & 1.85E-04 & 1.41E-05 & 1.76E-06 & 6.08E-07 \\
			& elapsed time & 41.89 & 341.21 & 3217.82 & 3232.38 \\
			
			Duo-CE & mean & 3.56E-02 & 1.99E-03 & 3.09E-04 & 3.61E-05 \\
			
			& standard deviation    & 7.03E-05 & 1.84E-06 & 1.08E-07 & 1.52E-08 \\
			& sd reduction ratio & 2.64  & 7.67  & 16.21 & 39.62 \\
			
			& time consumption ratio & 1.86  & 1.94  & 1.98  & 1.68 \\
			
			&  \textbf{efficiency ratio} & 1.93  & 5.51  & 11.53 & 30.60 \\
			& stage-1 time proportion & 0.13 & 0.06  & 0.03  & 0.05 \\
			
			Duo-V & mean  & 3.57E-02 & 1.99E-03 & 3.09E-04 & 3.61E-05 \\
			& standard deviation   & 6.86E-05 & 1.79E-06 & 1.06E-07 & 1.50E-08 \\
			
			& sd reduction ratio & 2.71  & 7.88  & 16.60 & 40.27 \\
			
			& time consumption ratio & 1.84  & 1.89  & 1.92  & 1.67 \\
			
			&  \textbf{efficiency ratio} & 2.00  & 5.73  & 11.98 & 31.12\\
			& stage-1 time proportion & 0.06  & 0.02  &  0.02 & 0.02 \\
			\hline
		\end{tabular}%
		
	}

	\label{tab:sird}%
\end{table}%

There are debates on the trade-off between controlling the spread of pandemics and lower economic growth. An accurate estimation is required to analyze such a trade-off.
To gain insight from our model, we shall examine the sensitivity of $\alpha$ on the probability of the event.\footnote{
	The other two parameters, $\beta$ and $\gamma$, represent the disease’s recovery rate and death rate. They will remain relatively stable if there is no breakthrough in treatment or significant mutations in the viruses.} 
$\alpha$ represents the intensity of the interaction between the infected and susceptible population, which determines the transmission speed. Moreover, its value is controllable by government policy, 
e.g.,  the restriction of social interactions.

\begin{figure}[!htbp]
	\centering
	\includegraphics[width=0.5\textwidth]{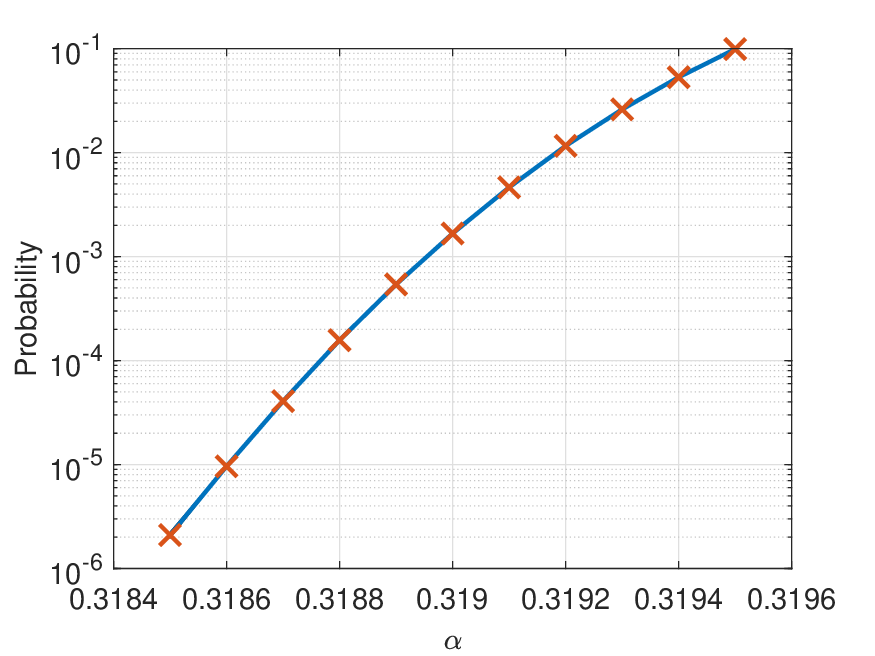}
	\caption{
		The estimated probability for the SIRD model under various parameters. When $\alpha$ increases from 0.3185 to 0.3195, the overflow event’s probability scale rises from $10^{-6}$ to $10^{-1}$. 
		The rest of the parameters are: $\beta = 0.1$, $\gamma = 0.00147$. The initial value is $N_0=5\times 10^6$, $I_0 = 100$. The critical threshold $c = bN_0$, $b = 0.337$.
	}
	\label{fig:sird alpha}
\end{figure}

Figure \ref{fig:sird alpha} depicts the overflow probability. The probability varies drastically when $\alpha$ is perturbed at around $10^{-4}$. 
as the probability grows nearly exponentially in $\alpha$. 
For example, when $\alpha=0.319$, the derivative is around 20.
Such sensitivity is not caused by the reproduction number $R_0$ being close to a critical value, as the range of $R_0$ is between $[3.139, 3.149]$. Thus, the overflow probability appears very sensitive to 
$\alpha$.

This observation has two implications: First, any marginal effort to limit the transmission and social interaction may have a significant potential impact on preventing the overflow of resources. 
Second, it calls for more accurate parameter estimation, as estimation errors in the transmission speed can significantly change the resulting overflow probability.

\subsection{CoVaR simulation in finance}
\label{subsec:covar}

To study the systemic risk during the 2007-2008 financial crisis, \cite{adrian2016covar} extends the definition of VaR by introducing CoVaR, which is
the quantile of one loss conditioned on another loss being at or lower than its $-VaR_q^i$ level. Here $VaR_q^i$ is defined as the $1-q$ quantile when $X^i$ stands for profit/return. 
More precisely, $CoVaR_q^{j|C(X^i)}$ is defined as the quantity such that
\begin{equation}
	\label{eq:covar def}
	\p(X^j\leq -CoVaR_q^{j|C(X^i)}|C(X^i)) = 1-q,
\end{equation}
where $q\in (0,1)$ and $C(X^i)$ is an event of $X^i$, e.g. $\{X^i \leq -VaR_q^i\}$ or $\{X^i = -VaR_q^i\}$. Therefore,
CoVaR characterizes the tail dependence and systemic co-risk between two random variables. 

Here, we would like to extend the definition of CoVaR 
from bivariate random variables to a bivariate process $\{(X^i_t,X^j_t),t\geq 0\}$. 
More precisely, $CoVaR_q^{j|C(X^i_{\cdot})}$ is defined to be the quantile of $X^j_T$ given the event $C(X^i_{\cdot})$ that depends on the path of the whole process $\{X^i_t\}_{t\geq 0}$, i.e.,
\begin{equation}
	\label{eq:covar general def}
	\p(X_T^j\leq -CoVaR_q^{j|C(X^i_{\cdot})}|C(X^i_{\cdot})) = 1-q.
\end{equation}
The conditional event $C(X^i_{\cdot})$ can be quite general, e.g., the level of terminal value $\{X^i_T \leq \bar{x}^i\}$ or an endogenous default event $\{\tau^i_b \leq T\}$, where $\tau^i_{b} = \inf\{n\geq 0:S_n^i \leq b\}$ is the first time that institute $i$'s value falls below a level $b$, and $T$ is a given finite time. 

We consider a VAR-GARCH model for the return of three institutions. For simplicity, we assume the initial condition is $y_0 = 0$ and $H_1 = W$. 
\begin{equation}\label{eqn:vargarch}
	\left\{
	\begin{aligned}
		& y_t =\mu + \rho y_{t-1}+z_t, \ z_t = H_t^{\frac{1}{2}}\epsilon_t, \\
		& H_{t+1} = W+A \odot z_tz_t^\top + B \odot H_t,
	\end{aligned}\right.
\end{equation} 
where $\epsilon_t$ is the standard normally distributed innovation. $W,A,B,H_t$ are symmetric $3\times 3$ matrices, $\mu\in \mathbb{R}^3, \rho\in \mathbb{R}^{3\times 3}$, and $\odot$ denotes the Hadamard product/entry-wise product. The accumulative return is $S_n = \sum_{i=1}^n y_i$. Among the three institutions, we index its first component by $S^0$ 
which is used in the conditional event; for example, this institution can simply be the whole financial market. 
The other two components are indexed by $S^1,S^2$ and are regarded as two individual institutes. 
The multivariate GARCH model is taken in the simple diagonal form, following the specification in \cite{adrian2016covar}. Other studies on CoVaR using GARCH models can be found in \cite{girardi2013systemic} and \cite{white2015var}. 
Note that the GARCH model does not fit the classical exponential tilting importance sampling framework, as the eigenvalues-eigenfunctions are unavailable. 

We want to estimate the CoVaR of $S_T^2$, conditioned on $\tau^0_{b_0} \leq T$, where $\tau^0_{b_0} = \inf\{n\geq 0:S_n^0 \leq b_0\}$ is the first passage time. 
We first estimate 
\begin{equation}
	\label{eq:estimate 2}
	\frac{\p_{0,W}(S^j_T \leq b, \tau^0_{b_0} \leq T)}{\p_{0,W}( \tau^0_{b_0} \leq T)},
\end{equation}
for $j=1,2$ and any $b$. Then we find $b$ such that the above conditional probability is $1-q$ by bisection. The conditional event means that institution 1's asset accumulative return falls below a threshold and can be interpreted as institution 1's default before $T$. Thus, we can investigate the risk exposure of one institution in the event of another institution's default. 
The event in the denominator is common in the importance sampling literature, while the joint probability in the numerator is not standard in the literature.

By taking $\{ (y_t,H_{t+1}),t\geq 0\}$ as the underlying Markov chain, 
$\{(y_t,H_{t+1},S_t),t\geq 0\}$ still belongs to the general Markov random walk framework. However, the following eigenvalue problem becomes difficult to solve. 
We can apply our proposed importance sampling method to estimate the numerator and denominator respectively. 
Thus, we only need to choose the link function via the Poisson equation, which can be solved analytically and satisfy the required regularity conditions (see the E-Companion \ref{appendix:vargarch} and \ref{appendix:verify assumptions} for details). Theorem \ref{thm:lan optimal linear} can be applied to justify that our choice of link function is asymptotically optimal within the LAN family. The solution to the Poisson equation \eqref{eq:poisson garch} for this VAR-GARCH model turns out to be a linear function of $y$, the first component of the Markov process $(y_t,H_{t+1})$.
Thus, it suffices to use a linear link function in $y'$, i.e., to take $k\left( (y,H),(y',H'),\eta\right) = \eta^\top y'$.

In the simulation, model parameters are taken as $\mu = 0$, $W = \frac{1}{360} \times \begin{pmatrix}
	0.2 & 0.1 & 0.01 \\
	0.1 & 0.4 & 0 \\
	0.01 & 0 & 0.9
\end{pmatrix}$, $\rho =\frac{1}{360} \times \begin{pmatrix}
	0.3 & 0.05 & 0.1 \\
	0.1 & 0.2 & 0.3 \\
	0.1 & 0.3 & 0.2
\end{pmatrix}$, $A = \frac{1}{360}\times \begin{pmatrix}
	0.0815 & 0.091 & 0.0203 \\
	0.0910 & 0.0632 & 0.0322 \\
	0.0203 & 0.0322 & 0.0958
\end{pmatrix}$ and $B =\frac{1}{360} \times \begin{pmatrix}
	0.193 & 0.1115 & 0.1112 \\
	0.1115 & 0.0971 & 0.1222 \\
	0.1112 & 0.1222 & 0.1831
\end{pmatrix}$.

The simulation results are presented in Table \ref{tab:garch stopping}. To illustrate the effectiveness of variance reduction, we choose $j=1$. The efficiency ratio is very significant, ranging from $75.57$ to $199.42$, indicating the good performance of our proposed method. 
Compared to other examples, the efficiency ratio in Table \ref{tab:garch stopping} seems more ``impressive''. This is because of the smaller joint probabilities of events. 
Indeed, we consider a conditional probability in CoVaR, which equals the joint probability divided by the marginal probability (around 0.01). 
While the conditional probability has a range from $10^{-2}$ to $10^{-4}$, 
the joint probability has a range from $10^{-4}$ to $10^{-6}$. We only compare to plain Monte Carlo because we did not find any literature addressing rare event simulation for the VAR-GARCH model.

\begin{table}[!htbp]
	\centering
	\caption{VAR-GARCH model to estimate $\p(S^1_T < b_1|\tau^0_{b_0}\leq T)$. Here $T=5$. The standard deviation and its reduction ratio are computed by empirical standard deviation by repeating 100 times. The time consumption is taken as the average time consumption.}
	
	{\small
		\begin{tabular}{llcccc}
			\hline
			
			&  $b_1, b_0$     & $-0.2, -0.12$ & $-0.25, -0.12$ & $-0.32, -0.12$ & $-0.37,-0.12$ \\
			\hline
			Plain MC & mean  & 2.91E-02 & 4.66E-03 & 1.59E-04 & 8.44E-06 \\
			& standard deviation    & 1.36E-03 & 5.15E-04 & 3.41E-05 & 2.26E-06 \\
			& elapsed time & 6.25  & 6.19  & 40.18 & 391.39 \\
			Duo-CE    & mean  & 2.91E-02 & 4.64E-03 & 1.64E-04 & 8.64E-06 \\
			& standard deviation    & 9.57E-05 & 1.66E-05 & 1.90E-07 & 3.31E-09 \\
			& sd reduction ratio      & 14.22 & 31.00 & 179.63 & 683.23 \\
			& time consumption ratio      & 1.44  & 1.44  & 2.06  & 2.19 \\
			& \textbf{efficiency ratio} & 11.86 & 25.79 & 125.10 & 461.27 \\
			& stage-1 time proportion & 0.16 & 0.11  & 0.15  & 0.15 \\
			Duo-V & mean  & 2.91E-02 & 4.64E-03 & 1.64E-04 & 8.64E-06 \\
			& standard deviation    & 9.08E-05 & 1.52E-05 & 1.89E-07 & 3.03E-09 \\
			& sd reduction ratio      & 14.99 & 33.88 & 180.06 & 744.47 \\
			& time consumption ratio      & 1.43  & 1.44  & 2.07  & 2.19 \\
			& \textbf{efficiency ratio} & 12.56 & 28.24 & 125.29 & 503.48 \\
			& stage-1 time proportion & 0.24  & 0.11  & 0.15  & 0.15 \\
			
			\hline
		\end{tabular}%
	}
	
	\label{tab:garch stopping}%
\end{table}%

\begin{figure}[!htbp]
	\centering
	\includegraphics[width=0.5\textwidth]{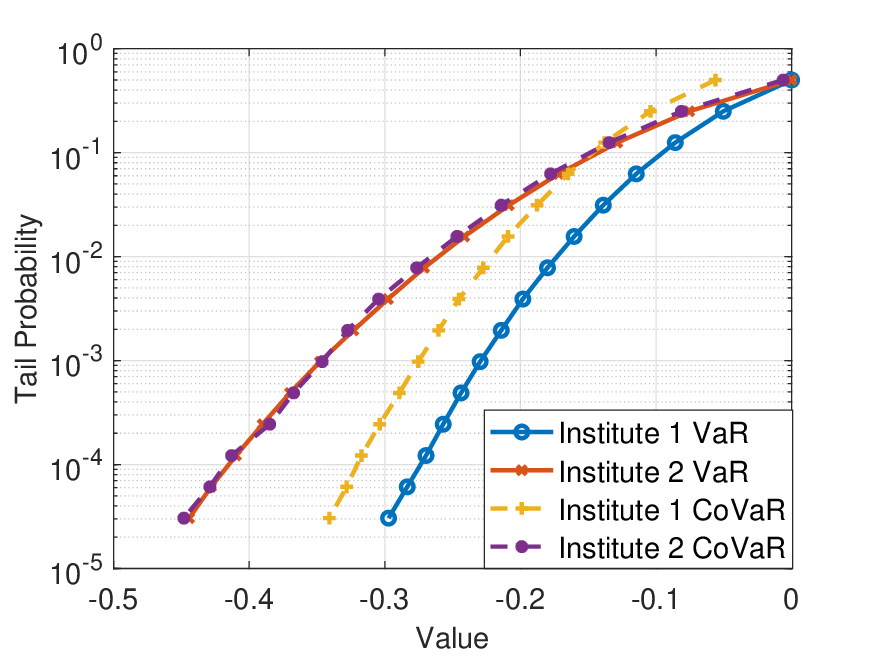}
	\caption{
		The estimated quantile for the VAR-GARCH model. It shows the quantile for the tail probability ranging from $2^{-1}$ to $2^{-15}$. The vertical axis is the tail probability $1-q$, and the horizontal axis is the corresponding conditional and unconditional $1-q$ quantiles of two different institutes $S^1_T,S_T^2$, that is, $-VaR_q^j,-CoVaR_q^j$ for $j=1,2$. The event conditioned on is fixed to be $\{\tau_{b_0}\leq T\}$, meaning that the institute 0 gets distressed before time $T$, where $b_0=-0.1$, $T=5$. Institute 1 significantly correlates with institute 0, so its VaR and CoVaR deviate. However, Institute 2 has little correlation with institute 0, so its VaR and CoVaR are almost identical.
	}
	\label{fig:quantile}
\end{figure}

Figure \ref{fig:quantile} depicts the patterns of the relation between VaR and 
CoVaR for the two institutions 1 and 2 conditioned on institution 0. 
In general, there is no clear link between
VaR and CoVaR. 

Our simulation results suggest two interesting observations.
First, at the fixed $q$ level, the relationship between CoVaR and VaR across different institutes can be very different. 
Cross-sectionally, institute 1 always has a higher VaR than institute 2. However, the CoVaR of institute 1 is larger only at smaller probabilities. In particular, at the 5\% risk level, institute 1 has higher VaR but lower CoVaR, suggesting that a safer institute can become much more vulnerable to systemic risk because of increased dependence on other distressed institutes. 
This is consistent with the empirical observation in \cite{adrian2016covar}, showing that the relationship between $VaR_q^j$ and $CoVaR^j_q$  is not significant at the fixed $q$ level.
Second, for varying $q$ levels, however, the CoVaR and VaR have similar trends within the same institution. 
For institute 2, its CoVaR and VaR are very close at all levels; for institute 1, CoVaR and VaR also exhibit a similar trend as the tail probability varies. 
This new observation suggests that an individual institute's co-risk may be well explained by its individual risk and an individual-specific factor measuring the strength of co-dependence. 
More precisely, our simulation results lead to an empirical hypothesis: at all risk levels $CoVaR_q^j \approx \beta^j VaR_q^j + \alpha^j$ for all $q$, rather than cross-sectionally $CoVaR_q^j \approx \beta_q VaR_q^j + \alpha_q$ for all $j$. Such a relation is discussed and derived in \cite{adrian2016covar} under the joint normal distribution for random variables. Here, we propose it for bivariate processes beyond random variables. 
The individual-specific feature $\beta^j$ can measure the risk in one institute exposed to the whole financial market.

\section{Some Remarks}
\label{sec:6}

Several limitations and future directions need to be  emphasized. Outside structured settings such as the LAN family, affine models, or the diffusion approximation developed in Section \ref{sec:diffusion}, the choice of a good link function remains partly heuristic and must often be judged empirically through variance reduction. Likewise, although the diffusion-based construction is guided by a PDE/Poisson approximation, we do not yet have a general theory linking approximation quality to logarithmic or near-logarithmic efficiency. It would also be interesting to compare the present framework more systematically with approximate-subsolution approaches and other state-dependent methods. On the computational side, adaptive first-stage optimization, warm starts, and ML-based parameterizations of link functions or related Poisson/HJB objects, as in \citet{qu2026deep,hult2024deep,nusken2021solving}, appear to be promising directions for extending the applicability of the framework.

\section*{Acknowledgment}
The authors thank two anonymous reviewers and the associate editor for constructive comments that help improve the paper. Jia is supported by the Start-up Fund at the Chinese University of Hong Kong and the Faculty Direct Fund 4055220. 

\bibliographystyle{informs2014}
\bibliography{demobib}

\ECSwitch


\ECHead{Electronic Companion to ``A General Framework for Importance Sampling with Markov Random Walks" by Cheng-Der Fuh, Yanwei Jia, and Steven Kou}

\section{Regularity Conditions}
\label{ec:technical conditions}

\subsection{Some Terminologies}
\label{appendix:regularity conditions}

A Markov chain $\{X_n,n \geq 0\}$ on a state space ${\cal X}$ is called {\it Harris recurrent} if there exist a recurrent set ${\cal R} \in {\cal B}({\cal X})$, a probability measure $\varphi$ on ${\cal R}$, a $\lambda > 0$ and an integer $n_0$ such that
\begin{equation}
	\label{min}
	P\{ X_n \in {\cal R}~{\rm for~some}~n \geq 1|X_0=x\} =1, ~~~P\{ X_{n_0} \in A |X_0=x\} \geq \lambda \varphi(A), 
\end{equation}
for all $x \in {\cal R}$ and $A \subset {\cal R}$.
It is known that under (\ref{min}), $X_n$ admits a regenerative scheme with i.i.d. inter-regeneration times for an augmented Markov chain, which is called the
``split chain;'' cf. \cite{meyn2012markov}.  Assumption \ref{ass:rigorous eigenvalues} guarantees that the Markov random walk $(X_n,S_n)$ admits a similar regenerative scheme.

Denote $\varrho_{\bar{a}}(0) = \varrho_{\bar{a}}$ and  let $\{\varrho_{\bar{a}}(j), j \geq 1\}$ be the times of consecutive visits to a recurrent state $\bar{a} \in {\cal X}$\footnote{It is known, cf. \cite{AthreyaNey1978} that if a Markov chain 
	$\{X_n, n \geq 0\}$ on a state space ${\cal X}$ with a countably generated $\sigma$-algebra is 
	Harris recurrent, then there exists an integer $n_0 > 1$ such that the sequence $\tilde{X}_n = X_{nn_0}, n = 0,1, 2, \ldots$ is a regenerative sequence of random variables. In fact, there exist random terms $\alpha_1,
	\alpha_2, \alpha_3, \ldots$ such that $\tilde{X}_{\alpha_i+j} : 0 \leq j \leq \alpha_{i+i} - \alpha_i -1,~ \alpha_{i+1}-\alpha_i\},~ i = 1, 2, \ldots$ are independent and identically distributed cycles.}. Let $\varrho= \varrho_{\bar{a}}$ be the first time $(>0)$ to reach the recurrent state $\bar{a}$ of the split chain, and define
$u(\alpha,\zeta)= E_{\nu} e^{\alpha S_\varrho-\zeta \varrho}$ for $\zeta \in {\mathbb R}$, where $\nu$ is an initial distribution on ${\cal X}$. Assume that
\begin{eqnarray}\label{mom}
	W:= \{(\alpha,\zeta):u(\alpha,\zeta) < \infty\}~{\rm is~an~open~subset~on}~{\mathbb R}^2.
\end{eqnarray}
\cite{ney1987markov1} shows that $D=\{\alpha:u(\alpha,\zeta) < \infty~{\rm for~some}~\zeta\}$ is an open set
and that for $\alpha \in D$, the transition kernel $\hat{P}_{\alpha}(x,A ) = E_{x}\{ e^{\alpha Y_1}\one_{\{X_1 \in A \}}\}$ has a maximal simple real
eigenvalue $e^{\Lambda(\alpha)}$, where $\Lambda(\alpha)$ is the unique solution
of the equation $u(\alpha,\Lambda(\alpha))=1$, with corresponding eigenfunction
$r(x,\alpha):= E_{x}\left[ \exp\{\alpha S_\varrho - \Lambda(\alpha) \varrho\}\right]$.


A Markov chain $\{X_n,n \geq 0\}$ on a state space ${\cal X}$ is called $V$-uniformly ergodic if there exist an invariant probability measure $\pi$, and a measurable function $V: {\cal X} \rightarrow [1,\infty)$, such that $\int V(x) \pi(\dd x) < \infty$, and
$$
\lim_{n \rightarrow \infty} \sup_{x \in {\cal X}} \bigg\{\frac{\big|E[h(X_n)|X_0=x]
	- \int h(x')\pi(\dd x')\big|}{V(x)}:|h| \leq V \bigg\} = 0, ~~~
\sup_{x \in {\cal X} } E_{x}\bigg\{\frac{V(X_1)}{V(x)} \bigg\} < \infty.
$$
\subsection{Regularity conditions}


\begin{assumption}
	\label{ass:convex}
	\begin{enumerate}[(i)]
		\item $\Theta$ is an open set containing 0, such that for a given $\theta \in \Theta $, $\E[e^{\theta^\top Y_1}|X_0=x,X_1=x'] < \infty$ for all $x,x' \in \mathcal{X}$.
		\item $H$ is an open set containing 0, such that for a given $\eta\in H$, $\E_x[e^{k(x,X_1,\eta)}] < \infty$ for all $x \in \mathcal{X}$. 
		\item When $\mathcal{X}$ is general (unbounded) state space, for all $(\theta,\eta)$ lying in a compact set, the functions $\psi(x,x',\theta),k(x,x',\eta),\phi(x,\eta)$ and their partial derivatives $\frac{\partial \psi}{\partial \theta}(x,x',\theta), \frac{\partial k}{\partial \eta}(x,x',\eta),\frac{\partial \phi}{\partial \eta}(x,\eta)$ all have linear growth in $x,x'$. There exist positive constant $L, C_0 > 0$ such that for any $\ell \leq L$, $\E_x\left[ e^{\ell |X_1|} \right] \leq e^{ C_0 \ell (1 + |x|)}$ for any $x\in\mathcal X$, and $\E_{\nu}\left[ e^{\ell |X_0|}  \right] < \infty$.
	\end{enumerate}
\end{assumption}

Assumption \ref{ass:convex} $(i)$ and $(ii)$ are necessary to guarantee the existence of two ``cumulant generating functions'' and hence necessary for the feasibility of the exponential tilting. This assumption is common in exponential tilting literature. The condition $(iii)$ is the growth and moment condition needed for (unbounded) general state Markov chain to guarantee the finiteness of quantities in our methods.

Next, we state two assumptions commonly imposed in the Markov random walk literature.

\begin{assumption}
	\label{ass:recurrent}
	The underlying Markov chain $\{X_n, n \geq 0\}$ is aperiodic, irreducible and
	Harris recurrent (i.e., \eqref{min} holds).
\end{assumption}

Assumption \ref{ass:recurrent} is a standard assumption of the underlying Markov chain $\{X_n, n \geq 0\}$, cf. \cite{ney1987markov1}, which also guarantees the existence of the stationary probability measure $\pi$. The Harris recurrent condition is defined in Section \ref{appendix:regularity conditions} (cf. \citealt{meyn2012markov}).

\begin{assumption}
	\label{ass:rigorous eigenvalues}
	When the additive part does not degenerate, i.e., $Y_n|X_{n-1},X_n$ is not deterministic, there exists a measurable function $h:\mathcal{X} \to \mathbb R_+$ and a probability measure $\mu$ on $\mathcal X \times\mathbb R^d$ such that $\int_{A\times B} p(x,\dd x') \rho(y|x,x')\dd y\geq  h(x)\mu(A\times B)$. 
\end{assumption}

Assumption \ref{ass:rigorous eigenvalues} is often called ``minorization condition'' (see, e.g., \citealt{ney1987markov1}).  It is a common assumption in the study of Markov random walk, and together with Assumption \ref{ass:convex}$(i)$ implies that $\{(X_n,S_n),n\geq 0\}$ is Harris recurrent and admits a regenerative scheme. More discussions on this assumption can be found in E-Companion \ref{appendix:regularity conditions}. The validity of Assumption \ref{ass:rigorous eigenvalues} 
may affect the theoretical properties of the algorithms but not the implementation of algorithms.

The next assumption is necessary for all LD-based importance sampling methods that require the existence of the eigenvalue and eigenfunction. 

\begin{assumption}
	\label{ass:eigenvalue exist}
	The solution to eigenvalue-eigenfunction problem \eqref{eq:eigenvalue} exists. Moreover, the positive eigenfunction $r(x,\alpha)$ is finite and analytical. 
\end{assumption}

There are sufficient conditions to ensure Assumption \ref{ass:eigenvalue exist}; 
see,  e.g., \eqref{mom} in E-Companion \ref{appendix:regularity conditions}, under which, Theorem 4.1 and Lemma 4.5 in \citet{ney1987markov1} show that the eigenvalue-eigenfunction problem \eqref{eq:eigenvalue} has a simple maximum eigenvalue whose eigenfunction is positive, finite, and analytical for $x$ in a ``full'' set of $\mathcal X$. Sufficient conditions to extend the existence to all $x\in \mathcal X$ are given in \citet{Chan2003}. 
However, we choose to directly state it as an assumption because finding eigenvalues and eigenfunctions is not required to implement our algorithm, and we only need this assumption to establish the linkage to existing LD-based, exponential tilting methods.

\begin{assumption}
	\label{ass:rigorous}
	The solution to the Poisson equation \eqref{eq:poisson 0} exists.
\end{assumption}

Assumption \ref{ass:rigorous} is required to establish the optimal link function in Theorem \ref{thm:lan optimal linear}. A sufficient condition for Assumption \ref{ass:rigorous} is Assumption \ref{ass:recurrent} and that the Markov chain $\{X_n, n \geq 0 \}$ is $V$-uniformly ergodic (to be defined in Section \ref{appendix:regularity conditions}), cf. Theorem 17.2 in \cite{meyn2012markov}.
For the examples we considered, either one can always find an explicit solution to the Poisson equation \eqref{eq:poisson 0}, or we do not need to solve \eqref{eq:poisson 0}. Typically, Assumption \ref{ass:rigorous} is a weaker condition than Assumption \ref{ass:eigenvalue exist}, 
as will be demonstrated in Lemma \ref{lemma:poisson}.

\begin{assumption}
	\label{ass:eigenvalue bounded}
	The eigenfunction in \eqref{eq:eigenvalue} satisfies $\limsup_{\alpha \to 0}\E_{\nu}\left[  r(X_0,\alpha)^2  \right] < \infty$ and $\liminf_{\alpha \to 0}\inf_{x\in \mathcal{X}} r(x,\alpha) > 0$.
\end{assumption}

Assumption \ref{ass:eigenvalue bounded} is required in Theorem \ref{thm:log efficiency} to show the proposed importance sampling method has logarithmic efficiency under asymptotic normal regimes. Commonly used sufficient conditions for Assumption \ref{ass:eigenvalue bounded} can be found, e.g., in \citet[Remark 2.3]{collamore2002importance} and \citet[Example 1]{Chan2003}. Note that Assumption \ref{ass:eigenvalue bounded} is mild because only the behavior at the neighborhood of $\alpha = 0$ is needed.


\section{Optimal Tilting (but not practical) under Exponential Martingale for Diffusion Processes}
\label{appendix_control}
Consider the stochastic model
$\dd X_t = b(X_t)\dd t + \sigma(X_t)\dd W_t,\ X_0 = x_0, $
and we want to estimate $\E_{x_0}[F(X_T)]$. Let $f(t,x) = \E[F(X_T)|X_t = x]$, which satisfies a partial differential equation (PDE) by the well-celebrated Feynman-Kac formula:
\[ \frac{\partial f}{\partial t} + b^\top(x) \frac{\partial f}{\partial x} + \frac{1}{2}\operatorname{trace}(\sigma\sigma^\top \frac{\partial^2 f}{\partial x^2})  = 0,\ f(T,x) = F(x). \]
We introduce the exponential martingale defined as 
$\dd L_t =  L_t U_t^\top\dd W_t,\  L_0 = 1$, where $U_t$ determines the associated change of measure, defined as $\frac{\dd \q}{\dd \p}|_{\f_T} = L_T$. $U_t$ will be treated as a control variable. Under $\q$, $W^{\q}_t = W_t - \int_0^t U_s\dd s$ is the standard Brownian motion, and $\dd X_t = [b(X_t) + \sigma(X_t)U_t]\dd t + \sigma(X_t)\dd W^{\q}_t$. 
Then $\E_{x_0}[F(X_T)] = \E_{x_0}^{\q}[F(X_T) \frac{1}{L_T} ]$. Applying It\^o's lemma to $f(t,X_t) \frac{1}{L_t}$, we obtain
\[ F(X_T) \frac{1}{L_T} - f(X_0) = \int_0^T \frac{1}{L_t}[\frac{\partial f}{\partial x}^\top(t,X_t)\sigma(X_t) - fU_t^\top]\dd W^{\q}_t.  \]

If we do importance sampling under $\q$, the variance of the estimator becomes 
\[ \operatorname{Var}^{\q}_{x_0}\big( F(X_T) \frac{1}{L_T} \big) = \E_{x_0}\big[ \int_0^T \frac{1}{L_t^2}|\sigma^\top(X_t)\frac{\partial f}{\partial x}(t,X_t) - fU_t|^2\dd t \big]. \]
Hence the optimal control has the form of $U^*(t,x) = \frac{1}{f(t,x)}\sigma^\top(x) \frac{\partial f}{\partial x}(t,x)$, is a function of time $t$ and state $x$ only. 
For our approximation, we only consider it as a function of $x$ that lies in a linear space spanned by some basis function of $x$.

On the other hand, for a stopping-time event associated with a first passage time $\tau_{\Gamma} = \inf\{t\geq 0:X_t\notin \Gamma \}$ for some region $\Gamma$, the value function becomes independent of $t$, i.e., $f(x) = \E[F(X_{\tau_{\Gamma}})|X_t = x]$, satisfying
\[ b^\top(x) \frac{\partial f}{\partial x} + \frac{1}{2}\operatorname{trace}(\sigma\sigma^\top \frac{\partial^2 f}{\partial x^2})  = 0,\forall x\in \Gamma, \ f(x) = F(x), \forall x\in \partial \Gamma, \]
where $\partial \Gamma$ stands for the boundary of the region $\Gamma$. The rest of the derivation is similar.


Recall in Section \ref{sec:diffusion}, the optimal $g^*$ actually depends on the value function of the event of interest, i.e., $f(t,x) = \E[F(X_{\tau})|X_t = x]$. In particular, $g^*(t,x) = \frac{1}{f(t,x)}\frac{\partial f}{\partial x}(t,x)$. Hence, the proposed approximation of the tilting family $B(x)\eta$ can also be viewed as an approximation to $\frac{1}{f(t,x)}\frac{\partial f}{\partial x}(t,x)$.

Given this approximation, an unbiased estimator is given by
\[ \begin{aligned}
	\E_{x_0}^{\p_{\eta}}\bigg[ F(X_{\tau})\exp\bigg\{\sum_{i=1}^{\tau} & -\eta^\top B^\top(X_{i-1}) X_{i}  + \eta^\top B^\top(X_{i-1})\left( X_{i-1} + b\left(X_{i-1} \right)\Delta t \right) \\
	& + \frac{1}{2}\Delta t\eta^\top B^\top(X_{i-1})\sigma\sigma^\top(X_{i-1}) B(X_{i-1}) \eta   \bigg\} \bigg].
\end{aligned} \]
The resulting tilting parameter $\eta$ is chosen to minimize
\[\begin{aligned}
	\E_{x_0}^{\p_{\eta}}\bigg[ F^2(X_{\tau})\exp\bigg\{\sum_{i=1}^{\tau} & -\eta^\top B^\top(X_{i-1}) X_{i}  + \eta^\top B^\top(X_{i-1})\left( X_{i-1} + b\left(X_{i-1} \right)\Delta t \right) \\
	& + \frac{1}{2}\Delta t\eta^\top B^\top(X_{i-1})\sigma\sigma^\top(X_{i-1}) B(X_{i-1}) \eta   \bigg\} \bigg].
\end{aligned}  \]

\subsection{Conventional Small-Noise Regime and Subsolutions to a Stochastic Control Problem}
The conventional regime to study the large deviation in diffusion processes is to scale ``$\dd W_t$" by ``$\sqrt{\epsilon} \dd W_t$", and to consider the SDE: $\dd X_t = b(X_t)\dd t + \sqrt{\epsilon} \sigma(X_t)\dd W_t$. Suppose that we want to estimate $\epsilon \log\E_{x_0}[\exp\{ \frac{1}{\epsilon} \log F(X_T) \}]$ and define $f(t,x) =  \E[\exp\{ \frac{1}{\epsilon} \log F(X_T) \}|X_t = x]$, which satisfies
\[ \frac{\partial f}{\partial t} + b^\top(x) \frac{\partial f}{\partial x} + \frac{\epsilon}{2}\operatorname{trace}(\sigma\sigma^\top \frac{\partial^2 f}{\partial x^2})  = 0,\ f(T,x) = \exp\{ \frac{1}{\epsilon} \log F(x) \}. \]
Then by introducing $g(t,x) = \epsilon \log f(t,x)$, we can derive that $g$ satisfies
\[ \frac{\partial g}{\partial t} + b^\top(x) \frac{\partial g}{\partial x} + \frac{1}{2} (\frac{\partial g}{\partial x})^\top  \sigma\sigma^\top \frac{\partial g}{\partial x} +  \frac{\epsilon}{2}\operatorname{trace}(\sigma\sigma^\top(x) \frac{\partial^2 g}{\partial x^2})  = 0 ,\ g(T,x) = F(x) . \]
Its leading order is a first-order nonlinear PDE: 
\begin{equation}
	\label{eq:sne first order pde}
	\frac{\partial u}{\partial t} + b(x)^\top \frac{\partial u}{\partial x} + \frac{1}{2} (\frac{\partial u}{\partial x})^\top \sigma\sigma^\top(x) \frac{\partial u}{\partial x} = 0,\ u(T,x) = F(x).
\end{equation}
Hence, $u\approx g$. From the conclusion in the optimal change-of-measure, the optimal control in this case should be
\[ U^*(t,x) = \frac{\sqrt\epsilon}{f(t,x)}\sigma^\top(x)\frac{
	\partial f}{\partial x}(t,x) = \frac{1}{\sqrt\epsilon} \sigma^\top(x) \frac{\partial g}{\partial x}(t,x) \approx  \frac{1}{\sqrt\epsilon} \sigma^\top(x) \frac{\partial u}{\partial x}(t,x). \]

The existing theory on the subsolution approach to the importance sampling for diffusion processes (e.g., \citealt{fouque2002variance,dupuis2012importance}) assert that it suffices to find a subsolution to \eqref{eq:sne first order pde} (still denoted as $u(t,x)$) and the above approximation would yield efficient importance sampling estimators. 

Similarly, for a stopping-time event associated with a first passage time $\tau_{\Gamma} = \inf\{t\geq 0:X_t\notin \Gamma \}$ for some region $\Gamma$, the value function becomes independent of $t$, i.e., $f(x) = \E[\exp\{ \frac{1}{\epsilon} F(X_{\tau_{\Gamma}}) \}|X_t = x]$, satisfying
\[ b^\top(x) \frac{\partial f}{\partial x} + \frac{1}{2}\operatorname{trace}(\sigma\sigma^\top \frac{\partial^2 f}{\partial x^2})  = 0,\forall x\in \Gamma, \ f(x) = \exp\{ \frac{1}{\epsilon}F(x)\}, \forall x\in \partial \Gamma. \]
Following the similar derivation, we obtain a first-order PDE: 
\begin{equation}
	\label{eq:sne first order pde stopping time}
	b(x)^\top \frac{\partial u}{\partial x} + \frac{1}{2} (\frac{\partial u}{\partial x})^\top \sigma\sigma^\top(x) \frac{\partial u}{\partial x} = 0, \forall x\in \Gamma, \ u(x) = F(x), \forall x\in \partial \Gamma .
\end{equation}

\subsection{Proposed Link Function in SIRD Model}
Recall that in Section \ref{sec:diffusion}, for general diffusion models (without requiring small-noise regime), we suggest the link function that equivalently approximates the tilting family by $U^*(t,x)\approx \sigma^\top(x) B(x) \eta$. This seems to imply we choose $\frac{\partial u}{\partial x}(t,x) = B(x) \eta$. Then a natural question is, is it a natural subsolution to \eqref{eq:sne first order pde} or \eqref{eq:sne first order pde stopping time}?

In the SIRD model in Section \ref{subsec:sir}, for simplicity, we denote $\Sigma(x) = \sigma\sigma^\top(x)$, and it is straightforward to verify that $\Sigma(X_t)$ is strictly positive definite with probability one. We use the form $B(x) = \Sigma^{-1}(x) M(x)$, where $M(x)\in \mathbb R^{3\times 5}$. In particular, we notice that there exists a $\eta_0 = (\alpha, \alpha,\beta,\gamma,\beta)^\top$ such that $M(x)\eta_0 = b(x)$. 

We plug-in the form $\frac{\partial u}{\partial x}(t,x) = B(x) \eta$ to PDE \eqref{eq:sne first order pde stopping time}, and obtain
\[ b(x)^\top \frac{\partial u}{\partial x} + \frac{1}{2} (\frac{\partial u}{\partial x})^\top \sigma\sigma^\top(x) \frac{\partial u}{\partial x} = \eta_0^\top M(x)^\top \Sigma^{-1}(x) M(x) \eta + \frac{1}{2}\eta^\top M(x)^\top \Sigma^{-1}(x) M(x) \eta .  \]
When $\eta = \delta \eta_0$ with $\delta\in \mathbb R$, we have $b(x)^\top \frac{\partial u}{\partial x} + \frac{1}{2} (\frac{\partial u}{\partial x})^\top \sigma\sigma^\top(x) \frac{\partial u}{\partial x} = \delta(1 + \frac{1}{2}\delta) \eta_0^\top M(x)^\top \Sigma^{-1}(x) M(x) \eta_0$. 

Moreover, for SIRD model, the stopping time is the exit time to the region $\Gamma = \{ (S,I,R)\in \mathbb R^3_+: I\leq c \}$, and the boundary condition is $F(x) = \one_{\{ x_2 = c \}}$, where $x_2$ stands for the second entry of the vector $x = (S,I,R)$. It is possible to construct a subsolution that satisfies $\frac{\partial u}{\partial x}(t,x) = B(x) \eta$ for a certain $\eta$. Since the importance sampling method only requires  $\frac{\partial u}{\partial x}(t,x)$, we do not give the construction for $u$.

\section{Notions of Logarithmic Efficiency}
\label{appendix:log efficiency}

Conventionally, in the study of rare event simulation, the properties of bounded relative error and logarithmic efficiency are often used as criteria. More specifically, let $\{A(\alpha)\}$ be a sequence of events of interest, indexed by $\alpha \in \mathbb{R}_+$ or $\alpha \in \mathbb{N}$, assume that $z(\alpha)= \p\left(A(\alpha)\right) \to 0$ as $\alpha \to \infty$, and for each $\alpha$ let $\hat Z(\alpha)$ be an unbiased estimator for $z(\alpha)$, i.e., $\E[\hat Z(\alpha)] = z(\alpha)$. 

A sequence of estimators $\{\hat Z(\alpha)\}$ is called to 
have \textit{logarithmic efficiency} if
\begin{eqnarray}
	\label{eq:log efficiency}
	\limsup_{\alpha \to \infty} \frac{\operatorname{Var}\left( \hat Z(\alpha) \right)}{z(\alpha)^{2-\varepsilon} }= 0,
\end{eqnarray}
for all $\varepsilon > 0$. See \citet[Chapter 6]{asmussen2007stochastic} for more introduction to these concepts.

Note that, we often call the event ``rare'', because we implicitly consider the limiting behavior of a sequence of events, whose probabilities tend to zero. Only when $z(\alpha)\to 0$, the notion of the logarithmic efficiency makes sense.

\subsection{Motivation for Definition under the Asymptotic Normal Regime}

{\bf Fixed-time events.}
We consider the simulation of the probability of a fixed-time event in the form of $\{ S_n > c_n \}$.
Under the large-deviation regime, the sequence of events of interests typically have the form of $A(n,\ell) = \{ \frac{S_n}{n}-\mu >  \ell \}$, indexed by $n$. Under the original probability $\p$, $\frac{S_n}{n} - \mu \to 0$ almost surely. Hence $z(n, \ell) = \p\left( A(n,\ell) \right) \to 0$, as $n \to \infty$ for any fixed $\ell$. Therefore, it naturally fits in the framework of rare event simulation. 
In contrast, under the asymptotic normal regime (see Definition \ref{def:normal regime}), the sequence of events of interests have the form of $A(n,\ell) = \{ \sqrt{\frac{n}{\sigma^2}}( \frac{S_n}{n}-\mu ) > \ell \}$, indexed by $n$. Under the original probability $\p$, $\sqrt{\frac{n}{\sigma^2}} ( \frac{S_n}{n} - \mu) \to \mathcal{N}(0, 1)$ in distribution for some $\sigma>0$. Therefore, $z(n,\ell) = \p\left( A(n,\ell) \right) \to \Phi(-\ell)\in (0,1)$, where $\Phi(\cdot)$ is the cumulative distribution function of a standard normal distribution. In other words, events 
under the asymptotic normal regime do not directly fit into the conventional framework of rare event simulation because the limiting probability is not zero. Consequently, the concept of bounded relative error and logarithmic efficiency are not discussed in the literature of moderate-deviation simulation, e.g., \citet{fuh2004efficient}.

The above discussion motivates us to redefine bounded relative error and logarithmic efficiency in simulation under the asymptotic normal regime 
by letting $\ell\to\infty$ (after first taking limit $n\to \infty$) to have a truly rare event. More precisely, a family of estimators $\{\hat Z(n,\ell)\}$ is called to 
have \textit{logarithmic efficiency} if
$$\limsup_{\ell \to \infty, n\to \infty, \ell/\sqrt{n} \to 0} \frac{\operatorname{Var}\left( \hat Z(n,\ell) \right)}{\left( \Phi(-\ell) \right)^{2-\varepsilon} }= 0,
$$
for all $\varepsilon > 0$.

{\bf First-passage-time events.}
Finally, we examine the simulation of the probability of another typical event involving the first-passage time in the form of $\{\tau_b < T\}$. In particular, 
under the asymptotic normal regime, the sequence of events of interests have the form of $A(b,\ell) = \{ \tau_b < \sqrt{\frac{b\sigma^2}{\mu^3}}\ell + \frac{b}{\mu} \} = \{ \frac{\tau_b - \frac{b}{\mu}}{\sqrt{ b\sigma^2/\mu^3 }} < \ell \}$, with $\mu > 0$, $b > 0$. It is well known that $z(b,\ell) = \p\left( A(b,\ell) \right) \to \Phi(\ell)$ as $b\to \infty$ for any given $\ell$, e.g., see \citet{fuh2004renewal}.
In this case, we are motivated to call a family of estimators $\{\hat Z(b,\ell)\}$ to 
have \textit{logarithmic efficiency} if
$$\limsup_{\ell \to -\infty,b\to \infty, \ell/\sqrt{b} \to 0}	 \frac{ \operatorname{Var}\left( \hat Z(b,\ell) \right)}{\left( \Phi(\ell) \right)^{2-\varepsilon} }= 0,
$$
for all $\varepsilon > 0$.

\subsection{Recovering Large Deviation Regime}
{\bf Fixed-time events.}
We consider the following limiting regime: $n\to \infty,\ell\to \infty,\ell/\sqrt{n} \to c > 0$ for the fixed-time events. Under these regimes, we say a family of estimators $\{\hat Z(n,\ell)\}$ has \textit{logarithmic efficiency} if
\begin{equation}
	\label{eq:log efficiency ld fixed n}
	\limsup_{\ell \to \infty, n\to \infty, \ell/\sqrt{n} \to c} \frac{\operatorname{Var}\left( \hat Z(n,\ell) \right)}{\left( z(n,\ell) \right)^{2-\varepsilon} }= 0,
\end{equation}
for all $\varepsilon > 0$. 

{\bf First-passage-time events.}
We consider the following limiting regime: $b\to \infty,\ell\to -\infty,\ell/\sqrt{b} \to c < 0$ for the first-passage-time events. Under these regimes, we say a family of estimators $\{\hat Z(b,\ell)\}$ has \textit{logarithmic efficiency} if
\begin{equation}
	\label{eq:log efficiency ld stopping}
	\limsup_{\ell \to -\infty,b\to \infty, \ell/\sqrt{b} \to c}	 \frac{ \operatorname{Var}\left( \hat Z(b,\ell) \right)}{\left( z(b,\ell) \right)^{2-\varepsilon} }= 0,
\end{equation}
for all $\varepsilon > 0$.

Under stronger conditions, we can show that the importance sampling estimator proposed in Theorem \ref{thm:lan optimal linear} has logarithmic efficiency in the above limiting regime. For simplicity, we make the following assumption.

\begin{assumption}
	\label{ass:uniform recurrent}
	There exists $\kappa \in \mathbb{N}$, $C_2 > C_1 > 0$, and a probability measure $\mu$ on $\mathcal X \times\mathbb R^d$ such that $C_1 \mu(A\times B) \leq \int_{A\times B} p(x,\dd x') \rho(y|x,x')\dd y \leq  C_2 \mu(A\times B)$. 
\end{assumption}
Note that Assumption \ref{ass:uniform recurrent} is called the uniform recurrent conditions, which imply Assumptions \ref{ass:rigorous eigenvalues} to \ref{ass:eigenvalue bounded}. The reason why we need a uniform condition because the limiting point of the ratio $\ell/\sqrt{n}$ (or $\ell/\sqrt{b}$) could lie anywhere, and we still need to guarantee that the eigenfunction at this point is bounded from both below and above.
\begin{theorem}
	\label{thm:ld log efficiency}
	Suppose Assumptions \ref{ass:convex}, \ref{ass:recurrent} and \ref{ass:uniform recurrent} hold. Then the importance sampling estimator proposed in Theorem \ref{thm:lan optimal linear} has logarithmic efficiency in the sense of \eqref{eq:log efficiency ld fixed n} and \eqref{eq:log efficiency ld stopping}.
\end{theorem}
\section{Calculation Details Related to the Tilting Probabilities in the Examples}
\label{appendix:details tilting}
\subsection{Heston Model}
\label{appendix:heston}
For simplicity, we denote $C = \frac{\sigma^2(1 - e^{-\kappa \Delta t})}{4\kappa}$, $c = \frac{4\kappa e^{-\kappa \Delta t}}{\sigma^2(1 - e^{-\kappa \Delta t})}$, and $d = \frac{4\kappa \alpha}{\sigma^2}$ in this subsection. Then
\[  X_{t_{k+1}}|X_{t_k} \sim C \chi^2_{d}\left( c X_{t_k}\right) . \]

We choose the link function as $k(x,x',\eta) = \eta x'$, and then the resulting tilting probability is
\[ \frac{\dd \p_{\theta,\eta}}{\dd \p} = \exp\{ \theta Y_{t_{k+1}} + \eta X_{t_{k+1}} -\phi(X_{t_k},\eta) - \psi(X_{t_k},X_{t_{k+1}},\theta) \}, \]
where 
\[ \phi(x,\eta) = \frac{c C\eta x}{1-2\eta C}  - \frac{d}{2}\log(1 - 2\eta C),  \]
\[ \psi(x,x',\theta) = \left( [\mu-\frac{1}{2}x-\frac{\rho}{\sigma}\kappa(\alpha-x)]\Delta t + \frac{\rho}{\sigma}(x' - x) \right) \theta + \frac{1}{2}x (1-\rho^2)\Delta t \theta^2 \]
are two cumulant functions. More precisely, under the proposed sampling distribution $\p_{\theta,\eta}$, the transition probability and conditional distribution are given by
\[ X_{t_{k+1}}|X_{t_k} \sim C \operatorname{Gamma}(\frac{d}{2}, \frac{1}{2} - \eta C, \frac{cX_{t_k}}{1 - 2\eta C}), \]
\[ Y_{t_{k+1}}|X_{t_{k+1}},X_{t_k} \sim \mathcal{N}\left( [\mu-\frac{1}{2}X_{t_k}-\frac{\rho}{\sigma}\kappa(\alpha-X_{t_k})]\Delta t + \frac{\rho}{\sigma}(X_{t_{k+1}}-X_{t_k}) +\theta X_{t_k}(1-\rho^2)\Delta t,X_{t_k}(1-\rho^2)\Delta t \right), \]
where we use $\operatorname{Gamma}(\alpha,\beta,\delta)$ to denote a ``non-central'' Gamma distribution, which has the same distribution as $\operatorname{Gamma}(\alpha + N,\beta)$, with $N\sim \operatorname{Poisson}(\frac{\delta }{2})$.

\subsection{SIRD Model}
\label{appendix:sird}
Recall the notations from the main text Section \ref{subsec:sir}. We further use $b_t$ to denote $b(X_t)$ and similarly to $\tilde{b}_t,\sigma_t,B_t$. Moreover, let $\Sigma_t = \sigma_t\sigma_t^\top$, and $\tilde{\theta}_t = \Sigma_t^{-1}(\tilde{b}_t - b_t) = B_t\eta$. 
Under the alternative measure, the process is sampled from 
\begin{equation}
	\begin{pmatrix}
		S_{t_{k+1}}\\
		I_{t_{k+1}}\\
		R_{t_{k+1}}\\
	\end{pmatrix} =
	\begin{pmatrix}
		S_{t_k} -\frac{\alpha_- S_{t_k} I_{t_k}}{N_{t_k}}\Delta t \\
		I_{t_k} + (\frac{\alpha_+ S_{t_k} I_{t_k}}{N_{t_k}} -\beta_- I_{t_k} - \gamma_- I_{t_k})\Delta t\\
		R_{t_k} + \beta_+ I_{t_k} \Delta t
	\end{pmatrix} + \sqrt{\Delta t}
	\begin{pmatrix}
		-\sqrt{\frac{\alpha S_{t_k} I_{t_k}}{N_{t_k}}} & 0 & 0\\
		\sqrt{\frac{\alpha S_{t_k} I_{t_k}}{N_{t_k}}} & -\sqrt{\beta I_{t_k}} & -\sqrt{\gamma I_{t_k}} \\
		0 & \sqrt{\beta I_{t_k}} & 0
	\end{pmatrix}
	\epsilon_{t_{k+1}}.
\end{equation}

The likelihood ratio  is 
\[ \frac{\dd \p^{\theta}_{t_k}}{\dd \p_{t_k}} = \exp\{ \tilde{\theta}_{t_k}^\top X_{t_{k+1}} - \tilde{\theta}_{t_k}^\top b_{t_k} - \frac{1}{2}\tilde{\theta}_{t_k}^\top \Sigma_{t_k} \tilde{\theta}_{t_k}   \}  = \exp\{ \eta^\top B_{t_k}^\top X_{t_{k+1}} - \eta^\top B_{t_k}^\top b_{t_k} - \frac{1}{2}\eta^\top B_{t_k}^\top \Sigma_{t_k}B_{t_k}\eta  \}. \]
The importance sampling estimator is
\[ \E^{\p^{\eta}}\bigg[\one_{\{\tau_c < \infty\}} \exp\{ -\sum_{k=0}^{\tau_c-1} \big(  \eta^\top B_{t_k}^\top X_{t_{k+1}} - \eta^\top B_{t_k}^\top b_{t_k} - \frac{1}{2}\eta^\top B_{t_k}^\top \Sigma_{t_k}B_{t_k}\eta   \big) \} \bigg] .\]
Then the second moment of the importance sampling estimator is 
$$ G(\eta) = \E^{\p}\bigg[\one_{\{\tau_c < \infty\}} \exp\big\{ -\sum_{k=0}^{\tau_c-1} \big( \eta^\top B_{t_k}^\top X_{t_{k+1}} - \eta^\top B_{t_k}^\top b_{t_k} - \frac{1}{2}\eta^\top B_{t_k}^\top \Sigma_{t_k}B_{t_k}\eta  \big) \big\} \bigg] .$$
The tilting parameter $\eta\in \mathbb{R}^5$ is selected by minimizing the above second moment (minimizing the variance) $G(\eta)$ of the estimator by stochastic approximation algorithm.

\subsection{VAR-GARCH Model}
\label{appendix:vargarch}
By taking $\{ (y_t,H_{t+1}),t\geq 0\}$ as the underlying Markov chain, 
$\{(y_t,H_{t+1},S_t),t\geq 0\}$ still belongs to the general Markov random walk framework. The following is the eigenvalue problem which  difficult to solve: 
\begin{eqnarray*}
	\label{eq:eigenvalue garch}
	&~& \E_{y_0,H_1}[e^{\theta^\top y_1} r(y_1, H_2)]\\
	&=& \E[e^{\theta^\top(\mu + \rho y_0 + H_1^{\frac{1}{2}}\epsilon)}r(\mu + \rho y_0 + H_1^{\frac{1}{2}}\epsilon,A \odot H_1^{\frac{1}{2}}\epsilon\epsilon^\top H_1^{\frac{1}{2}} + B \odot H_1)] = e^{\Lambda(\theta)}r(y_0,H_1). 
\end{eqnarray*}

In contrast, the Poisson equation \eqref{eq:poisson 0} has simpler form:  
$ \E_{y_0,H_1}[Y_1] - \E_{\pi}[Y_1] = \mu + \rho y - \E_{\pi}[Y_1] = \rho y_0 + \mu - \E_{\pi}[Y_1] =\rho y_0 + \mu - (I-\rho)^{-1}\mu = \rho y_0  -\rho(I-\rho)^{-1}\mu$. 
The function $g(y,H) = C_1 y$ with $C_1 = \rho(I - \rho)^{-1}$ is a solution because
\begin{equation}
	\label{eq:poisson garch}
	(I - \mathcal{P}_0)g (y,H) = C_1 y  - C_1 (\mu+\rho y) = C_1(I - \rho)y - C_1  \mu = \rho y - \rho(I - \rho)^{-1}\mu = \E_{y,H}[Y_1] - \E_{\pi}[Y_1]. 
\end{equation}
Therefore, the solution to the Poisson equation \eqref{eq:poisson garch} for this VAR-GARCH model turns out to be a linear function of $y$, the first component of the Markov process $(y_t,H_{t+1})$.
Thus, it suffices to use a linear link function in $y'$, i.e., to take $k\left( (y,H),(y',H'),\eta\right) = \eta^\top y'$.

\section{Verification of Regularity Conditions}
\label{appendix:verify assumptions}
We use the same notations as in E-Companion \ref{appendix:details tilting}.
\subsection{Heston Model}

We first verify Assumption \ref{ass:convex}. $(i)$ holds because the additive part is conditionally normally distributed and $\psi(x,x',\theta)$ exists for all $\theta$. $(ii)$ holds when $\eta < \frac{1}{2C}$. $(iii)$ holds because $\psi,k,\phi$ are all linear functions of $(x,x')$. The rest of the assumption in $(iii)$ also follows from $(ii)$. Assumption \ref{ass:recurrent} holds based on the properties of the CIR process; see, e.g., \citealt{jin2013positive}. Assumption \ref{ass:eigenvalue exist} holds because the eigenvalue and the eigenfunction can be explicitly calculated. Assumption \ref{ass:rigorous} holds because of Lemma \ref{lemma:poisson} and Assumption \ref{ass:eigenvalue exist}.

We mimic the argument in \citet[Example 2]{Chan2003} to justify Assumption \ref{ass:rigorous eigenvalues} holds. Consider two compact sets $C_X = [\epsilon, 1/\epsilon]$, $C_Y = [-1/\epsilon, 1/\epsilon]$, with $0 < \epsilon < 1$. Then for any subset $B_Y \subset C_Y$ and $x,x'\in C_X$, we have 
\[ \p(Y_1\in B_Y | X_0 = x, X_1 = x') = \int_{B_Y} \rho(y|x,x')\dd y \geq \rho_{\epsilon}\operatorname{meas}(B_Y), \]
where $\operatorname{meas}(B_Y)$ stands for the Lebesgue measure of the set $B_Y$, and $\rho_{\epsilon} = \min_{x,x'\in C_X, y \in C_Y}\rho(y|x,x') > 0$. Moreover, denote $p_{\epsilon} = \min_{x,x'\in C_X}p(x,x') > 0$, where $p(x,x')$ is the density function of $X_1\in \dd x'|X_0 = x$.

Then
\[
\begin{aligned}
	\p_x\left( (X_1,Y_1)\in A\times B \right)\geq & \int_{A\cap C_X} p(x,x') \p(Y_1\in B\cap C_Y | X_0 = x, X_1 = x')  \dd x' \\
	\geq & \rho_{\epsilon}\operatorname{meas}(C_Y) \frac{\operatorname{meas}(B\cap C_Y)}{\operatorname{meas}(C_Y)} \int_{A\cap C_X} p(x,x') \dd x' \\
	\geq & \one_{ \{x \in C_X\} }p_{\epsilon} \rho_{\epsilon}\operatorname{meas}(C_X)\operatorname{meas}(C_Y) \frac{\operatorname{meas}(B\cap C_Y)}{\operatorname{meas}(C_Y)} \frac{\operatorname{meas}(A\cap C_X)}{\operatorname{meas}(C_X)}.
\end{aligned}
\]
Hence Assumption \ref{ass:rigorous eigenvalues} holds with $h(x) = \one_{ \{x \in C_X\} }p_{\epsilon} \rho_{\epsilon}\operatorname{meas}(C_X)\operatorname{meas}(C_Y)$, and $\mu(A\times B) = \frac{\operatorname{meas}(B\cap C_Y)}{\operatorname{meas}(C_Y)} \frac{\operatorname{meas}(A\cap C_X)}{\operatorname{meas}(C_X)}$.

Finally, let us verify Assumption \ref{ass:eigenvalue bounded}. By the proof of Corollary \ref{coro:affine}, we know for this affine model, the eigenfunction has the form of $r(x,\theta) = e^{A(\theta) x}$, where $A(\theta)$ is the solution to an algebraic equation 
\[ \frac{c C (\frac{\rho}{\sigma}\theta + A)}{1 - 2 C (\frac{\rho}{\sigma}\theta + A)} + [(-\frac{1}{2} + \frac{\rho\kappa}{\sigma} )\theta \Delta t + \frac{1}{2}(1-\rho^2)\theta^2\Delta t ] =  \frac{\rho}{\sigma}\theta + A . \]
When $\theta = 0$, the above equation reduces to $2C A^2 + (c C-1) A = 0$, where $c C = e^{-\kappa \Delta t} < 1$ and $C > 0$. (Recall that in the simulation, $\Delta t$ is a fixed positive number) Hence it has a positive root. It implies that when $\theta$ is sufficiently small, we can always find a root $A(\theta) > 0$, and hence construct an eigenfunction $r(x,\theta) = e^{A(\theta) x}$.

In this example, the initial distribution is a fixed point so $\limsup_{\theta \to 0}\E_{\nu}\left[  r(X_0,\theta)^2  \right] = \limsup_{\theta \to 0}e^{2A(\theta) X_0} < \infty$, i.e., the first part in Assumption \ref{ass:eigenvalue bounded} holds. Moreover, since the state space is $\mathcal X = \mathbb{R}_+$, $\liminf_{\alpha \to 0}\inf_{x\in \mathcal{X}} r(x,\theta) =\liminf_{\alpha \to 0}\inf_{x \geq 0} e^{A(\theta) x} = 1$, i.e., the second part in Assumption \ref{ass:eigenvalue bounded} holds.

\subsection{SIRD Model}
Since the importance
sampling for \eqref{eq:stochastic sir} is based on the approximation in Section \ref{sec:diffusion}, we do not need to check Assumptions.

\subsection{VAR-GARCH Model}

We first verify Assumption \ref{ass:convex}. $(i)$ holds because the additive part degenerates to $S_1 = y_1$, i.e., $\psi(y,H,y'H',\theta) = \theta^\top y'$. $(ii)$ holds because we have chosen $k\left( (y,H),(y',H'),\eta\right) = \eta^\top y'$, and $\E_{y_0,H_1}[e^{\eta^\top y_1}] = e^{\eta^\top (\mu+\rho y_0) + \frac{1}{2}\eta^\top H_1 \eta}$. That is, $\phi(y,H,\eta) = \eta^\top (\mu+\rho y_0) + \frac{1}{2}\eta^\top H_1 \eta$. $(iii)$ holds because $\psi,k,\phi$ all are linear function in $(y,H,y',H')$. Moreover, consider $\ell_1,\ell_2 > 0$, we have
\[\begin{aligned}
	\E_{y_0,H_1}[e^{\ell_1 |y_1| + \ell_2 |H_2|}] \leq & \exp\{ \ell_1 (|\mu| + |\rho| |y_0|) + \ell_2 (|W| + |B \odot H_1|) \} \E_{z_1 \sim \mathcal{N}(0, H_1)}[ e^{\ell_1 |z_1| + \ell_2 |z_1^\top A z_1|} ] < \infty
\end{aligned}  \]
as long as $\ell_1,\ell_2$ are sufficiently small. When the above is finite, its logarithm has linear growth in $y_0$, $H_1$, and $\ell_1,\ell_2$. Assumption \ref{ass:recurrent} holds based on the properties of VAR-GARCH model (under suitable assumptions on the coefficients $\rho, A,B$), e.g., see \citet{meitz2008stability}. Assumption \ref{ass:rigorous eigenvalues} holds naturally since the additive part degenerates.

Assumption \ref{ass:rigorous} holds since we already find the solution to \eqref{eq:poisson 0} for this model in  \eqref{eq:poisson garch} explicitly.

\section{Two Additional Examples}
\label{appendix:two additional examples}

We present two additional examples where the Poisson equation has an analytical solution. 
The first  is an affine process in an autoregressive model. The second one is  a finite-state-space-affine process in a Markov switching model.
It is difficult to find  the eigenfunctions for the second example; thus, the LD-based simulation method cannot be applied easily.

\begin{example}
	\label{eg:ec affine}
	Consider a discrete-time affine model as described in \cite{le2010discrete}. $\{X_t\in \mathbb{R},t\geq 0\}$ is the underlying Markov process whose transition probability satisfies
	\[
	\log\E[e^{\eta X_{t+1}}|X_t] = (\alpha +\beta X_t)\eta +\frac{1}{2}(\iota+\gamma X_t)\sigma_x^2\eta^2 ,\]
	and the additive part $\{Y_t\in \mathbb{R},t\geq 1\}$ satisfies
	\[ \log\E[e^{\theta Y_{t+1}}|X_t,X_{t+1}] = (a_1X_t+a_2X_{t+1}+a_3)\theta + \frac{1}{2}  (b_1X_t+b_2X_{t+1}+b_3)^2\sigma_y^2\theta^2. \]
	
	Then $(X_t,S_t)$ is affine, with $\psi(x,x',\theta)=(a_1x+a_2x'+a_3)\theta+\frac{(b_1x+b_2x'+b_3)\theta^2}{2}\sigma_y^2$, $\phi(x,\eta)=(\alpha+\beta x)\eta+\frac{(\iota+\gamma x)\eta^2}{2}\sigma_x^2$. Suppose $X_t$ admits a stationary distribution $\pi$, then under the stationary distribution, we have $\E_{\pi}[X_1] = \alpha + \beta \E_{\pi}[X_0]$. Hence $\E_{\pi}[X_0] = \frac{\alpha}{1-\beta}$. In addition, $\E_{\pi}[Y_1] = a_1\E_{\pi}[X_0] + a_2\E_{\pi}[X_1] + a_3 = \frac{\alpha(a_1+a_2)}{1-\beta} + a_3$.
	
	Consider the Poisson equation
	\[ g(x) - \E_x[g(X_1)] = \E_x[Y_1] -\E_{\pi}[Y_1], \]
	where $\E_x[Y_1] -\E_{\pi}[Y_1] = a_1 x + a_2\E_x[X_1]  - \frac{\alpha(a_1+a_2)}{1-\beta} = (a_1 + a_2\beta)x - \frac{\alpha(a_1+ a_2\beta)}{1-\beta}$. 
	
	We notice that the solution to the Poisson equation is also a linear function in the form of $g(x) = \bar{A} x$. Given this form, $g(x) - \E_x[g(X_1)] = \bar{A} x - \bar{A} (\alpha + \beta x) = \bar{A}(1-\beta)x -\bar{A}\alpha$. Therefore, $\bar{A} = \frac{a_1+ a_2\beta}{1-\beta}$.

	On the other hand, consider the eigenvalue problem:
	\[ \E_x[e^{\theta Y_1} r(X_1)] = e^{\Lambda(\theta)} r(x) .\]
	We notice that one eigenfunction is given by $e^{Ax}$ with eigenvalue $e^{\Lambda(\theta)}$, where $A$ depends on $\theta$ and solves a quadratic function:
	\[ A = a_1\theta + \frac{b_1\theta^2\sigma_y^2}{2} + \beta(A+a_2\theta+\frac{b_2\theta^2\sigma_y^2}{2}) + \frac{\gamma\sigma_x^2}{2}(A+a_2\theta+\frac{b_2\theta^2\sigma_y^2}{2})^2,  \]
	and $\Lambda(\theta)=a_3\theta+\frac{b_3\theta^2\sigma_y^2}{2}+\alpha(A+a_2\theta+\frac{b_2\theta^2\sigma_y^2}{2})+\frac{\iota\sigma_x^2}{2}(A+a_2\theta+\frac{b_2\theta^2\sigma_y^2}{2})^2$.
	
	This is because 
	\[\begin{aligned}
		\E_x[e^{\theta Y_1} r(X_1)] = & \E_x\left[ \exp\left\{ AX_1 + \theta(a_1 x + a_2 X_1 + a_3) + \frac{\sigma_y^2\theta^2 (b_1 x + b_2 X_1 + b_3)}{2} \right\} \right] \\
		= & \E_x\left[ \exp\left\{ (A + \theta a_2 + \frac{\sigma_y^2 \theta^2 b_2}{2})X_1 \right\} \right] e^{(\theta a_1 + \frac{\sigma_y^2\theta^2b_1}{2}) x + \theta a_3 + \frac{\sigma_y^2\theta^2 b_3}{2}} \\
		= & \exp\bigg\{ (A + \theta a_2 + \frac{\sigma_y^2 \theta^2 b_2}{2})(\alpha + \beta x) + \frac{\sigma_x^2 (A + \theta a_2 + \frac{\sigma_y^2 \theta^2 b_2}{2})^2 }{2}(\iota + \gamma x) \\
		& + (\theta a_1 + \frac{\sigma_y^2\theta^2b_1}{2}) x + \theta a_3 + \frac{\sigma_y^2\theta^2 b_3}{2} \bigg\} \\
		= & \exp\left\{ \left(\theta a_1 + \frac{\sigma_y^2\theta^2b_1}{2}  + \beta  (A + \theta a_2 + \frac{\sigma_y^2 \theta^2 b_2}{2}) + \frac{\gamma \sigma_x^2 (A + \theta a_2 + \frac{\sigma_y^2 \theta^2 b_2}{2})^2 }{2} \right) x  \right\} \\
		& \times \exp\left\{  \left( \theta a_3 + \frac{\sigma_y^2\theta^2 b_3}{2} + \alpha (A + \theta a_2 + \frac{\sigma_y^2 \theta^2 b_2}{2}) +  \frac{\iota \sigma_x^2 (A + \theta a_2 + \frac{\sigma_y^2 \theta^2 b_2}{2})^2 }{2} \right)    \right\} \\
		= & e^{\Lambda(\theta) + A x} .
	\end{aligned}   \]
\end{example}

\begin{example}
	\label{eg:markov swtiching}
	
	Consider a general regime-switching $AR(1)$ model:
	\[ Y_{t+1} = \mu(X_{t+1}) + \beta(X_{t+1}) Y_t + \sigma(X_{t+1})\epsilon_{t+1}, \]
	where $X_t$ is an aperiodic, irreducible and positive recurrent Markov chain on a finite state space with transition matrix $P\in \mathbb{R}^{m\times m}$ and stationary distribution $\pi_x$, that is $\pi_x^\top P = \pi_x^\top$. We use $P[x,]$ to denote the $x$-th row in the transition matrix (which is a row vector). In addition, $\epsilon_{t+1}\sim \mathcal{N}(0,1)$ and is independent of other variables. Moreover, we denote ${\mu},\beta,{\sigma}^2$ by column vectors formed by arranging $\mu(X),\beta(X),\sigma^2(X)$ accordingly.

	First, under the joint stationary distribution $\pi$, $\E_{\pi}[Y_{t+1}] = \pi_x^\top {\mu} + \pi_x^\top {\beta}\E_{\pi}[Y_{t+1}]$. Hence the stationary mean for $Y$ is $\E_{\pi}[Y_1] = \frac{\pi_x^\top {\mu}}{1 - \pi_x^\top {\beta}}$.
	
	Consider the Poisson equation
	\[ g(x,y) - \E_{x,y}[g(X_1,Y_1)] = \E_{x,y}[Y_1] - \E_{\pi}[Y_1], \]
	where $\E_{x,y}[Y_1] - \E_{\pi}[Y_1] = P[x,]({\mu} + {\beta}y) - \frac{\pi_x^\top {\mu}}{1 - \pi_x^\top {\beta}}$, where $P[x,]$ denotes the $x$-th row of the transition matrix.
	
	We notice that the solution to the Poisson equation has the form of $g(x,y) = \bar{g}(x) + A(x) y$, where as a vector, $A$ satisfies $A - P diag\{{\beta}\} A = P {\beta}$, and $\bar{g}$ satisfies $\bar{g} - P\bar{g} = P{\mu} - \frac{\pi_x^\top {\mu}}{1 - \pi_x^\top {\beta}} + P diag\{{\mu} \}A$. In particular, when $\beta$ is a constant that does not depend on $X$, then we can verify that $A = (\frac{\beta}{1-\beta},\cdots, \frac{\beta}{1-\beta})^\top$ is the solution, 
	in which case $A$ is also a constant that does not depend on $X$.
	
	On the other hand, consider the eigenvalue problem:
	\[ \E_{x,y}[e^{\theta Y_1} r(X_1,Y_1) ] = e^{\Lambda(\theta)} r(x, y) .\]
	When $\beta$ is a constant that does not depend on $X$, we realize that the eigenfunction takes the form of $r(x,y) = \bar{r}(x) e^{\frac{\theta \beta}{1-\beta} y}$, and $e^{\Lambda(\theta)}$ is the largest eigenvalue associated with the matrix $P diag\{ \exp\{ \frac{\theta}{1-\beta}{\mu} + \frac{1}{2}\frac{\theta^2}{(1-\beta)^2}{\sigma}^2 \} \}$, and as a vector, $\bar{r}$ is the corresponding eigenvector. However, when $\beta$ is not a constant, the eigenvalue problem cannot be easily solved.

\end{example}

\section{The Example in \citet{djehiche2022importance}}
\label{appendix:example of markov intensity jump}

We review the problem studied in \citet{djehiche2022importance} as follows. We modified their notations and illustrate how this problem can be embedded into our framework, and contrast our approaches through the lens of the link functions under our framework.

The original problem in \citet{djehiche2022importance} concerns a (scaled by an index $n$) continuous-time $d$-dimensional pure jump process $X_t\in \mathbb R^d$, with $X_0 = 0$. The jump size takes values in $\{ \frac{1}{n}e_1,\cdots,\frac{1}{n}e_d\}$, where $e_j = (0,\cdots,0,1,0,\cdots,0)^\top$, with only the $j$-th entry being 1 while others being 0; each has intensity that is proportional to $\lambda_j(X_t) := a_j(w_j - X_{t,j}) e^{b \sum_{\ell =1}^d X_{t,\ell}}$, $j=1,\cdots,d$. The total jump intensity is $ R(X_t) := n e^{b \sum_{j=1}^d X_{t,j}} \sum_{j=1}^d a_j(w_j - X_{t,j})$. The event of interest is $\p_0( \tau_z \leq T )$, where $\tau_z = \inf\{t\geq 0: \sum_{j=1}^d X_{t,j} \geq z\}$, with $\sum_{j=1}^d w_j = 1$ and $z\in (0,1)$.

\citet{djehiche2022importance} find the asymptotically optimal importance sampling estimator based on the subsolution and LD principle when the model is homogeneous with $a_1=\cdots=a_d = a^*$. In particular, \citet{djehiche2022importance} suggest sampling under the distribution with $\bar \lambda_j(x) = \lambda_j(x) \left(1 + \frac{c}{a^* ( 1 - \sum_{\ell=1}^d x_{\ell}) e^{b \sum_{\ell=1}^d x_{\ell}}} \right)$, and $ \bar R(x) := R(x) \left(1 + \frac{c}{a^* ( 1 - \sum_{\ell=1}^d x_{\ell}) e^{b \sum_{\ell=1}^d x_{\ell}}} \right)$, for some $c\in \mathbb R$ that is determined by numerically solving a certain equation. When the model is inhomogeneous, \citet{djehiche2022importance} still suggest the same form of the sampling distribution by defining $a^* = \max_{1\leq \ell \leq d}\{a_{\ell}\}$, but there is no theoretical guarantee for such choices under the inhomogeneous model.

This continuous-time problem can be equivalently embedded into a discrete-time Markov random walk under our framework as follows. With abuse of notation, we denote $X_i\in \mathbb R^d$ with $X_0=0$ as a discrete-time Markov chain, where $\p(X_{i+1} = x + \frac{1}{n}e_j | X_i = x) = \frac{\lambda_j(x)}{ \sum_{\ell=1}^d \lambda_{\ell}(x) }$. Moreover, we define a sequence of random variables $Y_1,\cdots,Y_i,\cdots$, such that $Y_{i+1}|X_{i+1}=x',X_i = x \sim \operatorname{Exp}( R(x) )$, i.e., an exponential distribution with intensity $R(x)$. We define $S_m = \sum_{i=1}^m Y_i$, then $\{(X_i,S_i),i\geq 0\}$ forms a Markov random walk. Furthermore, denote $N = \lceil nz \rceil$, and we know $\sum_{j=1}^d X_{N-1,j} = \frac{N-1}{n} < z$ and $\sum_{j=1}^d X_{N,j} = \frac{N}{n} \geq z$. That is, $X_i$ takes $N$ steps to exceed the desired threshold. Therefore, the event of interest is equivalent to $\p_0( S_N \leq T)$. Since $Y_i$ are all positive random variables, the event of interest can also be written as $\p_0(\iota_T > N)$, where $\iota_T = \inf\{i\geq 0: S_i > T\}$. 

Under our framework, we consider the duo-exponential family for the Markov random walk $\{(X_i,S_i),i\geq 0\}$ through a link function $k(x,x',\eta)$. That is, the sampling transition distribution of  $\{(X_i,S_i),i\geq 0\}$ is 
\[ \p_{\theta,\eta}( x',s' | x, s ) = e^{\theta (s'-s) + k(x,x',\eta) - \psi(x,\theta) - \phi(x,\eta)}\p(x',s' | x, s) ,  \]
where $e^{\psi(x,\theta)} = \frac{R(x)}{R(x) - \theta}$, and $e^{\phi(x,\eta)} = \sum_{j=1}^d \frac{\lambda_j(x) e^{k(x,x+\frac{1}{n}e_j,\eta)}}{ \sum_{\ell=1}^d \lambda_{\ell}(x) } $. 

This implies that under $\p_{\theta,\eta}$, $Y_{i+1}|X_{i+1}=x',X_i=x \sim \operatorname{Exp}(R(x) - \theta)$, and $\p_{\theta,\eta}(X_{i+1} = x + \frac{1}{n}e_j | X_i = x) = e^{k(x,x+\frac{1}{n}e_j,\eta) - \phi(x,\eta)} \frac{\lambda_j(x)}{ \sum_{\ell=1}^d \lambda_{\ell}(x) } \propto e^{k(x,x+\frac{1}{n}e_j,\eta)} \lambda_j(x)$. As a special case, when $d=1$, the underlying Markov chain $\{X_i,i\geq 0\}$ becomes a deterministic process $X_n = \frac{i}{n}$, thus, the link function becomes unnecessary.

Note that the suggested change-of-measure by \citet{djehiche2022importance} is to multiply $R(x)$ and every $\lambda_j(x)$ by the same function of $x$. Hence, the transition probability of $\{X_i,i\geq 0\}$ is unchanged. Moreover, we note that $\bar R(x) = R(x) + nc$ under a homogeneous model. This is equivalent to our duo-exponential family with $\theta = -nc$ and $k(x,x',\eta) = 0$, a trivial link function. Moreover, searching over $c$ is equivalent to searching over $\theta$ while fixing the trivial link function.  

In general, our approach allows us specifying a suitable link function, and we have several options. (1) Using a trivial link function, then our tilting family includes the one suggested by \citet{djehiche2022importance} as a special case, and hence, optimizing over the unique tilting parameter $\theta$ should lead to superior finite-sample performance than the method by \citet{djehiche2022importance} (if the computational cost and error of the optimization stage are ignored). (2) Choosing the link function based on our Theorem 2 by solving a Poisson equation. Note that in this case, since $X_i$ can only take finitely many values (at the grid points that are multipliers of $\frac{1}{n}$ in each dimension), the Poisson equation and the eigenvalue problem is associated with a finite-dimensional matrix, but the dimension of this matrix grows as fast as $O\left( (n z)^d \right)$. Hence, it is computationally difficult to implement this procedure. (3) Other heuristic choices that do not require huge computational cost but may not have theoretical guarantee.

%

We implement the method in \citet{djehiche2022importance} and ours based on a trivial link function, and the results are presented in Table \ref{tab:markov jump}. We can see that with an optimization stage to search for the optimal tilting parameters, our approach can slightly improve the variance reduction of the method in \citet{djehiche2022importance}, and the overall efficiency is close. It suggests that a trivial link function can already induce comparable performance to a sophisticated method while the potential computational costs can be tremendous if a link function is involved. 

\begin{table}[H]
	\centering
	\caption{\textbf{Numerical results of different methods.} Estimation of $\p_0(\iota_T > N)$, where $N = \lceil nz \rceil$, and $\iota_T = \inf\{i\geq 0: S_i > T\}$. $\{(X_i,S_i),i\geq 0\}$ forms a Markov random walk with $\p(X_{i+1} = x + \frac{1}{n}e_j | X_i = x) = \frac{\lambda_j(x)}{ \sum_{\ell=1}^d \lambda_{\ell}(x) }$, $S_m = \sum_{i=1}^m Y_i$, where $Y_1,\cdots,Y_i,\cdots$, such that $Y_{i+1}|X_{i+1}=x',X_i = x \sim \operatorname{Exp}( R(x) )$. Here, $\lambda_j(X_t) := a_j(w_j - X_{t,j}) e^{b \sum_{\ell =1}^d X_{t,\ell}}$, $j=1,\cdots,d$, and total jump intensity is $ R(X_t) := n e^{b \sum_{j=1}^d X_{t,j}} \sum_{j=1}^d a_j(w_j - X_{t,j})$. In the simulation, $d=2$, $n=125$, $T = 5$, $b=5$, and $a_j = 0.01$ and $w_j = 1/d$ for $j=1,2$. For each column, the sample size is $10^6,10^6,10^6,10^7$, respectively. }
	\begin{tabular}{llcccc}
		\toprule
		& $z$     & 0.1 & 0.125  & 0.15 & 0.175 \\
		\midrule			
		Plain MC & mean  & 7.46E-02 & 1.36E-02 & 1.93E-03 & 2.18E-04  \\
		& standard deviation    & 2.63E-04 & 1.16E-04 & 4.39E-05 & 4.67E-06\\
		& elapsed time & 13.43 & 14.13 & 14.73 & 124.42 \\
		\citet{djehiche2022importance}    & sd reduction ratio & 2.35  & 4.07  & 7.97  & 16.04 \\
		& time ratio & 1.44  & 1.36  & 1.47  & 1.91   \\
		& \textbf{efficiency ratio} & 1.96  & 3.49  & 6.58  & 11.61  \\
		This paper & sd reduction ratio & 2.38  & 4.14  & 8.01  & 18.06  \\
		(trivial link function) & time ratio & 1.51  & 1.41  & 1.49  & 2.04  \\
		& \textbf{efficiency ratio} & 1.93  & 3.48  & 6.56  & 12.63 \\
		& stage-1 time proportion & 0.12  & 0.16  & 0.09  & 0.07   \\	
		\bottomrule
	\end{tabular}%
	\label{tab:markov jump}%
\end{table}%

\section{Optimizing Tilting Parameters via the Cross-Entropy Method}
\label{appendix:cross entropy}

In the main text Section \ref{sec:3.2}, we suggest a two-stage algorithm for duo-exponential tilting importance sampling. The first stage is a ``trial simulation" used to search for the optimal tilting parameters given a suitably chosen link function. We suggest using the stochastic gradient descent algorithm or other stochastic optimization algorithms to minimize the second moment of the importance sampling estimator $G(\theta,\eta)$ in \eqref{eq:2moment}, which effectively minimizes the variance of the importance sampling estimator. We describe the details for incorporating the cross-entropy (CE) method by \citet{rubinstein2004cross} into our duo-exponential tilting framework in this part. 

In short, the CE method aims to find the tilting distribution within the proposed family that is closest (in the sense of Kullback-Leibler distance) to the zero-variance distribution. For illustrative purpose, we consider the fixed-time event where $\tau$ is a deterministic time. Then the density function of the zero-variance distribution is $g^*(X_0,\cdots,X_{\tau},Y_1,\cdots,Y_{\tau})  = \frac{1}{C} F(S_{\tau}) \nu(X_0)p(X_0,X_1)\cdots p(X_{\tau-1}, X_{\tau})\rho(Y_1|X_0,X_1),\cdots \rho(Y_{\tau}|X_{\tau-1}, X_{\tau})$, where $C = \E_{\nu}\left[ F(S_{\tau}) \right]$ is an unknown constant. The CE method minimizes
\[\begin{aligned}
	\min_{\theta,\eta} & \int    g^*(x_0,\cdots,x_{\tau},y_1,\cdots,y_{\tau}) \log g^*(x_0,\cdots,x_{\tau},y_1,\cdots,y_{\tau}) \dd x_{0:\tau} \dd y_{1:\tau} \\
	& - \int \log \bigg[ \exp\left\{ \sum_{i=1}^{\tau} \theta^\top y_i + k(x_{i-1}, x_i,\eta) - \psi(x_{i-1}, x_i,\theta) - \phi(x_{i-1}, \eta)\right\} \\
	& \quad \quad \quad \quad \nu(x_0)p(x_0,x_1)\cdots p(x_{\tau-1}, x_{\tau})\rho(y_1|x_0,x_1),\cdots \rho(y_{\tau}|x_{\tau-1}, x_{\tau}) \bigg] g^*(x_0,\cdots,x_{\tau},y_1,\cdots,y_{\tau}) \dd x_{0:\tau} \dd y_{1:\tau} .
\end{aligned}  \] 

The objective function reduces to 
\[ \begin{aligned}
	& \max_{\theta,\eta}  \int \left[    \sum_{i=1}^{\tau} \theta^\top y_i + k(x_{i-1}, x_i,\eta) - \psi(x_{i-1}, x_i,\theta) - \phi(x_{i-1}, \eta) \right] F(S_{\tau}) \\
	& \quad \quad \quad \quad \nu(x_0)p(x_0,x_1)\cdots p(x_{\tau-1}, x_{\tau})\rho(y_1|x_0,x_1),\cdots \rho(y_{\tau}|x_{\tau-1}, x_{\tau})\dd x_{0:\tau} \dd y_{1:\tau} \\
	= &  \max_{\theta,\eta} \E_{\nu}\left[ F(S_{\tau})  \sum_{i=1}^{\tau} \left(  \theta^\top Y_i + k(X_{i-1}, X_i,\eta) - \psi(X_{i-1}, X_i,\theta) - \phi(X_{i-1}, \eta) \right)   \right] \\
	= & \max_{\theta,\eta} \E_{\nu}\left[ F(S_{\tau}) \left( \theta^\top S_{\tau} + \sum_{i=1}^{\tau}  k(X_{i-1}, X_i,\eta) - \sum_{i=1}^{\tau}\psi(X_{i-1}, X_i,\theta) - \sum_{i=1}^{\tau} \phi(X_{i-1}, \eta) \right)   \right] \\
	=:& \max_{\theta,\eta}  H(\theta,\eta) .
\end{aligned}\]

Therefore, to incorporate the CE method into our duo-exponential tilting framework, it suffices to replace $\min_{\theta,\eta} G(\theta,\eta)$ in the first stage in Algorithm \ref{algo:1} by $\max_{\theta,\eta}  H(\theta,\eta)$. Moreover, when solving this optimization problem, one may use a previously obtained tilting parameter to perform importance sampling, that is,
\[\begin{aligned}
	H(\theta,\eta) = &  \E_{\nu}\left[ F(S_{\tau}) \left( \theta^\top S_{\tau} + \sum_{i=1}^{\tau}  k(X_{i-1}, X_i,\eta) - \sum_{i=1}^{\tau}\psi(X_{i-1}, X_i,\theta) - \sum_{i=1}^{\tau} \phi(X_{i-1}, \eta) \right)   \right] \\
	= & \E_{\nu}^{\p_{\tilde\theta,\tilde\eta}}\bigg[ F(S_{\tau}) \left( \theta^\top S_{\tau} + \sum_{i=1}^{\tau}  k(X_{i-1}, X_i,\eta) - \sum_{i=1}^{\tau}\psi(X_{i-1}, X_i,\theta) - \sum_{i=1}^{\tau} \phi(X_{i-1}, \eta) \right) \\
	& \quad \quad \quad \quad  e^{-\tilde \theta^\top S_{\tau} + \sum_{i=1}^{\tau} - k(X_{i-1},X_i,\tilde\eta) + \psi(X_{i-1},X_{i},\tilde\theta)+\phi(X_{i-1},\tilde\eta)}  \bigg],
\end{aligned} \]
where $(\tilde\theta,\tilde\eta)$ is any tilting parameter, and $\p_{\tilde\theta,\tilde\eta}$ is the distribution defined in \eqref{eq:new prob}. Therefore, one may use the stochastic gradient algorithm to iteratively find the desired tilting parameter. In particular, our implementation follows the suggested procedure in \citet{de2005tutorial}.

\section{Proofs }
\vspace*{0.5cm}

\subsection{Proof of Lemma \ref{lemma:interchange order}}
\label{appendix:change order expectation differentiation}

We provide a detailed argument for interchanging the order between the expectation and differentiation. To this end, we present the following variant of the dominated convergence theorem:

\begin{lemma}
	\label{lemma:dominated convergence} 
	Let $f_n,g_n,h_n\in L^1$ such that $g_n\geq 0, h_n\geq 0$ almost surely for all $n$. Suppose $f_n\to f_{\infty}$ almost surely, and $|f_n| \leq g_n + h_n$.  In addition, we assume $g_n\to g_{\infty}$ in $L^1$ and almost surely, and $\E[h_n] \to 0$. Then $f_n\to f_{\infty}$ in $L^1$.
\end{lemma}

{\it Proof.} 
By Fatou's lemma, $\E[|f_{\infty}|]\leq \liminf_{n\to \infty} \E[|f_n|] \leq \E[|g_{\infty}|] < \infty$, hence $f_{\infty}\in L^1$.

Note that $|f_n - f_{\infty}| \leq g_n + h_n + |f_{\infty}|$, hence we apply Fatou's lemma again to $g_n + h_n + |f_{\infty}| - |f_n - f_{\infty}| \geq 0$ and obtain
\[ \E\left[\liminf_{n\to \infty}\left( g_n + h_n + |f_{\infty}| - |f_n - f_{\infty}|\right) \right] \leq \liminf_{n\to \infty} \E[g_n + h_n + |f_{\infty}| - |f_n - f_{\infty}|] .\]
That is,
\[ \E\left[ g_{\infty} + |f_{\infty}| + \liminf_{n\to \infty} h_n  \right]   \leq \E[g_{\infty}] + \E[|f_{\infty}|] - \limsup_{n\to\infty}\E[|f_n - f_{\infty}|] .\]
Therefore, $\limsup_{n\to\infty}\E[|f_n - f_{\infty}|] + \E[\liminf_{n\to \infty} h_n] \leq 0$. This proves $\E[|f_n -f_{\infty}|]\to 0$, i.e., $f_n\to f_{\infty}$ in $L^1$. $\hfill \Box$~

The following moment estimate is also needed in the proof.
\begin{lemma}
	\label{lemma:moment estimate}
	Under Assumptions \ref{ass:convex}, \ref{ass:recurrent}, and \ref{ass:rigorous eigenvalues}, for a fixed $n$, the following statements hold.
	\begin{enumerate}[(i)]
		\item When $\theta$ lies in the interior of $\Theta$ and sufficiently small, $\E_{\nu}\left[e^{\theta^\top S_n} |S_n|^q \right] < \infty$ for any $q\geq 0$.
		\item We denote $\tilde S_n = \sum_{i = 1}^n |X_i|$. For any $\ell$ sufficiently small and any $q \geq 0$, $\E_{\nu}[e^{\ell \tilde S_n} \tilde S_n^q] < \infty$. 
		\item When $\theta$ lies in the interior of $\Theta$ and $\theta,\eta$ sufficiently small, $\E_{\nu}\left[e^{\theta^\top S_n + \ell \tilde S_n} \left( |S_n|^q  +  \tilde S_n^{p} \right) \right] < \infty$ for any $p,q\geq 0$.
	\end{enumerate}
\end{lemma}
{\it Proof.} 

\begin{enumerate}[(i)]
	\item When the additive part does not degenerate, by Proposition 1(i) in \citet{fuh2004renewal}, under Assumptions \ref{ass:convex} and \ref{ass:rigorous eigenvalues}, and when $\theta$ is sufficiently small, $\E_{\nu}\left[ e^{|\theta|  |S_n|}\right] < \infty$. Hence $\p_{\nu}(|S_n| > t) \leq \E_{\nu}\left[ e^{|\theta|  |S_n|}\right] e^{-|\theta| t}$. Then for any $q >0$,
	\[ \E_{\nu}[|S_n|^q] = \int_0^{\infty} q t^{q-1}\p_{\nu}(|S_n|  > t)\dd t \leq q \E_{\nu}\left[ e^{|\theta| |S_n|} \right] \int_0^{\infty} t^{q-1} e^{-|\theta| t}\dd t < \infty . \] 
	Since $\theta$ lies in the interior of $\Theta$, there exists $v>1$ such that $v\theta \in \Theta$ and $\E_{\nu}\left[ e^{v |\theta|  |S_n|} \right] < \infty$. By H\"older's inequality, 
	\[ \E_{\nu}\left[ e^{\theta^\top S_n} |S_n|^q  \right] \leq \left( \E_{\nu}\left[ e^{v |\theta|  |S_n|} \right] \right)^{1/v} \left( \E_{\nu}\left[  |S_n|^{qu} \right]   \right)^{1/u} < \infty, \]
	where $\frac{1}{v} + \frac{1}{u} = 1$.
	
	When the additive part degenerates, it reduces to $(ii)$.
	
	\item We only need to prove the case when $\mathcal X$ is unbounded. When $\ell < \frac{L}{1 + C_0 + \cdots + C_0^n}$, by Assumption \ref{ass:convex}(iii), we have
	\[ \E_{\nu}\left[ e^{\ell \tilde S_n} \right] = \E_{\nu}\left[ e^{\ell \tilde S_{n-1}} \E_{\nu}\left[ e^{\ell |X_n|} \Big| \f_{n-1}\right]  \right] \leq e^{C_0 \ell} \E_{\nu}\left[ e^{\ell \tilde S_{n-2} + (\ell + C_0 \ell)|X_{n-1}| }   \right] \leq \cdots \leq e^{C} \E_{\nu}\left[ e^{\ell (1 + C_0 + \cdots + C_0^n) |X_0|} \right] ,  \]
	for some finite constant $C$. Hence $\E_{\nu}\left[ e^{\ell \tilde S_n} \right] < \infty$ for any $\ell$ sufficiently small.  Moreover, for any $q > 0$, for the same reason, $ \E[\tilde S_n^q] < \infty $. Finally, $\E_{\nu}[e^{\ell \tilde S_n} \tilde S_n^q] < \infty$ follows by  H\"older's inequality.

	\item The conclusion follows from (i) and (ii) by H\"older's inequality.
\end{enumerate}
$\hfill \Box$~

\subsubsection*{Fixed-time events}

We first consider the case of a finite-state Markov random walk. 
\[ G(\theta,\eta) = \E_{\nu}[ \one_{\{ S_n>c \}} e^{-\theta S_{n} + \sum_{i=1}^{n} - k(X_{i-1},X_i,\eta) + \psi(X_{i-1},X_{i},\theta)+\phi(X_{i-1},\eta)}] . \]
Fix a sufficiently small $\delta > 0$, such that $\{ (\theta',\eta'): |\theta'-\theta| + |\eta' - \eta| < 2\delta \} \subset \Theta \times H$. Because $X_i$ only takes finite many values, we denote
\[\begin{aligned}
	C_{\delta} = & \sup_{x,x',|\theta'-\theta| < \delta} |\frac{\partial }{\partial \theta}\psi(x,x',\theta')| \vee \sup_{x,x',|\eta'-\eta| < \delta} |\frac{\partial }{\partial \eta}k(x,x',\eta')| \vee \sup_{x,x',|\eta'-\eta| < \delta} |\frac{\partial }{\partial \eta}\phi(x,x',\eta')| \\
	& \vee \sup_{x,x',|\theta'-\theta| < \delta} |\psi(x,x',\theta')| \vee \sup_{x,x',|\eta'-\eta| < \delta} |k(x,x',\eta')| \vee \sup_{x,x',|\eta'-\eta| < \delta} |\phi(x,x',\eta')|. 
\end{aligned} \]
We have
\[\begin{aligned}
	& \bigg|  \sum_{i=1}^{n} - k(X_{i-1},X_i,\eta + \Delta\eta) + k(X_{i-1},X_i,\eta) + \psi(X_{i-1},X_{i},\theta+\Delta \theta)+\phi(X_{i-1},\eta+\Delta \eta) \\
	& \quad \quad - \psi(X_{i-1},X_{i},\theta)-\phi(X_{i-1},\eta) \bigg| \\
	\leq & nC_{\delta}(2|\Delta\eta| + |\Delta\theta|) .
\end{aligned}  \]

We consider the case where $|\Delta\theta|\vee |\Delta\eta| < \frac{1}{4nC_{\delta}} \wedge \delta$. The difference has an estimate:
\[ \begin{aligned}
	& \Big| \one_{\{ S_n>c \}} e^{-(\theta+\Delta \theta) S_{n} + \sum_{i=1}^{n} - k(X_{i-1},X_i,\eta+\Delta\eta) + \psi(X_{i-1},X_{i},\theta+\Delta\theta)+\phi(X_{i-1},\eta+\Delta \eta)} \\
	& - \one_{\{ S_n>c \}} e^{-\theta^\top S_{n} + \sum_{i=1}^{n} - k(X_{i-1},X_i,\eta) + \psi(X_{i-1},X_{i},\theta)+\phi(X_{i-1},\eta)} \Big| \\
	\leq & \one_{\{ S_n>c \}} e^{-\theta S_{n} + \sum_{i=1}^{n} - k(X_{i-1},X_i,\eta) + \psi(X_{i-1},X_{i},\theta)+\phi(X_{i-1},\eta)} \\
	& \times \bigg\{ \left| e^{-\Delta\theta S_n} - 1  \right| e^{\sum_{i=1}^{n} - k(X_{i-1},X_i,\eta + \Delta\eta) + k(X_{i-1},X_i,\eta) + \psi(X_{i-1},X_{i},\theta+\Delta \theta)+\phi(X_{i-1},\eta+\Delta\eta) - \psi(X_{i-1},X_{i},\theta)-\phi(X_{i-1},\eta)} \\
	& + \left| e^{\sum_{i=1}^{n} - k(X_{i-1},X_i,\eta + \Delta\eta) + k(X_{i-1},X_i,\eta) + \psi(X_{i-1},X_{i},\theta+\Delta \theta)+\phi(X_{i-1},\eta+\Delta\eta) - \psi(X_{i-1},X_{i},\theta)-\phi(X_{i-1},\eta)}  -1  \right| \bigg\} \\
	\leq & \one_{\{ S_n>c \}} e^{-\theta S_{n} + 3nC_{\delta}}  \bigg\{ \one_{\{ |\Delta\theta| |S_n| \leq 1  \}} C |\Delta\theta| |S_n|  + \one_{\{ |\Delta\theta| |S_n| > 1  \}} \left[ e^{-\Delta\theta S_n} + 1 \right] + n C C_{\delta} (2|\Delta\eta| + |\Delta\theta|) \bigg\} \\
	\leq & \tilde C e^{-\theta S_{n}}\left( |S_n| + 1\right)(|\Delta\theta| + |\Delta\eta|) + \tilde C \one_{\{ |\Delta\theta| |S_n| > 1  \}} \left[ e^{-(\theta+\Delta\theta)S_n} + e^{-\theta S_n} \right] . 
\end{aligned} \]
Hence
\begin{eqnarray}
	&~& \frac{1}{|\Delta\theta| + |\Delta \eta|}\Big| \one_{\{ S_n>c \}} e^{-(\theta+\Delta \theta) S_{n} + \sum_{i=1}^{n} - k(X_{i-1},X_i,\eta+\Delta\eta) + \psi(X_{i-1},X_{i},\theta+\Delta\theta)+\phi(X_{i-1},\eta+\Delta \eta)} \nonumber \\
	&~& - \one_{\{ S_n>c \}} e^{-\theta^\top S_{n} + \sum_{i=1}^{n} - k(X_{i-1},X_i,\eta) + \psi(X_{i-1},X_{i},\theta)+\phi(X_{i-1},\eta)} \Big| \nonumber \\
	&\leq & \tilde C e^{-\theta S_{n}}\left( |S_n| + 1\right) + \frac{\tilde C}{|\Delta \theta|}\one_{\{ |\Delta\theta| |S_n| > 1  \}} \left[ e^{-(\theta+\Delta\theta)S_n} + e^{-\theta S_n} \right]. \label{rate of convergence} 
\end{eqnarray} 

To analyze \eqref{rate of convergence}, we first note that 
under Assumptions \ref{ass:convex}, \ref{ass:recurrent}, and \ref{ass:rigorous eigenvalues}, Lemma \ref{lemma:moment estimate} holds, then $\E[ e^{-\theta S_{n}}\left( |S_n| + 1\right) ] < \infty$. Next we need to show that $\frac{1}{|\Delta\theta|}\E\left[\one_{\{ |\Delta\theta| |S_n| > 1  \}} \left[ e^{-(\theta+\Delta\theta)S_n} + e^{-\theta S_n} \right] \right] \to 0$ as $\Delta\theta\to 0$.

Here we have 
\[ \frac{1}{|\Delta\theta|}\E\left[ \one_{\{ |\Delta\theta| |S_n| > 1  \}} e^{-\theta S_n} \right] \leq \E\left[ \one_{\{ |\Delta\theta| |S_n| > 1  \}}|S_n| e^{-\theta S_n} \right] \to 0 \]
due to $\E[ e^{-\theta S_{n}}  |S_n|  ] < \infty$ and the monotone convergence theorem.

The last term requires to estimate the probability $\p\left( |S_n| > \frac{1}{|\Delta\theta|} \right)$. Under Assumptions \ref{ass:convex}, \ref{ass:recurrent}, and \ref{ass:rigorous eigenvalues}, from the proof of Lemma \ref{lemma:moment estimate}, it is known that $\p\left( |S_n| > \frac{1}{|\Delta\theta|} \right) \approx e^{-\frac{K n}{|\Delta \theta|}}$ for some $K >0$.  Then by the H\"older's inequality, 
\[\begin{aligned}
	\frac{1}{|\Delta \theta|} \E\left[ \one_{\{ |S_n|>\frac{1}{|\Delta\theta|}\}} e^{-(\theta+\Delta \theta) S_n}  \right] \leq & \frac{1}{|\Delta \theta|} \left( \p\left( |S_n| > \frac{1}{|\Delta\theta|} \right) \right)^{1/q} \left( \E\left[ 
	e^{-p(\theta+\Delta \theta) S_n} \right] \right)^{1/p} ,
\end{aligned} \]
where $\frac{1}{p} + \frac{1}{q} = 1$. We may pick $p > 1$ but sufficiently close to 1 such that $-p(\theta+\Delta \theta) \in \Theta$ for any $|\Delta \theta| < \delta$. Hence $\frac{1}{|\Delta\theta|}\E\left[\one_{\{ |\Delta\theta| |S_n| > 1  \}}  e^{-(\theta+\Delta\theta)S_n}  \right]  \to 0$ as $\Delta\theta\to 0$.

When the state space of the underlying Markov chain is unbounded, we have Assumption \ref{ass:convex} $(iii)$ on the growth condition.

Note that on the event $\{ \tilde S_n \leq \frac{1}{24C_{\delta} (|\Delta \theta| + |\Delta \eta|)}  \}$, we have 
\[\begin{aligned}
	& \Big| - k(X_{i-1},X_i,\eta + \Delta\eta) + k(X_{i-1},X_i,\eta) + \psi(X_{i-1},X_{i},\theta+\Delta \theta)+\phi(X_{i-1},\eta+\Delta\eta) \\
	& \quad \quad - \psi(X_{i-1},X_{i},\theta)-\phi(X_{i-1},\eta) \Big| \\
	\leq & 3C_{\delta}(n + 2\tilde S_n) (|\Delta \theta| \vee |\Delta \eta|)  
	\leq 3n C_{\delta}(|\Delta \theta| \vee |\Delta \eta|) + 6 C_{\delta} \tilde S_n (|\Delta \theta| \vee |\Delta \eta|)  \leq 1 .
\end{aligned}  \]

We consider the similar estimate on the difference:
\[ \begin{aligned}
	& \Big| \one_{\{ S_n>c \}} e^{-(\theta+\Delta \theta) S_{n} + \sum_{i=1}^{n} - k(X_{i-1},X_i,\eta+\Delta\eta) + \psi(X_{i-1},X_{i},\theta+\Delta\theta)+\phi(X_{i-1},\eta+\Delta \eta)} \\
	& - \one_{\{ S_n>c \}} e^{-\theta^\top S_{n} + \sum_{i=1}^{n} - k(X_{i-1},X_i,\eta) + \psi(X_{i-1},X_{i},\theta)+\phi(X_{i-1},\eta)} \Big| \\
	\leq & \one_{\{ S_n>c \}} e^{-\theta S_{n} + \sum_{i=1}^{n} - k(X_{i-1},X_i,\eta) + \psi(X_{i-1},X_{i},\theta)+\phi(X_{i-1},\eta)} \\
	& \times \bigg\{ \left| e^{-\Delta\theta S_n} - 1  \right| e^{\sum_{i=1}^{n} - k(X_{i-1},X_i,\eta + \Delta\eta) + k(X_{i-1},X_i,\eta) + \psi(X_{i-1},X_{i},\theta+\Delta \theta)+\phi(X_{i-1},\eta+\Delta\eta) - \psi(X_{i-1},X_{i},\theta)-\phi(X_{i-1},\eta)} \\
	& + \left| e^{\sum_{i=1}^{n} - k(X_{i-1},X_i,\eta + \Delta\eta) + k(X_{i-1},X_i,\eta) + \psi(X_{i-1},X_{i},\theta+\Delta \theta)+\phi(X_{i-1},\eta+\Delta\eta) - \psi(X_{i-1},X_{i},\theta)-\phi(X_{i-1},\eta)}  -1  \right| \bigg\} \\
	\leq & \one_{\{ S_n>c \}} e^{-\theta S_{n} + 3nC_{\delta} + 6C_{\delta} \tilde S_n}  \bigg\{ \one_{\{ |\Delta\theta| |S_n| \leq 1  \}} C |\Delta\theta| |S_n|  + \one_{\{ |\Delta\theta| |S_n| > 1  \}} \left[ e^{-\Delta\theta S_n} + 1 \right]\bigg\}  \\
	& + \one_{\{ S_n>c \}} e^{-\theta S_{n} + \sum_{i=1}^{n} - k(X_{i-1},X_i,\eta) + \psi(X_{i-1},X_{i},\theta)+\phi(X_{i-1},\eta)} \one_{\{ \tilde S_n \leq \frac{1}{24C_{\delta} (|\Delta \theta| + |\Delta \eta|)}  
		\}}  3 C C_{\delta}(n + 2\tilde S_n) (|\Delta \theta| + |\Delta \eta|) \\
	& + \one_{\{ S_n>c \}} e^{-\theta S_n} \one_{\{ \tilde S_n > \frac{1}{24C_{\delta} (|\Delta \theta| + |\Delta \eta|)} \}} \Big| e^{\sum_{i=1}^{n} - k(X_{i-1},X_i,\eta+\Delta \eta) + \psi(X_{i-1},X_{i},\theta + \Delta \theta)+\phi(X_{i-1},\eta+\Delta \eta)} \\
	& - e^{\sum_{i=1}^{n} - k(X_{i-1},X_i,\eta) + \psi(X_{i-1},X_{i},\theta)+\phi(X_{i-1},\eta)} \Big| \\
	\leq & \tilde C e^{-\theta S_{n}+6C_{\delta} \tilde S_n }\left( |S_n| + \tilde S_n + 1\right)(|\Delta\theta| + |\Delta\eta|) + \tilde C \one_{\{ |\Delta\theta| |S_n| > 1  \}} \left[ e^{-(\theta+\Delta\theta)S_n} + e^{-\theta S_n} \right] e^{6C_{\delta} \tilde S_n} \\
	& + \tilde C \one_{\{ \tilde S_n > \frac{1}{24C_{\delta} (|\Delta \theta| + |\Delta \eta|)} \}} e^{-\theta S_n} e^{6C_{\delta}\tilde S_n} . 
\end{aligned} \]

It suffices to show 
\[ \E\left[ e^{-\theta S_{n}+6C_{\delta} \tilde S_n }\left( |S_n| + \tilde S_n + 1\right) \right] < \infty, \]
and when $|\Delta \theta| , |\Delta \eta| \to 0$,  
\[ \frac{1}{|\Delta \theta|} \E\left[ \one_{\{  |S_n| > \frac{1}{|\Delta\theta|}  \}} \left[ e^{-(\theta+\Delta\theta)S_n} + e^{-\theta S_n} \right] e^{6C_{\delta} \tilde S_n} \right] \to 0,\ \frac{1}{|\Delta \theta| + |\Delta \eta|} \E\left[ \one_{\{ \tilde S_n > \frac{1}{24C_{\delta} (|\Delta \theta| + |\Delta \eta|)} \}} e^{-\theta S_n} e^{6C_{\delta}\tilde S_n}  \right] \to 0  .\]

The first expectation is finite due to the H\"older's inequality and Lemma \ref{lemma:moment estimate}. The second limit holds by the same argument for the finite state-space case. The last limit holds because
\[\frac{1}{|\Delta \theta| + |\Delta \eta|} \E\left[ \one_{\{ \tilde S_n > \frac{1}{24C_{\delta} (|\Delta \theta| + |\Delta \eta|)} \}} e^{-\theta S_n} e^{6C_{\delta}\tilde S_n}  \right] \leq 24C_{\delta} \E\left[ \one_{\{ \tilde S_n > \frac{1}{24C_{\delta} (|\Delta \theta| + |\Delta \eta|)} \}} \tilde S_n e^{-\theta S_n} e^{6C_{\delta}\tilde S_n}  \right] \to 0 \]
by the monotone convergence theorem and the fact $\E\left[  \tilde S_n e^{-\theta S_n} e^{6C_{\delta}\tilde S_n} \right] <\infty$.

\subsubsection*{First-passage-time events}
To make the derivation transparent, we first consider the i.i.d. case. Note that for given $\mu > 0$, the second moment has the form 
\[ G(\theta) = \E[\one_{\{\tau_b < T\}} e^{-\theta S_{\tau_b} + \tau_b\psi(\theta)}] = \E[\one_{\{\tau_b < T\}} e^{-\theta R(b) + \tau_b\psi(\theta)}] e^{-\theta b} , \]
where $R(b) = S_{\tau_b} - b$ is the overshoot.

Define $\tilde G(\theta) = \E[\one_{\{\tau_b < T\}} e^{-\theta R(b) + \tau_b\psi(\theta)}]$.
It suffices to show 
\[ \tilde G'(\theta) = \E\left[ \left(-R(b) + \tau_b \psi'(\theta) \right) \one_{\{\tau_b < T\}} e^{-\theta R(b) + \tau_b\psi(\theta)} \right] , \]
for each $\theta$ such that $\theta,-\theta \in \Theta$.

For a given constant $C_1 > 0$, we consider $\delta > 0$ such that for any $\theta'\in B_{\delta}(\theta)$, $|\psi(\theta') - \psi(\theta) | \leq  C_1 |\theta' - \theta|$, and $|\psi(\theta')| < C_1|\psi(\theta)|$.

Assume $|\Delta \theta| < \delta$, and consider the difference
\[\begin{aligned}
	& \frac{1}{|\Delta \theta|} \left| \one_{\{\tau_b < T\}} e^{-(\theta + \Delta \theta) R(b) + \tau_b\psi(\theta + \Delta \theta)} - \one_{\{\tau_b < T\}} e^{-\theta R(b) + \tau_b\psi(\theta)} \right|\\
	\leq & \frac{1}{|\Delta \theta|} \one_{\{\tau_b < T\}} e^{-\theta R(b) + \tau_b\psi(\theta)}  \left| e^{-\Delta \theta R(b)}e^{\tau_b [\psi(\theta + \Delta \theta) - \psi(\theta)]} -1 \right| \\
	\leq & \frac{1}{|\Delta \theta|} \one_{\{\tau_b < T\}} e^{-\theta R(b) + \tau_b\psi(\theta)} \left\{ \left| \left( e^{-\Delta \theta R(b)} - 1 \right) e^{\tau_b [\psi(\theta + \Delta \theta) - \psi(\theta)]} \right| + \left| e^{\tau_b [\psi(\theta + \Delta \theta) - \psi(\theta)]} -1 \right|  \right\} \\
	\leq & \frac{1}{|\Delta \theta|}\one_{\{\tau_b < T\}} e^{-\theta R(b) + \tau_b\psi(\theta)} \left\{ \left| e^{-\Delta \theta R(b)} -1 \right| e^{ C_1 |\Delta \theta|}     +  \left| e^{\tau_b [\psi(\theta + \Delta \theta) - \psi(\theta)]} -1 \right| \right\}     \\
	\leq & \one_{\{\tau_b < T\}} e^{-\theta R(b) + \tau_b\psi(\theta)} \one_{\{ |\Delta \theta R(b)| \leq 1,\  \tau_b C_1 |\Delta \theta| \leq 1  \}} \left\{ R(b) e^{ C_1 \delta} + C C_1 \tau_b \right\} \\
	& + \frac{1}{|\Delta \theta|} \one_{\{\tau_b < T\}} \one_{\{ |\Delta \theta R(b)| > 1,\text{ or } \tau_b C_1 |\Delta \theta| > 1  \}} \left| e^{-(\theta + \Delta \theta) R(b) + \tau_b\psi(\theta+\Delta \theta)}  + e^{-\theta R(b) + \tau_b\psi(\theta)} \right| \\
	\leq & \one_{\{\tau_b < T\}} e^{-\theta R(b) + \tau_b\psi(\theta)}  \left\{ R(b) e^{ C_1 \delta} + CC_1 \tau_b  \right\}  \\
	& +  \one_{\{ |\Delta \theta R(b)| > 1,\text{ or } \tau_b C_1  |\Delta \theta| > 1  \}} \one_{\{\tau_b < T\}} \frac{\left| e^{-(\theta + \Delta \theta) R(b) + \tau_b\psi(\theta+\Delta \theta)}  + e^{-\theta R(b) + \tau_b\psi(\theta)} \right|}{|\Delta \theta|} \\
	\leq & \tilde C e^{-\theta R(b)} \left[ R(b) + 1 \right] + \tilde C \one_{\{ |\Delta \theta R(b)| > 1,\text{ or }\tau_b C_1  |\Delta \theta| > 1  \}} \frac{\left| e^{-(\theta + \Delta \theta) R(b) }  + e^{-\theta R(b)} \right|}{|\Delta \theta|}. 
\end{aligned}  \]

It suffices to prove $\E\left[  e^{-\theta R(b)} [R(b)  + 1] \right] < \infty$, and $\E\left[  \one_{\{ |\Delta \theta R(b)| > 1,\text{ or }\tau_b C_1   |\Delta \theta| > 1  \}} \frac{\left| e^{-(\theta + \Delta \theta) R(b)}  + e^{-\theta R(b)} \right|}{|\Delta \theta|}   \right] \to 0$ as $\Delta \theta \to 0$. For the latter, we also need to estimate the probability 
\[ \begin{aligned}
	& \p\left( |\Delta \theta R(b)| > 1,\text{ or } \tau_b C_1 |\Delta \theta| > 1 \right)\\
	= & \p\left( R(b) > \frac{1}{|\Delta \theta|},\text{ or } \tau_b > \frac{1}{C_1 |\Delta\theta|} \right) \\
	\leq & \p\left( R(b) > \frac{1}{|\Delta \theta|} \right) + \p\left( \tau_b > \frac{1}{C_1 |\Delta\theta|} \right) \\
	\leq & |\Delta \theta|^2 \left(  \E[R(b)^2 + C_1^2\tau_b^2] \right) .
\end{aligned}  \]

Next, we consider the case of a finite-state Markov random walk. For given $\mu > 0$,
\begin{eqnarray*} 
	G(\theta,\eta) &=& \E_{\nu}[ \one_{\{ \tau_b < T \}} e^{-\theta S_{\tau_b} + \sum_{i=1}^{\tau_b} - k(X_{i-1},X_i,\eta) + \psi(X_{i-1},X_{i},\theta)+\phi(X_{i-1},\eta)}] \\
	&=& \E_{\nu}[ \one_{\{ \tau_b < T \}} e^{-\theta R(b) + \sum_{i=1}^{\tau_b} - k(X_{i-1},X_i,\eta) + \psi(X_{i-1},X_{i},\theta)+\phi(X_{i-1},\eta)}] e^{-\theta b}. 
\end{eqnarray*}
Fix a sufficiently small $\delta > 0$, such that $\{ (\theta',\eta'): |\theta'-\theta| + |\eta' - \eta| < 2\delta \} \subset \Theta \times H$. Because $X_i$ only takes finite many values, we denote
\[\begin{aligned}
	C_{\delta} = & \sup_{x,x',|\theta'-\theta| < \delta} |\frac{\partial }{\partial \theta}\psi(x,x',\theta')| \vee \sup_{x,x',|\eta'-\eta| < \delta} |\frac{\partial }{\partial \eta}k(x,x',\eta')| \vee \sup_{x,x',|\eta'-\eta| < \delta} |\frac{\partial }{\partial \eta}\phi(x,x',\eta')| \\
	& \vee \sup_{x,x',|\theta'-\theta| < \delta} |\psi(x,x',\theta')| \vee \sup_{x,x',|\eta'-\eta| < \delta} |k(x,x',\eta')| \vee \sup_{x,x',|\eta'-\eta| < \delta} |\phi(x,x',\eta')|. 
\end{aligned} \]
On the event $\{\tau_b < T\}$, we have
\[\begin{aligned}
	& \bigg|  \sum_{i=1}^{\tau_b} - k(X_{i-1},X_i,\eta + \Delta\eta) + k(X_{i-1},X_i,\eta) + \psi(X_{i-1},X_{i},\theta+\Delta \theta)+\phi(X_{i-1},\eta+\Delta\eta) \\
	& ~ - \psi(X_{i-1},X_{i},\theta)-\phi(X_{i-1},\eta) \bigg| 
	\leq  TC_{\delta}(2|\Delta\eta| + |\Delta\theta|) .
\end{aligned}  \]

We consider the case where $|\Delta\theta|\vee |\Delta\eta| < \frac{1}{4TC_{\delta}} \wedge \delta$. The difference has an estimate:
\[ \begin{aligned}
	& \Big| \one_{\{ \tau_b < T \}} e^{-(\theta+\Delta \theta) R(b) + \sum_{i=1}^{\tau_b} - k(X_{i-1},X_i,\eta+\Delta\eta) + \psi(X_{i-1},X_{i},\theta+\Delta\theta)+\phi(X_{i-1},\eta+\Delta \eta)} \\
	& - \one_{\{ \tau_b < T \}} e^{-\theta^\top R(b) + \sum_{i=1}^{\tau_b} - k(X_{i-1},X_i,\eta) + \psi(X_{i-1},X_{i},\theta)+\phi(X_{i-1},\eta)} \Big| \\
	\leq & \one_{\{ \tau_b< c\}} e^{-\theta R(b) + \sum_{i=1}^{\tau_b} - k(X_{i-1},X_i,\eta) + \psi(X_{i-1},X_{i},\theta)+\phi(X_{i-1},\eta)} \\
	& \times \bigg\{ \left| e^{-\Delta\theta R(b)} - 1  \right| e^{\sum_{i=1}^{\tau_b} - k(X_{i-1},X_i,\eta + \Delta\eta) + k(X_{i-1},X_i,\eta) + \psi(X_{i-1},X_{i},\theta+\Delta \theta)+\phi(X_{i-1},\eta+\Delta\eta) - \psi(X_{i-1},X_{i},\theta)-\phi(X_{i-1},\eta)} \\
	& + \left| e^{\sum_{i=1}^{\tau_b} - k(X_{i-1},X_i,\eta + \Delta\eta) + k(X_{i-1},X_i,\eta) + \psi(X_{i-1},X_{i},\theta+\Delta \theta)+\phi(X_{i-1},\eta+\Delta\eta) - \psi(X_{i-1},X_{i},\theta)-\phi(X_{i-1},\eta)}  -1  \right| \bigg\} \\
	\leq & \one_{\{ \tau_b < T \}} e^{-\theta R(b) + 3TC_{\delta}}  \bigg\{ \one_{\{ |\Delta\theta| R(b) \leq 1  \}} C |\Delta\theta| R(b)  + \one_{\{ |\Delta\theta| R(b) > 1  \}} \left[ e^{-\Delta\theta R(b)} + 1 \right] + TC C_{\delta} (2|\Delta\eta| + |\Delta\theta|) \bigg\} \\
	\leq & \tilde C e^{-\theta R(b)}\left( R(b) + 1\right)(|\Delta\theta| + |\Delta\eta|) + \tilde C \one_{\{ |\Delta\theta| R(b) > 1  \}} \left[ e^{-(\theta+\Delta\theta) R(b)} + e^{-\theta R(b)} \right]. 
\end{aligned} \]
Hence
\begin{eqnarray}
	&~& \frac{1}{|\Delta\theta| + |\Delta \eta|}\Big| \Big| \one_{\{ \tau_b < T \}} e^{-(\theta+\Delta \theta) R(b) + \sum_{i=1}^{\tau_b} - k(X_{i-1},X_i,\eta+\Delta\eta) + \psi(X_{i-1},X_{i},\theta+\Delta\theta)+\phi(X_{i-1},\eta+\Delta \eta)} \nonumber \\
	&~& - \one_{\{ \tau_b < T \}} e^{-\theta R(b) + \sum_{i=1}^{\tau_b} - k(X_{i-1},X_i,\eta) + \psi(X_{i-1},X_{i},\theta)+\phi(X_{i-1},\eta)} \Big| \nonumber \\
	&\leq & \tilde C e^{-\theta R(b)}\left( R(b) + 1\right) + \frac{\tilde C}{|\Delta \theta|}\one_{\{ |\Delta\theta| R(b) > 1  \}} \left[ e^{-(\theta+\Delta\theta) R(b)} + e^{-\theta R(b)} \right]. \label{rate of convergence for the overshoot}
\end{eqnarray} 

To analyze \eqref{rate of convergence for the overshoot}, we first note that 
under Assumptions \ref{ass:convex}, \ref{ass:recurrent}, and \ref{ass:rigorous eigenvalues}, by Lemma 2 in \cite{fuh2004renewal}, we have 
$\sup_{b \geq 0} E_\pi R(b) \leq E_\pi(S_1^+)^2/E_\pi S_1,$ which implies that
$\E[ e^{-\theta R(b)}\left( R(b) + 1\right) ] < \infty$. Next, we will
analyze the second term in \eqref{rate of convergence for the overshoot}. By Markov inequality, we have $\frac{1}{|\Delta\theta|} \p\left( R(b) > \frac{1}{|\Delta\theta|} \right) \leq  E_\pi R(b) < 
\infty$, via Lemma 2 in \cite{fuh2004renewal}.
Applying H\"older's inequality, we have, under Assumptions \ref{ass:convex}, \ref{ass:recurrent}, and \ref{ass:rigorous eigenvalues}, for $\theta > 0$, $\Delta\theta > 0$,  $\frac{1}{|\Delta\theta|}\E\left[\one_{\{ |\Delta\theta| R(b) > 1  \}} \left[ e^{-(\theta+\Delta\theta) R(b)} + e^{-\theta R(b)} \right] \right] \to 0$ as $\Delta\theta\to 0$.

When the state space of the underlying Markov chain is unbounded, we have Assumption \ref{ass:convex} $(iii)$ on the growth condition.

Note that on the event $\{\tau_b < T, \  \tilde S_T \leq \frac{1}{24C_{\delta} (|\Delta \theta| + |\Delta \eta|)}  \}$, we have 
\[\begin{aligned}
	& \sum_{i=1}^{\tau_b} \Big| - k(X_{i-1},X_i,\eta + \Delta\eta) + k(X_{i-1},X_i,\eta) + \psi(X_{i-1},X_{i},\theta+\Delta \theta)+\phi(X_{i-1},\eta+\Delta\eta) \\
	& \quad \quad  - \psi(X_{i-1},X_{i},\theta)-\phi(X_{i-1},\eta) \Big| \\
	\leq & 3C_{\delta}(T+2\tilde S_T) (|\Delta \theta| \vee |\Delta \eta|)  
	\leq 3TC_{\delta}(|\Delta \theta| \vee |\Delta \eta|) + 6 C_{\delta} \tilde S_T (|\Delta \theta| \vee |\Delta \eta|)  \leq 1 .
\end{aligned}  \]

The difference can be bounded by
\[ \begin{aligned}
	& \Big| \one_{\{ \tau_b < T \}} e^{-(\theta+\Delta \theta) R(b) + \sum_{i=1}^{\tau_b} - k(X_{i-1},X_i,\eta+\Delta\eta) + \psi(X_{i-1},X_{i},\theta+\Delta\theta)+\phi(X_{i-1},\eta+\Delta \eta)} \\
	& - \one_{\{ \tau_b < T \}} e^{-\theta^\top R(b) + \sum_{i=1}^{\tau_b} - k(X_{i-1},X_i,\eta) + \psi(X_{i-1},X_{i},\theta)+\phi(X_{i-1},\eta)} \Big| \\
	\leq & \one_{\{ \tau_b< c\}} e^{-\theta R(b) + \sum_{i=1}^{\tau_b} - k(X_{i-1},X_i,\eta) + \psi(X_{i-1},X_{i},\theta)+\phi(X_{i-1},\eta)} \\
	& \times \bigg\{ \left| e^{-\Delta\theta R(b)} - 1  \right| e^{\sum_{i=1}^{\tau_b} - k(X_{i-1},X_i,\eta + \Delta\eta) + k(X_{i-1},X_i,\eta) + \psi(X_{i-1},X_{i},\theta+\Delta \theta)+\phi(X_{i-1},\eta+\Delta\eta) - \psi(X_{i-1},X_{i},\theta)-\phi(X_{i-1},\eta)} \\
	& + \left| e^{\sum_{i=1}^{\tau_b} - k(X_{i-1},X_i,\eta + \Delta\eta) + k(X_{i-1},X_i,\eta) + \psi(X_{i-1},X_{i},\theta+\Delta \theta)+\phi(X_{i-1},\eta+\Delta\eta) - \psi(X_{i-1},X_{i},\theta)-\phi(X_{i-1},\eta)}  -1  \right| \bigg\} \\
	\leq & \one_{\{ \tau_b < T \}} e^{-\theta R(b) + 3TC_{\delta} + 6 C_{\delta} \tilde S_T}  \bigg\{ \one_{\{ |\Delta\theta| R(b) \leq 1  \}} |\Delta\theta| R(b)   + \one_{\{ |\Delta\theta| R(b) > 1  \}} \left[ e^{-\Delta\theta R(b)} + 1 \right] \bigg\} \\
	& + \one_{\{ \tau_b < T \}} e^{-\theta R(b) + \sum_{i=1}^{\tau_b} - k(X_{i-1},X_i,\eta) + \psi(X_{i-1},X_{i},\theta)+\phi(X_{i-1},\eta) } \one_{\{ \tilde S_T \leq \frac{1}{24 C_{\delta} (|\Delta \theta| + |\Delta \eta|)}  \}}  3C C_{\delta}(T+2\tilde S_T) (|\Delta \theta| + |\Delta \eta|)   \\
	& + \one_{\{ \tau_b < T \}} e^{-\theta R(b)} \one_{\{ \tilde S_T >\frac{1}{24 C_{\delta} (|\Delta \theta| + |\Delta \eta|)}  \}} \Big| e^{\sum_{i=1}^{\tau_b} - k(X_{i-1},X_i,\eta+\Delta \eta) + \psi(X_{i-1},X_{i},\theta + \Delta \theta)+\phi(X_{i-1},\eta+\Delta \eta)} \\
	& - e^{\sum_{i=1}^{n} - k(X_{i-1},X_i,\eta) + \psi(X_{i-1},X_{i},\theta)+\phi(X_{i-1},\eta)} \Big|  \\
	\leq & \tilde C e^{-\theta R(b) + 6 C_{\delta} \tilde S_T }\left(\tilde S_T +  R(b) + 1\right)(|\Delta\theta| + |\Delta\eta|) + \tilde C \one_{\{ |\Delta\theta| R(b) > 1  \}} \left[ e^{-(\theta+\Delta\theta) R(b)} + e^{-\theta R(b)} \right]e^{6 C_{\delta} \tilde S_T} \\
	& + \tilde C  \one_{\{ \tilde S_T >\frac{1}{24 C_{\delta} (|\Delta \theta| + |\Delta \eta|)}  \}} e^{-\theta R(b) + 6 C_{\delta} \tilde S_T }  . 
\end{aligned} \]

It suffices to show 
\[ \E\left[ e^{-\theta R(b) +6C_{\delta} \tilde S_T }\left( R(b) + \tilde S_T + 1\right) \right] < \infty, \]
and when $|\Delta \theta| , |\Delta \eta| \to 0$,  
\[\begin{aligned}
	& \frac{1}{|\Delta \theta|} \E\left[ \one_{\{  R(b) > \frac{1}{|\Delta\theta|}  \}} \left[ e^{-(\theta+\Delta\theta) R(b)} + e^{-\theta R(b)} \right] e^{6C_{\delta} \tilde S_T} \right] \to 0,\\
	&  \frac{1}{|\Delta \theta| + |\Delta \eta|} \E\left[ \one_{\{ \tilde S_T > \frac{1}{24C_{\delta} (|\Delta \theta| + |\Delta \eta|)} \}} e^{-\theta R(b)} e^{6C_{\delta}\tilde S_T}  \right] \to 0  .
\end{aligned} \] 
The first two conditions hold due to the same argument as above and the H\"older's inequality.

The last limit holds because
\[\frac{1}{|\Delta \theta| + |\Delta \eta|} \E\left[ \one_{\{ \tilde S_T > \frac{1}{24C_{\delta} (|\Delta \theta| + |\Delta \eta|)} \}} e^{-\theta R(b)} e^{6C_{\delta}\tilde S_T}  \right] \leq 24C_{\delta} \E\left[ \one_{\{ \tilde S_T > \frac{1}{24C_{\delta} (|\Delta \theta| + |\Delta \eta|)} \}} \tilde S_T e^{-\theta R(b)} e^{6C_{\delta}\tilde S_T}  \right] \to 0 \]
by the monotone convergence theorem and the fact $\E\left[  \tilde S_T e^{-\theta R(b)} e^{6C_{\delta}\tilde S_T} \right] <\infty$, by Lemma \ref{lemma:moment estimate}.

\subsection{Proof of Theorem \ref{thm:link}}\label{ec:proof link}
Under Assumption  \ref{ass:eigenvalue exist},	$r(x;\alpha)$ is non-zero and well-defined. 

Recall that $\{(X_n,S_n),n\geq 0\}$ is a Markov chain, for any classical exponential tilting probability $\p_{\theta^*}$, the transition probability is given by \eqref{eq:ld prob}. Moreover, based on the definition of the eigenvalue and eigenfunction in \eqref{eq:eigenvalue}, we have
\[ e^{\Lambda(\theta^*)}r(x,\theta^*) =  \E_x[e^{{\theta^*}^\top Y_1}r(X_1,\theta^*)] = \E_x\big[r(X_1,\theta^*)\E_x[e^{ {\theta^*}^\top Y_1}|X_1]\big] = \E_x[r(X_1,\theta^*)e^{\psi(x,X_1,\theta^*)}]. \]

With $k(x,x',\eta) = \psi(x,x',\eta) + \log r(x',\eta)$, let us compute
\[ \phi(x,\eta) = \log\E_x[e^{\psi(x,X_1,\eta) + \log r(X_1,\eta)}] = \Lambda(\eta) + \log r(x,\eta) .\]
Therefore, the transition probability $\p_{\theta,\eta}$ defined in \eqref{eq:new prob}, with $\theta=\eta = \theta^*$, is 
\[ 
\begin{aligned}
	\p_{\theta^*,\theta^*}(\dd x',\dd s'|x,s) = e^{\log r(X_1,\theta^*) + \theta^\top (s'-s) - \Lambda(\theta^*) - \log r(x,\theta^*)}p(x,\dd x')\rho(s'-s|x,x')\dd s'.
\end{aligned}
\]
This coincides with \eqref{eq:ld prob}.
$\hfill \Box$

\subsection{Proof of Corollary \ref{coro:affine}}

Here, we consider the affine model and directly verify the proposed forms are the eigenvalues and the associated eigenfunction. Hence only Assumption \ref{ass:convex} is required.

Under the affine structure, $r(x,\theta) = A(\theta)^\top x$ is an eigenfunction if and only if 
\[  
\begin{aligned}
	& E_x[e^{\theta^\top Y_1}r(X_1,\theta)] = \E_x\big[r(X_1,\theta)\E_x[e^{\theta^\top Y_1}|X_1]\big] = \E[e^{A(\theta)^\top X_1+D_0(\theta)^\top X_1 + D_1(\theta)^\top x + D_2(\theta)}]\\
	= & e^{\big[ C_1\big(A(\theta) + D_0(\theta)\big) + D_1(\theta)\big]^\top x + C_2\big(A(\theta) + D_0(\theta)\big) + D_2(\theta)}=   e^{\Lambda(\theta)}e^{A(\theta)^\top x}.
\end{aligned}
\]
This is equivalent to $C_1\big(A(\theta) + D_0(\theta)\big) + D_1(\theta) = A(\theta)$ and $ C_2\big(A(\theta) + D_0(\theta)\big) + D_2(\theta) = \Lambda(\theta)$. For any given $\theta$, $D_0(\theta),D_1(\theta),D_2(\theta)$ are three constants, and $C_1(\cdot),C_2(\cdot)$ are two functions. Solving this system of algebraic equations leads to the eigenfunction and eigenvalue. 

Moreover, in this case, the classical exponential tilting family is 
\[ 	\p_{\theta}(\dd x',\dd s'|x,s) = e^{A(\theta)^\top (x'-x)-\Lambda(\theta)+\theta^\top (s'-s)}p(x,\dd x')\rho(s'-s|x,x')\dd s'. \]
On the other hand, the duo-exponential tilting distribution with link function $k(x,x',\eta) = \eta^\top x'$ is 
\[ 	\p_{\theta,\eta}(\dd x',\dd s'|x,s) = e^{\eta^\top x' + \theta^\top (s'-s) - C_1(\eta)^\top x - C_2(\eta) - D_0(\theta)^\top x' - D_1(\theta)^\top x - D_2(\theta) }p(x,\dd x')\rho(s'-s|x,x')\dd s'. \]
By direct comparison, we could see that when $\eta = A(\theta) + D_0(\theta)$, probability $	\p_{\theta,\eta}(\dd x',\dd s'|x,s)$ coincides with $\p_{\theta}(\dd x',\dd s'|x,s)$. Therefore, the proposed duo-exponential tilting family contains the classical exponential tilting as a proper subset. $\hfill \Box$

\subsection{Proof of Lemma \ref{thm:optimal foc}}
\label{ec:proof foc}
First, under Assumption \ref{ass:convex}, we claim $\psi(x,x',\theta)$ and $\phi(x,\eta)$ are convex functions of $\theta$ and $\eta$, respectively, and hence $\Theta$ and $H$ are both convex sets. This is a well-known result for cumulant generating functions. For example, it can be proved by 
H\"older's inequality: for given $\iota > 0$
\[\begin{aligned}
	e^{\phi(x,  \iota \eta_1 + (1-\iota)\eta_2 )}  = & \E_x[e^{(\iota\eta_1 + (1-\iota)\eta_2)^\top \tilde{k}(x,X_1)}] = \E_x[(e^{\eta_1^\top \tilde{k}(x,X_1)})^{\iota} (e^{\eta_2^\top \tilde{k}(x,X_1)})^{1-\iota} ] \\
	\leq &  \big( \E_x[e^{\eta_1^\top \tilde{k}(x,X_1)}] \big)^{\iota} \big( \E_x[e^{\eta_2^\top \tilde{k}(x,X_1)}] \big)^{1-\iota}  = e^{\iota\phi(x,\eta_1)+(1-\iota)\phi(x,\eta_2)} .
\end{aligned} \]
Hence $\phi\big( \iota\eta_1 + (1-\iota)\eta_2,x \big) \leq \iota\phi(x,\eta_1)+(1-\iota)\phi(x,\eta_2)$. That is, $\phi$ is convex in $\eta$.

Therefore, inside the expectation, for any trajectory, 
$F(X_{\tau}, S_{\tau}) \exp\{ -\theta^\top S_{\tau} + \sum_{i=1}^{\tau} - \eta^\top \tilde{k}(X_{i-1},X_i) + \phi(X_{i-1},\eta) + \psi( X_{i-1},X_i, \eta)  \}$ is a composition between exponential function and a convex function and hence remains convex in terms of $(\theta,\eta)$. After taking the expectation, $G(\theta,\eta)$ is also convex. $\hfill \Box$

\subsection{Proof of Theorem \ref{thm:lan optimal linear}}
\label{ec:proof poisson}
We first present a lemma about the optimal log-likelihood ratio for a joint normal distribution.
\begin{lemma}
	\label{lemma:joint normal optimal}
	Under the LAN family and asymptotic normal regime, the second moment of the Monte-Carlo estimator \eqref{eq:2moment} is asymptotically equivalent to $\E[\bar{F}^2(N) e^{-N_{\tilde{\p}}}]$, where $(N,N_{\tilde{\p}})$ follows a joint normal distribution. Suppose 
	\[ \begin{pmatrix}
		N \\ N_{\tilde{\p}}
	\end{pmatrix} \sim \mathcal{N} \bigg(
	\begin{pmatrix}
		\mu_N \\
		\mu_{\tilde{\p}}
	\end{pmatrix},
	\begin{pmatrix}
		\Sigma_N & \Sigma_{LN}^\top \\
		\Sigma_{LN} & \sigma^2_{{\tilde{\p}}}
	\end{pmatrix} \bigg) .\]
	Then the minimum of $\E[\bar{F}^2(N) e^{-N_{\tilde{\p}}}]$ is achieved when $\sigma^2_{\tilde{\p}} = \Sigma_{LN}^\top\Sigma_N^{-1}\Sigma_{LN}$. 
\end{lemma}

{\it Proof.} First, we note that under the LAN family assumption of the likelihood ratio and asymptotic normality of the statistic of interest $\bar{F}^2(S_\tau)$, 
by using a standard joint central limit theorem, we have that the second moment of the Monte-Carlo estimator \eqref{eq:2moment} is asymptotically equivalent to $\E[\bar{F}^2(N) e^{-N_{\tilde{\p}}}]$. (3.6) and (3.7) in
\cite{fuh2007estimation} presents the simple case in finite state Markov chains. Second,
since $N_{\tilde{\p}}$ is the limit of a log-likelihood ratio, $\E[e^{N_{\tilde{\p}}}] = 1$. It means $\mu_{\tilde{\p}} = -\frac{1}{2}\sigma_{\tilde{\p}}^2$. 

We calculate the objective by conditioning on $N$ first. Since 
\[ N_{\tilde{\p}}|N \sim \mathcal{N}\left( \mu_{\tilde{\p}} + \Sigma_{LN}^\top\Sigma_N^{-1}(N - \mu_N), \sigma^2_{\tilde{\p}} - \Sigma_{LN}^\top\Sigma_N^{-1}\Sigma_{LN}  \right),\]
we have
\[ \E[\bar{F}^2(N) e^{-N_{\tilde{\p}}}] = \E[\bar{F}^2(N) \E[e^{-N_{\tilde{\p}}}|N] ] = \E[\bar{F}^2(N) e^{-\Sigma_{LN}^\top\Sigma_N^{-1}(N - \mu_N)} ]e^{\sigma^2_{\tilde{\p}} - \frac{1}{2}\Sigma_{LN}^\top\Sigma_N^{-1}\Sigma_{LN}} .\]
Hence it is a decreasing function in $\sigma_{\tilde{\p}}^2$, and the joint normality implies that $\sigma_{\tilde{\p}}^2 \geq \Sigma_{LN}^\top\Sigma_N^{-1}\Sigma_{LN}$. This proves our claim. Moreover, when $\sigma^2_{\tilde{\p}} = \Sigma_{LN}^\top\Sigma_N^{-1}\Sigma_{LN}$, there exist a vector $\bar{\theta}$ and a scalar $c$, such that $N_{\tilde{\p}} = \bar{\theta}^\top N + c$. $\hfill \Box$


Now let us first prove an intermediary result. We use the following notion to denote the log-likelihood ratio up to time $n$ induced by our proposed change of measure \eqref{eq:new prob}:
\[ LR_n(\theta,\eta) := \sum_{i=1}^n\left[ \theta^\top Y_i + k(X_{i-1}, X_i,\eta) - \psi(X_{i-1}, X_i,\theta)- \phi(X_{i-1},\eta)  \right]. \]
\begin{lemma}
	\label{lemma:lan optimal}
	Under Assumptions \ref{ass:convex}, \ref{ass:recurrent}, and \ref{ass:rigorous}, the asymptotic optimal tilting distribution within the LAN family for an event in the asymptotic normal regime is achieved by choosing the link function according to $k(x,x',\eta) = \psi(x,x',\eta) + \eta^\top g(x')$, where $g$ satisfies \eqref{eq:poisson 0}, and the tilting parameter satisfies $\theta= O(\frac{1}{\sqrt{n}})$, $\eta= O(\frac{1}{\sqrt{n}})$, and $\theta-\eta = o(\frac{1}{\sqrt{n}})$.

\end{lemma}

{\it Proof.} 
Assumption \ref{ass:convex} guarantees the existence of $\psi$ and Assumption \ref{ass:rigorous} assumes the existence of $g$. 

Given this link function $k(x,x',\eta) = \psi(x,x',\eta) + \eta^\top g(x')$, based on the definition, the cumulant function is 
\[ e^{\phi(x,\eta)} = \E[e^{k(X_0,X_1,\eta)}|X_0=x] = \E[e^{\psi(X_0,X_1,\eta) + \eta^\top g(X_1)}|X_0 = x] = \E[e^{\eta^\top Y_1 + \eta^\top g(X_1)}|X_0 = x] .\]
Therefore, $\frac{\partial \phi}{\partial \eta}(x,0) = \E_x[Y_1 + g(X_1)]$. Since $g$ solves \eqref{eq:poisson 0}, $g(x) - \E_x[g(X_1)] = \E_x[Y_1] - \E_{\pi}[Y_1]$, we obtain $\frac{\partial \phi}{\partial \eta}(x,0) = g(x) + \E_{\pi}[Y_1]$. 

Now let us check the log-likelihood ratio when $\theta=\eta = -\frac{\bar{\theta}}{\sqrt{n}}$:
\[\begin{aligned}
	LR_n(-\frac{\bar{\theta}}{\sqrt{n}},-\frac{\bar{\theta}}{\sqrt{n}}) = & -\frac{\bar{\theta}^\top}{\sqrt{n}}\sum_{i=1}^n[Y_i + g(X_i)] - \sum_{i=1}^n\phi(X_{i-1},-\frac{\bar{\theta}}{\sqrt{n}}) \\
	= & -\frac{\bar{\theta}^\top}{\sqrt{n}}\sum_{i=1}^n[Y_i + g(X_i) - \frac{\partial \phi}{\partial \eta}(X_{i-1},0)] - \frac{1}{2n}\sum_{i=1}^n \bar{\theta}^\top\frac{\partial^2 \phi}{\partial \eta^2}(X_{i-1},0) \bar{\theta} + o(1) \\
	= & -\frac{\bar{\theta}^\top}{\sqrt{n}}\sum_{i=1}^n\big[Y_i + g(X_i) - g(X_{i-1}) - \E_{\pi}[Y_1]\big] - \frac{1}{2n}\sum_{i=1}^n \bar{\theta}^\top\frac{\partial^2 \phi}{\partial \eta^2}(X_{i-1},0) \bar{\theta} + o(1) \\
	= & -\frac{\bar{\theta}^\top}{\sqrt{n}}\sum_{i=1}^n\big[Y_i - \E_{\pi}[Y_1]\big] -\frac{\bar{\theta}^\top}{\sqrt{n}}[g(X_n) - g(X_0)] - \frac{1}{2n}\sum_{i=1}^n \bar{\theta}^\top\frac{\partial^2 \phi}{\partial \eta^2}(X_{i-1},0) \bar{\theta} + o(1) \\
	= & \underbrace{-\frac{\bar{\theta}^\top}{\sqrt{n}}\big( S_n - n\E_{\pi}[Y_1]\big)}_{\text{central limit theorem}} \underbrace{-\frac{\bar{\theta}^\top}{\sqrt{n}}[g(X_n) - g(X_0)]}_{\to 0} \underbrace{- \frac{1}{2n}\sum_{i=1}^n \bar{\theta}^\top\frac{\partial^2 \phi}{\partial \eta^2}(X_{i-1},0) \bar{\theta}}_{\text{law of large numbers}} + o(1).
\end{aligned}\] 
By the central limit theorem in Theorem 17.5.3 of \cite{meyn2012markov}, Assumption \ref{ass:recurrent}, and Slutsky's theorem, $LR_n(-\frac{\bar{\theta}}{\sqrt{n}},-\frac{\bar{\theta}}{\sqrt{n}})$ converges to a normal distribution. 

Next, we check the asymptotic mean and variance of the log likelihood ratio. For simplicity, we denote by $m\in \mathbb{R}^d,C\in \mathbb{R}^{d\times d}$ the asymptotic mean and variance of $S_n$. In particular, $m = \E_{\pi}[Y_1]$, $C = \lim_{n\to 0^+}\frac{1}{n}\operatorname{Var}_{\pi}(S_n)$. The asymptotic mean of $LR_n(-\frac{\bar{\theta}}{\sqrt{n}},-\frac{\bar{\theta}}{\sqrt{n}}) $ is determined by the law of large numbers term that converges to $-\frac{1}{2}\bar{\theta}^\top\E_{\pi}[\frac{\partial^2 \phi}{\partial \eta^2}(X_0,0) ]\bar{\theta}$. The asymptotic variance of $LR_n(\bar{\theta},\bar{\theta}) $ is determined by the central limit theorem term, which equals to $\bar{\theta}^\top C\bar{\theta}$. It remains to verify $C = \E_{\pi}[\frac{\partial^2 \phi}{\partial \eta^2}(X_0,0) ]$. 

Denote $M_n = S_n + g(X_n) -nm - g(X_0) = \sum_{i=1}^n z_i$, with $z_n = Y_n + g(X_n) - g(X_{n-1}) - m$. Notice that $M_n$ is a martingale under any $\p_x$, and $\E_x[M_n] = 0$. Hence $\operatorname{Var}_x(\frac{1}{\sqrt{n}} M_n) = \frac{1}{n}\sum_{i=1}^n \E_x[z_iz_i^\top] =\frac{1}{n}\sum_{i=1}^n \E_x[\frac{\partial^2 \phi}{\partial \eta^2}(X_{i-1},0)]$. Recall that $\frac{1}{\sqrt{n}} M_n \to \mathcal{N}(m,C)$ in distribution under any $\p_x$. Therefore $\operatorname{Var}_x(\frac{1}{\sqrt{n}} M_n) \to C$ as $n\to \infty$ for any $x$. In particular, we take $x$ follows the stationary distribution and obtain $\E_{\pi}[\frac{\partial^2 \phi}{\partial \eta^2}(X_0,0) ]= C$. This proves that our tilting family also belongs to LAN family. 

Moreover, this normal distribution is a linear transformation of $S_n$. Therefore, by Lemma \ref{lemma:joint normal optimal}, the proposed tilting probability is asymptotically optimal within the LAN family. This is because the joint normal distributions for the log-likelihood ratio and the event considered in Lemma \ref{lemma:joint normal optimal} are exhaustive for the limiting distributions of all twisted probabilities in the LAN family (see Definitions \ref{def:lan} and \ref{def:normal regime}). Lemma \ref{lemma:joint normal optimal} shows that the proposed tilting probability minimizes the asymptotic variance and hence is asymptotically optimal within the LAN family.

Lastly, we relax the condition to $\theta =-\frac{\bar{\theta}_n}{\sqrt{n}}$, $\eta = -\frac{\bar{\eta}_n}{\sqrt{n}}$, and $\lim_{n \to \infty}\bar\theta_n = \lim_{n \to \infty}\bar\eta_n = \bar\theta < \infty$. In this case, the log-likelihood ratio becomes
\[\begin{aligned}
	&LR_n(-\frac{\bar{\theta}_n}{\sqrt{n}},-\frac{\bar{\eta}_n}{\sqrt{n}}) \\
	= & -\frac{\bar{\eta}_n^\top}{\sqrt{n}}\sum_{i=1}^n[Y_i + g(X_i)] - \sum_{i=1}^n\phi(X_{i-1},-\frac{\bar{\eta}_n}{\sqrt{n}}) + \sum_{i=1}^n[ -\frac{\bar{\theta}_n^\top - \bar{\eta}_n^\top}{\sqrt{n}} Y_i - \psi(X_{i-1},X_i,-\frac{\bar{\theta}_n}{\sqrt{n}}) + \psi(X_{i-1},X_i,-\frac{\bar{\eta}_n}{\sqrt{n}})  ] \\
	= & -\frac{\bar{\eta}_n^\top}{\sqrt{n}}\sum_{i=1}^n[Y_i + g(X_i)] - \sum_{i=1}^n\phi(X_{i-1},-\frac{\bar{\eta}_n}{\sqrt{n}}) -\frac{\bar{\theta}_n^\top - \bar{\eta}_n^\top}{\sqrt{n}} \sum_{i=1}^n[Y_i - \frac{\partial \psi}{\partial \theta}(X_{i-1},X_i,0)] \\
	& - \frac{1}{2}\sum_{i=1}^n[\frac{1}{n} \bar{\theta}_n^\top\frac{\partial^2 \psi}{\partial \theta^2}(X_{i-1},X_i,0)\bar{\theta}_n- \frac{1}{n}\bar{\eta}_n^\top\frac{\partial^2 \psi}{\partial \theta^2}(X_{i-1},X_i,0)\bar{\eta}_n + o(\frac{1}{n})] \\
	=: & I_1 + I_2 + I_3,
\end{aligned}\] 
where 
\[ \begin{aligned}
	I_1 = & -\frac{\bar{\eta}_n^\top}{\sqrt{n}}\sum_{i=1}^n[Y_i + g(X_i)] - \sum_{i=1}^n\phi(X_{i-1},-\frac{\bar{\eta}_n}{\sqrt{n}}) \\
	= & -\frac{\bar{\eta}_n^\top}{\sqrt{n}}\sum_{i=1}^n\big[Y_i - \E_{\pi}[Y_1]\big] -\frac{\bar{\eta}_n^\top}{\sqrt{n}}[g(X_n) - g(X_0)] - \frac{1}{2n}\sum_{i=1}^n \bar{\eta}_n^\top\frac{\partial^2 \phi}{\partial \eta^2}(X_{i-1},0) \bar{\eta}_n + o(1) \\
	= & -\frac{\bar{\theta}^\top}{\sqrt{n}}\sum_{i=1}^n\big[Y_i - \E_{\pi}[Y_1]\big] -\frac{\bar{\theta}^\top}{\sqrt{n}}[g(X_n) - g(X_0)] - \frac{1}{2n}\sum_{i=1}^n \bar{\theta}^\top\frac{\partial^2 \phi}{\partial \eta^2}(X_{i-1},0) \bar{\theta} + o(1)
\end{aligned} \]
by the above argument, 
\[ I_2 =  -\frac{\bar{\theta}_n^\top- \bar{\eta}_n^\top}{\sqrt{n}} \sum_{i=1}^n[Y_i - \frac{\partial \psi}{\partial \theta}(X_{i-1},X_i,0)]  = (\bar{\theta}_n^\top- \bar{\eta}_n^\top) O(1) =  o(1) , \]
and 
\[ I_3 = - \frac{1}{2}\sum_{i=1}^n[\frac{1}{n} \bar{\theta}_n^\top\frac{\partial^2 \psi}{\partial \theta^2}(X_{i-1},X_i,0)\bar{\theta}_n - \frac{1}{n}\bar{\eta}_n^\top\frac{\partial^2 \psi}{\partial \theta^2}(X_{i-1},X_i,0)\bar{\eta}_n + o(\frac{1}{n})] = o(1) .\]

Combining all the above together, we show that asymptotically $LR_n(-\frac{\bar{\theta}_n}{\sqrt{n}},-\frac{\bar{\eta}_n}{\sqrt{n}})$ converges to the same distribution as $LR_n(-\frac{\bar{\theta}}{\sqrt{n}},-\frac{\bar{\theta}}{\sqrt{n}})$. The proof is completed. $\hfill \Box$~\\

Finally, let us prove Theorem \ref{thm:lan optimal linear}. Recall that Assumptions \ref{ass:convex}, \ref{ass:recurrent}, and \ref{ass:rigorous} guarantee the existence of $\psi$ and $g$.

Given the link function \eqref{eq:md choose k} with $\theta = -\frac{\bar{\theta}}{\sqrt{n}}$, $\eta = -\frac{\bar{\eta}}{\sqrt{n}}$, the log-likelihood ratio is 
\[
\begin{aligned}
	& LR_n(-\frac{\bar{\theta}}{\sqrt{n}},-\frac{\bar{\eta}}{\sqrt{n}}) \\
	= & -\frac{\bar{\theta}^\top}{\sqrt{n}} S_n + \sum_{i=1}^n \big\{ -\psi(X_{i-1},X_i,-\frac{\bar{\theta}}{\sqrt{n}}) - \frac{\bar{\eta}^\top}{\sqrt{n}}[ \frac{\partial \psi}{\partial \theta}(X_{i-1},X_i,0) + \Delta(X_i,0)] - \phi(X_{i-1}, - \frac{\bar{\eta}}{\sqrt{n}}) \big\} \\
	= & -\frac{\bar{\theta}^\top}{\sqrt{n}} S_n + \sum_{i=1}^n \big\{ \frac{\bar{\theta}^\top - \bar{\eta}^\top}{\sqrt{n}}\frac{\partial \psi}{\partial \theta}(X_{i-1},X_i,0)  - \frac{\bar{\eta}^\top}{\sqrt{n}} \Delta(X_i,0) + \frac{\bar{\eta}^\top}{\sqrt{n}} \frac{\partial \phi}{\partial \eta}(X_{i-1},0)\\
	& + \frac{1}{2n}[-\bar{\theta}^\top \frac{\partial^2 \psi}{\partial \theta^2}(X_{i-1},X_i,0)\bar{\theta}  -\bar{\eta}^\top \frac{\partial^2 \phi}{\partial \eta^2}(X_{i-1},0)\bar{\eta} ] + o(\frac{1}{n}) \big\} \\
	= & \frac{1}{\sqrt{n}}\sum_{i=1}^n \big\{ -\bar{\theta}^\top Y_i + (\bar{\theta}^\top - \bar{\eta}^\top)\frac{\partial \psi}{\partial \theta}(X_{i-1},X_i,0)  - \bar{\eta}^\top \Delta(X_i,0) + \bar{\eta}^\top \frac{\partial \phi}{\partial \eta}(X_{i-1},0) \big\}\\
	& - \frac{1}{2n}\sum_{i=1}^n[ \bar{\theta}^\top \frac{\partial^2 \psi}{\partial \theta^2}(X_{i-1},X_i,0)\bar{\theta}   + \bar{\eta}^\top \frac{\partial^2 \phi}{\partial \eta^2}(X_{i-1},0)\bar{\eta} ] + o(1). 
\end{aligned} \]
In this case, 
\[ \E_x[e^{\eta^\top[ \frac{\partial \psi}{\partial \theta}(X_0,X_1,0)+\Delta(X_1,0)  ]}] = e^{\phi(x,\eta)},\ \E[e^{\theta^\top Y_1}|X_0,X_1] = e^{\psi(X_0,X_1,\theta)}, \]
hence 
\[ \E_x[ \frac{\partial \psi}{\partial \theta}(X_0,X_1,0)+\Delta(X_1,0) ] = \frac{\partial \phi}{\partial \eta}(x,0),\ \E[Y_1|X_0,X_1] = \frac{\partial \psi}{\partial \theta}(X_0, X_1,0) . \]
\[ \operatorname{Var}_x[ \frac{\partial \psi}{\partial \theta}(X_0,X_1,0)+\Delta(X_1,0) ] = \frac{\partial^2 \phi}{\partial \eta^2}(x,0),\ \operatorname{Var}[Y_1|X_0,X_1] = \frac{\partial^2 \psi}{\partial \theta^2}(X_0, X_1,0) . \]

Under the stationary distribution $\pi$ of the underlying Markov chain, then jointly, 
$\{(X_{i-1},X_i,Y_i), i \geq 0\}$ is also a Markov chain. Following the similar argument in the proof of Lemma \ref{lemma:lan optimal},  $LR_n(-\frac{\bar{\theta}}{\sqrt{n}},-\frac{\bar{\eta}}{\sqrt{n}})$ converges to a normal distribution.  Moreover,
we can find the asymptotic mean of $LR_n(-\frac{\bar{\theta}}{\sqrt{n}},-\frac{\bar{\eta}}{\sqrt{n}})$ is $-\frac{1}{2}\E_{\pi}[\bar{\theta}^\top \frac{\partial^2 \psi}{\partial \theta^2}(X_{0},X_1,0)\bar{\theta}   + \bar{\eta}^\top \frac{\partial^2 \phi}{\partial \eta^2}(X_{0},0)\bar{\eta}]$, and the asymptotic variance is $\E_{\pi}[\bar{\theta}^\top \frac{\partial^2 \psi}{\partial \theta^2}(X_{0},X_1,0)\bar{\theta}   + \bar{\eta}^\top \frac{\partial^2 \phi}{\partial \eta^2}(X_{0},0)\bar{\eta}]$.  This proves that our tilting family also belongs to LAN family.

In particular, by choosing $\theta = \eta = -\frac{\bar{\theta}}{\sqrt{n}}$, 
we will claim that $LR_n(-\frac{\bar{\theta}}{\sqrt{n}},-\frac{\bar{\theta}}{\sqrt{n}}) = \widebar{LR}_n(-\frac{\bar{\theta}}{\sqrt{n}}) + o(1)$, where
\[\widebar{LR}_n(\theta) = \theta^\top [S_n - n\Lambda'(0) + \sum_{i=1}^n \big(\Delta(X_{i},0) - \Delta(X_{i-1},0)\big) ] - \frac{n}{2}\theta^\top\Lambda''(0)\theta.\]
To see this, consider
\[\begin{aligned}
	& LR_n(-\frac{\bar{\theta}}{\sqrt{n}},-\frac{\bar{\theta}}{\sqrt{n}}) - \widebar{LR}_n(-\frac{\bar{\theta}}{\sqrt{n}})  -   \frac{1}{2}\bar{\theta}^\top\Lambda''(0)\bar{\theta} \\
	= & \sum_{i=1}^n \bigg[  -\frac{\bar{\theta}^\top}{\sqrt{n}}\Delta(X_{i-1},0) - \frac{\bar{\theta}^\top}{\sqrt{n}} \Lambda'(0)-\psi(X_{i-1},X_i,-\frac{\bar{\theta}}{\sqrt{n}}) - \frac{\bar{\theta}^\top}{\sqrt{n}}\frac{\partial \psi}{\partial \theta}(X_{i-1},X_i,0) - \phi(X_{i-1},-\frac{\bar{\theta}}{\sqrt{n}}) \bigg]  \\
	= & \sum_{i=1}^n \bigg[ -\frac{1}{2n}\bar{\theta}^\top[\frac{\partial^2 \psi}{\partial \theta^2}(X_{i-1},X_i,0) + \frac{\partial^2 \phi}{\partial \eta^2}(X_{i-1},0)]\bar{\theta} + \frac{\bar{\theta}^\top}{\sqrt{n}}[ \frac{\partial \phi}{\partial \eta}(X_{i-1},0) - \Delta(X_{i-1},0) - \Lambda'(0) ]    \bigg].
\end{aligned}\]
By definition of $\phi$, we have
\[ e^{\phi(x,\eta)} = \E\big[e^{\eta^\top [\frac{\partial \psi}{\partial \theta}(X_{0},X_1,0) + \Delta(X_1,0)]}|X_0 = x\big] ,\]
hence $\frac{\partial \phi}{\partial \eta}(x,0)  = \E[\frac{\partial \psi}{\partial \theta}(X_{0},X_1,0) + \Delta(X_1,0)|X_0 = x]$. On the other hand, from (i) we get
\[ \Lambda'(0) + \Delta(x,0) = \E[Y_1 + \Delta(X_1,0)|X_0 = x] = \E[\frac{\partial \psi}{\partial \theta}(X_{0},X_1,0) + \Delta(X_1,0)|X_0 = x]. \]
Therefore, $\Lambda'(0) + \Delta(x,0) = \frac{\partial \phi}{\partial \eta}(x,0)$.  

Note that $\widebar{LR}_n(-\frac{\bar{\theta}}{\sqrt{n}}) = -\frac{\bar{\theta}^\top}{\sqrt{n}}[S_n - n\Lambda'(0)] -\frac{\bar{\theta}^\top}{\sqrt{n}}[\Delta(X_n,0) - \Delta(X_0,0)] + c = -\frac{\bar{\theta}^\top}{\sqrt{n}}[S_n - n\Lambda'(0)] + c  +o(1) $. Therefore, by Lemma \ref{lemma:joint normal optimal}, the proposed tilting probability is asymptotically optimal.

Lastly, we relax the condition to $\theta =-\frac{\bar{\theta}_n}{\sqrt{n}}$, $\eta = -\frac{\bar{\eta}_n}{\sqrt{n}}$, and $\lim_{n \to \infty}\bar\theta_n = \lim_{n \to \infty}\bar\eta_n = \bar\theta < \infty$. The rest of the proof follows the same argument as in the proof of Lemma \ref{lemma:lan optimal}. $\hfill \Box$~\\

\subsection{Proof of Lemma \ref{lemma:poisson}}
Under Assumption \ref{ass:eigenvalue exist} $r(x;\alpha)$ is analytic.

We start from the definition of eigenvalue and eigenfunction in \eqref{eq:eigenvalue}. Notice that $\Lambda(0) = 0$ and $r(x,0)$ is normalized to be 1. We consider the Taylor expansion of both $r(x,\theta)$ and $\Lambda(\theta)$ at any point $\theta_0$:
\[ e^{\Lambda(\theta)} \approx e^{\Lambda(\theta_0)} + e^{\Lambda(\theta_0)}\Lambda'(\theta_0)^\top (\theta - \theta_0) + o(\theta - \theta_0),  \]
\[ r(x,\theta) \approx r(x,\theta_0) + \frac{\partial r}{\partial \theta}(x,\theta_0)^\top (\theta - \theta_0) + o(\theta - \theta_0). \]
Therefore, under Assumption \ref{ass:convex}, \eqref{eq:eigenvalue} becomes
\[ 
\begin{aligned}
	& \big[ e^{\Lambda(\theta_0)} + e^{\Lambda(\theta_0)}\Lambda'(\theta_0)^\top (\theta - \theta_0) + o(\theta - \theta_0) \big] \big[ r(x,\theta_0) + \frac{\partial r}{\partial \theta}(x,\theta_0)^\top (\theta - \theta_0) + o(\theta - \theta_0)  \big]\\
	= & \E_x\bigg[ \big( e^{\theta_0^\top Y_1} + e^{\theta_0^\top Y_1}Y_1^\top(\theta - \theta_0) + o(\theta-\theta_0)\big) \big[ r(X_1,\theta_0) + \frac{\partial r}{\partial \theta}(X_1,\theta_0)^\top(\theta - \theta_0) + o(\theta - \theta_0)  \big]  \bigg] .
\end{aligned}
\]
The leading order terms are canceled. Then comparing $O(\theta-\theta_0)$ term leads to 
\[ 
e^{\Lambda(\theta_0)}\frac{\partial r}{\partial \theta}(x,\theta_0) + e^{\Lambda(\theta_0)}\Lambda'(\theta_0)r(x,\theta_0) = \E_x\big[ e^{\theta_0^\top Y_1}\frac{\partial r}{\partial \theta}(X_1,\theta_0) + e^{\theta_0 Y_1}Y_1r(X_1,\theta_0)  \big].
\]
Denote $\Delta(x,\theta) = \frac{\partial r}{\partial \theta}(x,\theta)$, then the above equality becomes
\[ \Delta(x,\theta_0) + \Lambda'(\theta_0) = \E_x\big[e^{\theta_0^\top Y_1 - \Lambda(\theta_0)}\frac{r(X_1,\theta_0)}{r(x,\theta_0)} \big(\Delta(X_1,\theta_0) + Y_1  \big) \big] . \]
Recall the definition of the probability $\p_{\theta}$ in \eqref{eq:eigenvalue}, we obtain
\[ \E^{\p_{\theta_0}}_x[\Delta(X_1,\theta_0) - \Delta(x,\theta_0) + Y_1] = \Lambda'(\theta_0). \]
Moreover, since the right hand side does not depend on the initial state $x$, we have taken the initial state as any arbitrary distribution $\nu$. Then 
\[ \E^{\p_{\theta_0}}_{\nu}[\Delta(X_1,\theta_0) - \Delta(x,\theta_0) + Y_1] = \Lambda'(\theta_0). \]
In particular, if we take the initial distribution to be the stationary distribution under $\p_{\theta}$, denoted by $\pi_{\theta}$, then $\E^{\p_{\theta_0}}_{\pi_{\theta}}[\Delta(X_1,\theta_0) - \Delta(x,\theta_0)] = 0$, hence $\E_{\pi_{\theta}}^{\p_{\theta}}[Y_1] = \Lambda'(\theta)$.

In particular, take $\theta_0 = 0$ to obtain
\[\E_x[\Delta(X_1,0) - \Delta(x,0)] = \Lambda'(0) - \E_x[Y_1] = \E_{\pi}[Y_1] - \E_x[Y_1],  \]
where $\pi$ is the stationary distribution of the underlying Markov chain $\{X_n,n \geq 0\}$. The left hand side can be represented as $(\mathcal{P}_0 - I)\Delta(x,0)$, where $I$ is the identity operator and $\mathcal{P}_0$ is the operator defined in \eqref{eq:operator}. $\hfill \Box$~\\

\subsection{Proof of Theorem \ref{thm:log efficiency}}
\label{appendix:proof log efficiency}
\begin{enumerate}[(i)]
	\item We first consider the fixed-time event. 
	
	Under the asymptotic normal regime, by the Borjesson-Sundberg bound on the Mills ratio, we have 
	\[  \Phi(-\ell) \geq \frac{\ell}{1 + \ell^2}\varphi(\ell) = \frac{\ell}{(1+\ell^2)\sqrt{2\pi\sigma^2}} e^{-\frac{\ell^2}{2}}  . \]
	To prove the bounded relative or logarithmic efficiency, it suffices to find suitable upper bound for the variance of the estimator.

	When $S_n = \sum_{k=1}^n Y_k$, and $\{(X_n,S_n),n\geq 0\}$ forms a Markov random walk as described in Section \ref{sec:2}. The classical exponential tilting family is given by \eqref{tilt}, with $r(\cdot,\theta)$ and $\Lambda(\theta)$ eigenfunction and eigenvalue respectively. Note that under Assumptions \ref{ass:convex} and \ref{ass:eigenvalue exist},  $\Lambda(\theta)$ is analytic for $\theta \in \Theta$. 
	
	Without loss of generality, we assume $\ell>0$, and $\sqrt{n} \frac{S_n}{n} \to \mathcal{N}(0,1)$
	in the later development, and restrict to the one-dimensional case $S_n\in \mathbb{R}$. 
	Here we assume the stationary mean of $S_1$ is $0$ and the stationary variance of $S_1$ is $1$. 
	Under Assumptions \ref{ass:convex} to \ref{ass:eigenvalue exist}, by Theorem 17.0.1 of \cite{meyn2012markov}, we have $ \frac{S_n}{\sqrt{n}} \to \mathcal{N}(0,1)$.
	
	By Theorem 4.1
	of \cite{ney1987markov1}, we have the relation between the eigenvalue and the mean and variance of the limiting distribution as
	\[ 0 = \E_{\pi}[Y_1] = \Lambda'(0),\ 1 = \Lambda''(0) . \] 
	And we still can define the Legendre transformation of $\Lambda$ by 
	\[ \hat\Lambda(c) = \sup_{\theta\geq 0}\{ c\theta - \Lambda(\theta) \} = c {\Lambda'}^{-1}(c) - \Lambda\left( {\Lambda'}^{-1}(c) \right).\]

	Recall that the importance sampling estimator in the classical exponential tilting, $\hat Z(n,\ell;\theta)$, is to sample under $\p_{\theta}$, with $\theta > 0$ and is given by $\one_{\{ \sqrt{n}\frac{S_n}{n} > \ell  \}} \frac{r(X_0,\theta)}{r(X_n,\theta)} e^{-\theta S_n +  n\Lambda(\theta)}$. It has variance
	\[\begin{aligned}
		\operatorname{Var}\left( \hat Z(n,\ell;\theta) \right) = & \E_{\nu}^{\p_{\theta}}[ \one_{\{ \sqrt{n}\frac{S_n}{n} > \ell  \}} e^{-2\theta S_n + 2n\Lambda(\theta) + 2 \log r(X_0,\theta) - 2\log r(X_n,\theta)} ] - \left( z(n,x) \right)^2 \\
		\leq & \E_{\nu}^{\p_{\theta}}[ \one_{\{ \sqrt{n}\frac{S_n}{n} > \ell  \}} e^{-2\theta S_n +2 n\Lambda(\theta) + 2 \log r(X_0,\theta) - 2\log r(X_n,\theta) } ] \\
		\leq &  \exp\left\{ -2n \left[ \frac{\ell}{\sqrt{n}} \theta -\Lambda(\theta) \right] \right\} \E_{\nu}^{\p_{\theta}}\left[ \left( \frac{r(X_0,\theta)}{r(X_n,\theta)} \right)^2  \right]  .
	\end{aligned}  \]
	By picking $\theta_{n,\ell}^* = {\Lambda'}^{-1}(\frac{\ell}{\sqrt{n}})$, we have 
	\[ \operatorname{Var}\left( \hat Z(n,\ell;\theta_{n,\ell}^*) \right) \leq  \exp\left\{-2n \hat{\Lambda}(\frac{\ell}{\sqrt{n}})  \right\} \E_{\nu}^{\p_{\theta_{n,\ell}^*}}\left[ \left( \frac{r(X_0,\theta_{n,\ell}^*)}{r(X_n,\theta_{n,\ell}^*)} \right)^2  \right] .\]
	
	Moreover, by Taylor's expansion, we have 
	\[ \theta_{n,\ell}^* = \left[ \frac{1}{\Lambda''\left( {\Lambda'}^{-1}(0) \right)} \frac{\ell}{\sqrt{n}} + o(\frac{\ell}{\sqrt n})  \right] = \frac{\ell}{\sqrt n} + o(\frac{\ell}{\sqrt n}) \to 0 .\]
	Hence
	\[-2n\hat{\Lambda}(\frac{\ell}{\sqrt{n}}) = - 2n \left[ \hat{\Lambda}(0) + \hat\Lambda'(0)\frac{\ell}{\sqrt{n}} + \frac{1}{2}\hat\Lambda''(0)\frac{\ell^2}{n} + o(\frac{\ell^2}{n})  \right] . \]
	In addition, Assumption \ref{ass:eigenvalue bounded} implies $\E_{\nu}^{\p_{\theta_{n,\ell}^*}}\left[ \left( \frac{r(X_0,\theta_{n,\ell}^*)}{r(X_n,\theta_{n,\ell}^*)} \right)^2  \right]  \leq C$ for some constant $C$ as $\theta^*_{n,\ell}\to 0$.

	Note that as $\ell\to \infty, n \to \infty, \ell/\sqrt{n}\to 0$, $\theta^*_{n,\ell}\to 0$. Therefore, 
	\[\begin{aligned}
		& \limsup_{\ell\to \infty, n \to \infty, \ell/\sqrt{n}\to 0} \frac{\operatorname{Var}\left( \hat Z(n,\ell;\theta_{n,\ell}^*) \right)}{\left( \Phi(-\ell) \right)^{2-\epsilon}  } \\
		\leq &  \limsup_{\ell\to \infty, n \to \infty, \ell/\sqrt{n}\to 0} C \left( \frac{(1+\ell^2)\sqrt{2\pi}}{\ell}\right)^{2-\epsilon} e^{-2n[\hat{\Lambda}(\frac{\ell}{\sqrt{n}}) - \frac{1}{2}\frac{\ell^2}{n} + \frac{\epsilon}{4} \frac{\ell^2}{n} ]} \\
		= &  \limsup_{\ell\to \infty, n \to \infty, \ell/\sqrt{n}\to 0} C \left( \frac{(1+\ell^2)\sqrt{2\pi}}{\ell}\right)^{2-\epsilon} e^{-2n[o(\frac{\ell^2}{n}) + \frac{\epsilon}{4} \frac{\ell^2}{n} ]}  = 0 ,
	\end{aligned} \]
	where the last limit equals to zero because for sufficiently large $n,\ell$, we have $n[o(\frac{\ell^2}{n}) + \frac{\epsilon}{4} \frac{\ell^2}{n} ] \geq \frac{\epsilon}{8}\ell^2$. Hence the sequence of estimator $\{ \hat Z(n,\ell;\theta_{n,\ell}^*) \}$ has logarithmic efficiency. 
	
	Next, we show the sequence of estimator $\{ \hat Z(n,\ell;\theta_{n,\ell}^*) \}$ belongs to the LAN family (based on Definition \ref{def:lan}). The log-likelihood ratio is 
	\begin{eqnarray}
		&~& \theta^*_{n,\ell} S_n - n \Lambda(\theta^*_{n,\ell}) +  \log r(X_n,\theta^*_{n,\ell}) - \log r(X_0,\theta^*_{n,\ell}) \nonumber \\
		&= & {\Lambda'}^{-1}(\frac{\ell}{\sqrt{n}}) S_n - n \Lambda\left( {\Lambda'}^{-1}(\frac{\ell}{\sqrt{n}}) \right) + \log r\left( X_n, {\Lambda'}^{-1}(\frac{\ell}{\sqrt{n}}) \right) - \log r\left( X_0, {\Lambda'}^{-1}(\frac{\ell}{\sqrt{n}}) \right) \nonumber \\
		&= & \left[ \frac{1}{\Lambda''\left( {\Lambda'}^{-1}(0) \right)} \frac{\ell}{\sqrt{n}} + o(\frac{1}{\sqrt n})  \right] S_n  - n \left[ \frac{1}{2}\Lambda''(0) \left[  \frac{1}{\Lambda''\left( {\Lambda'}^{-1}(0) \right)} \frac{\ell}{\sqrt{n}} + o(\frac{1}{\sqrt n}) \right]^2 +  o(\frac{1}{n}) \right] \nonumber \\
		& ~&+ \left[ \frac{\partial r}{\partial \theta}(X_n, 0) -\frac{\partial r}{\partial \theta}(X_0, 0) \right]  \left[ \frac{1}{\Lambda''\left( {\Lambda'}^{-1}(0) \right)} \frac{\ell}{\sqrt{n}} + o(\frac{1}{\sqrt n})  \right] \nonumber \\
		&= & \ell \frac{1}{\sqrt n} S_n + O(\frac{1}{\sqrt n}) - \frac{\ell^2}{2} + o(1) \to \mathcal{N}( -\frac{\ell^2}{2}, \ell^2 )~{\rm as}~ n \to \infty. \label{an}
	\end{eqnarray}
	Note that under Assumptions \ref{ass:convex} to \ref{ass:eigenvalue exist}, the asymptotic normality in \eqref{an}  comes from Theorem 17.0.1 of \cite{meyn2012markov}.

	Now suppose there is another sequence of estimator $\{ \hat Z_{opt}(n,\ell) \}$ belonging to the LAN family and achieving the smallest asymptotic variance. Then it means 
	\[ \limsup_{\ell\to \infty, n \to \infty, \ell/\sqrt{n}\to 0} \frac{\operatorname{Var}\left( \hat Z_{opt}(n,\ell) \right)}{\left( \Phi(-\ell) \right)^{2-\epsilon}  } \leq \limsup_{\ell\to \infty, n \to \infty, \ell/\sqrt{n}\to 0} \frac{\operatorname{Var}\left( \hat Z(n,\ell;\theta_{n,\ell}^*) \right)}{\left( \Phi(-\ell) \right)^{2-\epsilon}  } . \]
	Hence $\{ \hat Z_{opt}(n,\ell) \}$ also has logarithmic efficiency.
	
	\item Next we consider the first-passage time event.

	When $S_n = \sum_{k=1}^n Y_k$, and $\{(X_n,S_n),n\geq 0\}$ forms a Markov random walk as described in Section \ref{sec:2}. Recall that the importance sampling estimator in the classical exponential tilting, $\hat Z(b,\ell;\theta)$, is to sample under $\p_{\theta}$, with $\theta > 0$ and is given by $\one_{\{ \tau_b < \sqrt{\frac{b\sigma^2}{\mu^3}}\ell + \frac{b}{\mu}  \}} \frac{r(X_0,\theta)}{r(X_{\tau_b},\theta)} e^{-\theta S_{\tau_b} +  \tau_b \Lambda(\theta)}$, where $\theta\geq 0$. Since $\Lambda'(0) = \mu > 0$, $\Lambda''(0) = \sigma^2 <\infty$, we know when $\theta > 0$ is sufficiently close to 0, $\Lambda(\theta) > 0$.
	
	In this case, the importance sampling estimator in the classical exponential tilting has variance
	\[\begin{aligned}
		\operatorname{Var}\left( \hat Z(b,\ell;\theta) \right) = & \E_{\nu}^{\p_{\theta}}[ \one_{\{ \tau_b < \sqrt{\frac{b\sigma^2}{\mu^3}}\ell + \frac{b}{\mu} \}} e^{-2\theta S_{\tau_b} + 2\tau_b\Lambda(\theta) + 2 \log r(X_0,\theta) - 2\log r(X_{\tau_b},\theta)} ] - \left( z(n,b) \right)^2 \\
		\leq & \E_{\nu}^{\p_{\theta}}[ \one_{\{ \tau_b < \sqrt{\frac{b\sigma^2}{\mu^3}}\ell + \frac{b}{\mu} \}} e^{-2\theta b +2 [\sqrt{\frac{b\sigma^2}{\mu^3}}\ell + \frac{b}{\mu}]\Lambda(\theta) + 2 \log r(X_0,\theta) - 2\log r(X_{\tau_b},\theta) } ] \\
		\leq &  \exp\left\{ -2\left( \sqrt{\frac{b\sigma^2}{\mu^3}}\ell + \frac{b}{\mu} \right) \left[ \frac{b}{\sqrt{\frac{b\sigma^2}{\mu^3}}\ell + \frac{b}{\mu}} \theta - \Lambda(\theta) \right] \right\} \E_{\nu}^{\p_{\theta}}\left[ \left( \frac{r(X_0,\theta)}{r(X_{\tau_b},\theta)} \right)^2  \right]  .
	\end{aligned}  \]
	By picking $\theta_{b,\ell}^* = {\Lambda'}^{-1}(\frac{b}{\sqrt{\frac{b\sigma^2}{\mu^3}}\ell + \frac{b}{\mu}}) = {\Lambda'}^{-1}(\frac{\mu}{\sigma \ell/\sqrt{b \mu} + 1}) = \frac{1}{\sigma^2}\left( - \frac{\sqrt{\mu} \sigma}{\sqrt{b}}\ell + \frac{\sigma^2}{2b}\ell + o(\frac{1}{b})  \right) + O(\frac{\ell}{b})$, we have 
	\[ \operatorname{Var}\left( \hat Z(b,\ell;\theta_{b,\ell}^*) \right) \leq  \exp\left\{-2\left( \sqrt{\frac{b\sigma^2}{\mu^3}}\ell + \frac{b}{\mu} \right) \hat{\Lambda}\left(\frac{\mu}{\sigma \ell/\sqrt{b \mu} + 1}  \right)  \right\} \E_{\nu}^{\p_{\theta_{b,\ell}^*}}\left[ \left( \frac{r(X_0,\theta_{b,\ell}^*)}{r(X_{\tau_b},\theta_{b,\ell}^*)} \right)^2  \right] .\]
	
	Moreover, by Taylor's expansion,
	\[\begin{aligned}
		\hat{\Lambda}(\frac{\mu}{\sigma \ell/\sqrt{b \mu} + 1}) = & {\Lambda'}^{-1}(\frac{\mu}{\sigma \ell/\sqrt{b \mu} + 1}) \frac{\mu}{\sigma \ell/\sqrt{b \mu} + 1} - \Lambda\left( {\Lambda'}^{-1}(\frac{\mu}{\sigma \ell/\sqrt{b \mu} + 1})  \right) \\
		= & {\Lambda'}^{-1}(\frac{\mu}{\sigma \ell/\sqrt{b \mu} + 1}) \left[\mu - \frac{\sqrt{\mu} \sigma}{\sqrt{b}}\ell + \frac{\sigma^2}{2b}\ell + o(\frac{1}{b})  \right] \\
		& - \mu {\Lambda'}^{-1}(\frac{\mu}{\sigma \ell/\sqrt{b \mu} + 1}) - \frac{\sigma^2}{2}\left( {\Lambda'}^{-1}(\frac{\mu}{\sigma \ell/\sqrt{b \mu} + 1}) \right)^2 \\
		= & \left[ \frac{1}{\sigma^2}\left( - \frac{\sqrt{\mu} \sigma}{\sqrt{b}}\ell + \frac{\sigma^2}{2b}\ell + o(\frac{1}{b})  \right) + O(\frac{1}{b})\right] \left[ - \frac{\sqrt{\mu} \sigma}{\sqrt{b}}\ell + \frac{\sigma^2}{2b}\ell + o(\frac{1}{b})  \right] \\
		& - \frac{\sigma^2}{2}\left[ \frac{1}{\sigma^2}\left( - \frac{\sqrt{\mu} \sigma}{\sqrt{b}}\ell + \frac{\sigma^2}{2b}\ell + o(\frac{1}{b})  \right) + O(\frac{1}{b})\right]^2 \\
		= & \frac{\mu}{2b} \ell^2 + o(\frac{\ell^2}{b}) .
	\end{aligned}  \]
	Hence
	\[ -2\left( \sqrt{\frac{b\sigma^2}{\mu^3}}\ell + \frac{b}{\mu}\right) \hat{\Lambda}(\frac{\mu}{\sigma \ell/\sqrt{b \mu} + 1}) = -2\left( \sqrt{\frac{b\sigma^2}{\mu^3}}\ell + \frac{b}{\mu}\right) \left(\frac{\mu}{2b} \ell^2 + o(\frac{1}{b}) \right) = -\ell^2 + O(\frac{\ell}{\sqrt b}) . \]
	In addition, Assumption \ref{ass:eigenvalue bounded} implies $\E_{\nu}^{\p_{\theta_{b,\ell}^*}}\left[ \left( \frac{r(X_0,\theta_{b,\ell}^*)}{r(X_{\tau_b},\theta_{b,\ell}^*)} \right)^2  \right] \leq C$ for some constant $C$ as $\theta^*_{b,\ell}\to 0$.
	
	
	Note that as $\ell\to-\infty, b\to \infty, \ell/\sqrt{b}\to 0$, $\theta^*_{b,\ell}\to 0$. Therefore, 
	\[\begin{aligned}
		& \limsup_{\ell\to -\infty, n \to \infty, \ell/\sqrt{n}\to 0} \frac{\operatorname{Var}\left( \hat Z(n,\ell;\theta_{b,\ell}^*) \right)}{\left( \Phi(\ell) \right)^{2-\epsilon}  } \\
		\leq &  \limsup_{\ell\to -\infty, n \to \infty, \ell/\sqrt{n}\to 0} C \left( \sqrt{2\pi \ell}\right)^{2-\epsilon} e^{-2n[\hat{\Lambda}(\frac{\ell}{\sqrt{n}}) - \frac{1}{2}\frac{\ell^2}{n} + \frac{\epsilon}{4} \frac{\ell^2}{n} ]} \\
		= &  \limsup_{\ell\to -\infty, n \to \infty, \ell/\sqrt{n}\to 0} C \left( \sqrt{2\pi \ell}\right)^{2-\epsilon} e^{-2n[o(\frac{\ell^2}{n}) + \frac{\epsilon}{4} \frac{\ell^2}{n} ]}  = 0 ,
	\end{aligned} \]
	for the same reason as in the fixed-time event case. Hence the sequence of estimator $\{ \hat Z(b,\ell;\theta_{b,\ell}^*) \}$ has logarithmic efficiency. 
	
	Next, we show the sequence of estimator $\{ \hat Z(b,\ell;\theta_{b,\ell}^*) \}$ belongs to the LAN family (based on Definition \ref{def:lan} (ii)). Denote $R(b) = S_{\tau_b} - b$ as the overshoot.  The log-likelihood ratio is 
	\begin{eqnarray}
		& ~&\theta_{b,\ell}^* S_{\tau_b} - \tau_b \Lambda(\theta_{b,\ell}^*) +  \log r(X_{\tau_b},\theta^*_{b,\ell}) - \log r(X_0,\theta^*_{b,\ell}) \nonumber \\
		&=&  {\Lambda'}^{-1}(\frac{\mu}{\sigma \ell/\sqrt{b \mu} + 1}) [b + R(b)] - \tau_b \Lambda\left( {\Lambda'}^{-1}(\frac{\mu}{\sigma \ell/\sqrt{b \mu} + 1}) \right)\nonumber \\
		&~& + \left[ \frac{\partial r}{\partial \theta}(X_{\tau_b}, 0) -\frac{\partial r}{\partial \theta}(X_0, 0) \right]  \left[ \frac{1}{\sigma^2}\left( - \frac{\sqrt{\mu} \sigma}{\sqrt{b}}\ell + \frac{\sigma^2}{2b}\ell + o(\frac{1}{b})  \right) + O(\frac{1}{b})  \right]\nonumber \\
		&= & {\Lambda'}^{-1}(\frac{\mu}{\sigma \ell/\sqrt{b \mu} + 1}) [b + R(b)]  - \left[ \frac{\tau_b - b/\mu}{\sqrt{b\sigma^2/\mu^3}} \sqrt{\frac{b\sigma^2}{\mu^3}}  + \frac{b}{\mu} \right] \bigg[ \mu {\Lambda'}^{-1}(\frac{\mu}{\sigma \ell/\sqrt{b \mu} + 1}) \nonumber \\
		&~& + \frac{\sigma^2}{2}\left(  {\Lambda'}^{-1}(\frac{\mu}{\sigma \ell/\sqrt{b \mu} + 1}) \right)^2 + o(\frac{1}{b})\bigg] 
		+ \left[ \frac{\partial r}{\partial \theta}(X_{\tau_b}, 0) -\frac{\partial r}{\partial \theta}(X_0, 0) \right]  \left[ -\frac{\sqrt\mu}{\sigma\sqrt b}\ell + O(\frac{1}{b})  \right]\nonumber \\
		&= &  \left[ -\frac{\sqrt\mu}{\sigma\sqrt b}\ell + O(\frac{1}{b})  \right] R(b)  - \frac{b\sigma^2}{2\mu} \left( \frac{\mu}{b\sigma^2}\ell^2 + o(\frac{1}{b}) \right) - \frac{\tau_b - b/\mu}{\sqrt{b\sigma^2/\mu^3}} \sqrt{\frac{b\sigma^2}{\mu^3}} \left(\frac{\mu}{\sigma^2}(-\frac{\sqrt{\mu}\sigma}{\sqrt b}\ell) + O(\frac{1}{b})  \right)\nonumber \\
		&~& + \left[ \frac{\partial r}{\partial \theta}(X_{\tau_b}, 0) -\frac{\partial r}{\partial \theta}(X_0, 0) \right]  \left[ -\frac{\sqrt\mu}{\sigma\sqrt b}\ell + O(\frac{1}{b})  \right]\nonumber \\
		&= & R(b)O(\frac{1}{b}) - \frac{1}{2}\ell^2 + \frac{\tau_b - b/\mu}{\sqrt{b\sigma^2/\mu^3}}\ell + O(\frac{1}{b}). \label{an1}
	\end{eqnarray}
	To handle the asymptotic behavior in \eqref{an1}, we first have that 
	under Assumptions \ref{ass:convex} to \ref{ass:eigenvalue exist}, Theorems 2 and 3 of \cite{fuh2004renewal} show that $\{\tau < n, S_n < \gamma n^{1/2} \}$, for a suitable chosen $\gamma$, satisfies the asymptotic normality.
	Note that $\p_\pi (\tau < n) = \p_\pi (S_n \geq b) + \p_\pi(\tau < n, S_n < b)$. By using the central limit theorem of $\p_\pi(S_n \geq b)$ for a suitable chosen $b$, 
	we have $\frac{\tau_b - b/\mu}{\sqrt{b\sigma^2/\mu^3}} \to \mathcal{N}(0,1)$ as $b\to \infty$.
	Next, we consider the overshoot $R(b)$. Under Assumptions \ref{ass:convex} to \ref{ass:eigenvalue exist}, by Lemma 2 of \cite{fuh2004renewal}, we have $\sup_{b \geq 0} E_\pi R(b) \leq E_\pi (S_1^+)^2/E_\pi S_1,$  thus $R(b)O(\frac{1}{b}) \to 0$ in probability as $b \to \infty$. Therefore, $\theta_{b,\ell}^* S_{\tau_b} - \tau_b \psi(\theta_{b,\ell}^*) \to \mathcal{N}(-\frac{1}{2}\ell^2, \ell^2)$. 
	
	Now suppose there is another sequence of estimator $\{ \hat Z_{opt}(b,\ell) \}$ belonging to the LAN family and achieving the smallest asymptotic variance. Then it means 
	\[\limsup_{\ell\to -\infty, n \to \infty, \ell/\sqrt{n}\to 0} \frac{\operatorname{Var}\left( \hat Z_{opt}(b,\ell) \right)}{\left( \Phi(\ell) \right)^{2-\epsilon}  } \leq \limsup_{\ell\to -\infty, n \to \infty, \ell/\sqrt{n}\to 0} \frac{\operatorname{Var}\left( \hat Z(n,\ell;\theta_{b,\ell}^*) \right)}{\left( \Phi(\ell) \right)^{2-\epsilon}  }  . \]
	Hence $\{ \hat Z_{opt}(b,\ell) \}$ also has logarithmic efficiency.
	$\hfill \Box$~\\

\end{enumerate}			

\subsection{Proof of Theorem \ref{thm:ld log efficiency}}
We only illustrate the fixed-time event here, and the proof for the first-passage-time event is similar. 

From the proof of Theorem \ref{thm:log efficiency}, we have 
\[ \operatorname{Var}\left( \hat Z(n,\ell;\theta_{n,\ell}^*) \right) \leq  \exp\left\{-2n \hat{\Lambda}(\frac{\ell}{\sqrt{n}})  \right\} \E_{\nu}^{\p_{\theta_{n,\ell}^*}}\left[ \left( \frac{r(X_0,\theta_{n,\ell}^*)}{r(X_n,\theta_{n,\ell}^*)} \right)^2  \right] ,\]
where $\theta_{n,\ell}^* = {\Lambda'}^{-1}(\frac{\ell}{\sqrt{n}})$. Under Assumption \ref{ass:uniform recurrent}, by \citet[Example 1]{chan2011sequential}, the eigenfunction $r(x,\theta)$ is uniformly positive and bounded for all $\theta\in \Theta$. Hence, $\operatorname{Var}\left( \hat Z(n,\ell;\theta_{n,\ell}^*) \right) \leq C \exp\left\{-2n \hat{\Lambda}(\frac{\ell}{\sqrt{n}})  \right\} $.

On the other hand, let us examine a lower bound for $z(n,\ell)$. Note that
\[ \begin{aligned}
	z(n,\ell) = & \E_{\nu}^{\p_{\theta_{n,\ell}^*}}[ \one_{\{ \sqrt{n}\frac{S_n}{n} > \ell  \}} e^{-\theta_{n,\ell}^* S_n + n\Lambda(\theta_{n,\ell}^*) }\frac{r(X_0,\theta_{n,\ell}^*)}{ r(X_n,\theta_{n,\ell}^*)} ] \\
	\geq & C' e^{n\Lambda(\theta_{n,\ell}^*) - \theta_{n,\ell}^*\sqrt{n}\ell }  \E_{\nu}^{\p_{\theta_{n,\ell}^*}}[ \one_{\{ 1 > \frac{S_n - \sqrt{n} \ell}{\sqrt{n}} > 0  \}} e^{-\theta_{n,\ell}^* (S_n - \sqrt{n}\ell) }] \\
	\geq & C' e^{n\Lambda(\theta_{n,\ell}^*) - \theta_{n,\ell}^*\sqrt{n}\ell -  \theta_{n,\ell}^*\sqrt{n} } = C'\exp\{ -n \hat{\Lambda}(\theta_{n,\ell}^*) -  \theta_{n,\ell}^*\sqrt{n} \} .
\end{aligned} \]

Recall that $\theta_{n,\ell}^* = {\Lambda'}^{-1}(\frac{\ell}{\sqrt{n}}) \to {\Lambda'}^{-1}(c)$, therefore,
\[\begin{aligned}
	& \limsup_{\ell \to \infty, n\to \infty, \ell/\sqrt{n} \to c}	 \frac{ \operatorname{Var}\left( \hat Z(n,\ell) \right)}{\left( z(n,\ell) \right)^{2-\varepsilon} } \\
	\leq & \limsup_{\ell \to \infty, n\to \infty, \ell/\sqrt{n} \to c} \frac{C}{{C'}^2}\exp\{ -\epsilon n[ \hat{\Lambda}(\theta_{n,\ell}^*) - \frac{2-\epsilon}{\epsilon}\frac{1}{\sqrt{n}} \theta_{n,\ell}^* ] \} = 0 .
\end{aligned}   \]

From the proof of Theorem \ref{thm:log efficiency}, $\hat Z(n,\ell;\theta_{n,\ell}^*)$ belongs to the same parametric family as $\{ \hat Z_{opt}(n,\ell) \}$, which minimizes the variance for every $(n,\ell)$. Then it means 
\[ \limsup_{\ell\to \infty, n \to \infty, \ell/\sqrt{n}\to 0} \frac{\operatorname{Var}\left( \hat Z_{opt}(n,\ell) \right)}{\left( z(n,\ell) \right)^{2-\epsilon}  } \leq \limsup_{\ell\to \infty, n \to \infty, \ell/\sqrt{n}\to 0} \frac{\operatorname{Var}\left( \hat Z(n,\ell;\theta_{n,\ell}^*) \right)}{\left( z(n,\ell) \right)^{2-\epsilon}  } . \]
Hence $\{ \hat Z_{opt}(n,\ell) \}$ also has logarithmic efficiency.
\end{document}